\documentclass[twoside,12pt]{article}
\usepackage{graphicx,amssymb,amsmath,epsfig,cite}
\usepackage{array}
\usepackage{bm}
\usepackage{subfigure}
\usepackage{wrapfig}

\newcommand{\be}{\begin{equation}}
\newcommand{\ee}{\end{equation}}
\newcommand{\bea}{\begin{eqnarray}}
\newcommand{\eea}{\end{eqnarray}}

\newcommand{\CN}{{\cal N}}
\newcommand{\bear}{\begin{eqnarray}}
\newcommand{\eear}{\end{eqnarray}}

\newcommand{\del}{\partial}
\newcommand{\dd}{{\rm d}}
\newcommand{\eee}{{\rm e}}

\renewcommand{\theequation}
{\arabic{section}.\arabic{equation}}

\makeatletter
\@addtoreset{equation}{section}
\makeatother

\makeatletter
\def\eqnarray{ \stepcounter{equation} \let\@currentlabel=\theequation
 \global\@eqnswtrue
 \global\@eqcnt\z@
 \tabskip\@centering
 \let\\=\@eqncr
 $$\halign to \displaywidth\bgroup\@eqnsel\hskip\@centering
 $\displaystyle\tabskip\z@{##}$&\global\@eqcnt\@ne
 \hfil$\displaystyle{{}##{}}$\hfil
 &\global\@eqcnt\tw@$\displaystyle\tabskip\z@{##}$\hfil
 \tabskip\@centering&\llap{##}\tabskip\z@\cr}
\makeatother

\begin{document}

\title{
\hspace*{145mm}{\normalsize \tt APCTP-Pre2012-007} \\
Holographic QCD: Past, Present, and Future}
\author{
Youngman Kim,$^{1,2,3}$ Ik Jae Shin,$^{1,2}$ and Takuya Tsukioka$^{1,2}$\\
\\
$^1$Asia Pacific Center for Theoretical Physics, \\
Pohang, Gyeongbuk 790-784, Korea \\
$^2$ Institute for Basic Science,\\
Yuseong-daero 1689-gil 70, Yuseong-gu, Daejeon 305-811, Korea
  \\
$^3$ Department of Physics, Pohang University of Science and
Technology, \\
Pohang, Gyeongbuk 790-784, Korea\\}

\date{\hfill}

\maketitle
\begin{abstract}
At the dawn of a new theoretical tool based on the
AdS/CFT correspondence for nonperturbative aspects of quantum chromodynamics,
we give an interim review on the new tool, holographic QCD,
with some of its accomplishment.
 We try to give an  A-to-Z picture of the holographic QCD,
from string theory to a few selected top-down holographic QCD models
with one or two physical applications in each model.
We may not attempt to collect diverse results from various
holographic QCD model studies.
\end{abstract}
\tableofcontents

\section{Introduction}

Quantum chromodynamics (QCD) is widely believed to be
the fundamental theory of the strong interaction.
Despite its tremendous success in high energy hadronic phenomenology
based on perturbation theory, QCD is not directly applicable
to every physical system governed by the strong interaction.
Especially, at low energies due to the largeness of the coupling
constant involved, the standard perturbation method is not viable in most cases.
There have been many interesting and (partly if not fully) qualified
approaches for the strong interaction at low energy.
In this article, we review one of them which is flourishing these days
in describing QCD (or its cousins) in a nonperturbative regime.

Based on the AdS/CFT correspondence (or the gauge/gravity duality,
its generalized version)
\cite{Maldacena:1997re, Gubser:1998bc,Witten:1998qj},
interesting and important attempts have been made to understand
nonperturbative aspects of QCD under the name ``holographic QCD'',
see~\cite{ babington,Brodsky:2003px, Kruczenski:2003uq,Sakai:2004cn,Erlich:2005qh,
DaRold:2005zs,Hirn:2005vk,Brodsky:2006uqa} for some prototypical models.
There are in general two different routes to arrive at the holographic
dual description of QCD.
One way is, so-called, a top-down approach based on stringy
D-brane configurations.
The other way is a bottom-up approach, in which a 5D holographic QCD
model is constructed from QCD.

In this interim review, for those who are not experts
in holographic QCD, we review holographic QCD models starting from
a concise but essential summary of string theory including
the AdS/CFT correspondence.
Then, we will introduce some frequently used top-down models and
present a few selected sample works for QCD (or more precisely
QCD-like) phenomena in each model.
We will not attempt to show innumerable outcomes out of holographic
QCD models, though some of them are quite interesting and influential,
for instance, the universal viscosity to entropy ratio~\cite{Policastro:2001yc}.
For more on holographic QCD, we may refer to
\cite{Erdmenger:2007cm,McGreevy:2009xe,CasalderreySolana:
2011us,deTeramond:2012rt} and \cite{Kim:2011ey} for a summary
focused on results from bottom-up models.

After the introduction of basic string theory and the AdS/CFT
correspondence, we will start with the D3/D7 model.
In this D3/D7 model, we will show a few results on the probe of
the quark-gluon plasma (QGP).
We will then describe D4/D8 model with a touch on holographic
baryons and nuclear forces.
Lastly, we make a brief summary of the D4/D6 model and discuss
nuclear symmetry energy and nuclear to strange matter transition.
To demonstrate the usefulness and versatility of the holographic 
QCD approach and its recent contributions to QCD (or QCD-like) phenomenology,
we briefly mention some of recent results from various holographic
studies in section \ref{sampleW}.
Some remarks will close this write-up.

\section{Basics of string theory}

We start by giving some basic aspects about general relativity and string theory.
In this chapter,
we shall introduce main ingredients and set up the stage for the next sections.
We start with recalling the Einstein's general relativity.
The standard text books can be found in~\cite{string}
for string theory and in~\cite{gravity} for gravity theory.

\subsection{Einstein gravity}

The idea of Einstein gravity is that the spacetime geometry is dynamical.
We work in $D$-dimensional spacetime with metric $G_{MN}(x)$,
\begin{equation}
\dd s^2=G_{MN}\dd x^M\dd x^N,
\label{riemann_metric}
\end{equation}
whose signature is $(- + \cdots +)$.
The dynamics of spacetime can be determined by
\begin{equation}
\label{eh}
S_{\rm EH}
=\frac{1}{16\pi G_{\rm N}^{(D)}}
\!\int_{\cal M}\!\dd^D\!x\sqrt{-G}R
=\frac{1}{2\kappa^2}\!\int_{\cal M}\!\dd^D\!x\sqrt{-G}R,
\end{equation}
where $G=\det G_{MN}$ is the determinant of the metric, and $R(x)$
is the scalar curvature.
The action (\ref{eh}) is called as the Einstein-Hilbert action which
contains second derivative terms of the metric $G_{MN}(x)$.
 $G_{\rm N}^{(D)}(=\kappa^2/(8\pi))$ is a $D$-dimensional
Newton constant and its mass dimension is mass$^{(2-D)}$
which has a negative power for $D>2$.
Depending on the spacetime topologies
and setups to be considered, one could add several possible
diffeomorphism invariant terms such as the boundary terms,
the cosmological constant, and more higher derivative contributions
to the Einstein-Hilbert action.
We define $D$-dimensional Planck
length $l^{(D)}_{\rm P}$ as
$l_{\rm P}^{(D)}=(8\pi G_{\rm N}^{(D)})^{1/(D-2)}$ in
the dimensional analysis of the action (\ref{eh}).

The couplings to matter fields $\psi(x)$ are given through
the metric $G_{MN}(x)$ and  the Christoffel
symbol $\Gamma^M{}_{KL}(x)$\footnote{
For fermions, we need to introduce the local Lorentz frame
defined by vielbein, and the minimal coupling is given through the spin connections.}
which is defined by the metric.
Introducing a matter action as $S_{\rm matter}[\psi, G_{MN}]$,
the variation of the total action $S_{\rm EH}+S_{\rm matter}$ with
respect to the metric $G^{MN}(x)$ gives the Einstein equation
\begin{equation}
\label{einstein}
	R_{MN}-\frac{1}{2}G_{MN}R=8\pi G_{\rm N}^{(D)}T_{MN},
\end{equation}
where $R_{MN}(x)$ is the Ricci tensor and $T_{MN}(x)$
is the energy momentum tensor defined by
\begin{equation}
T_{MN}\equiv-\frac{2}{\sqrt{-G}}
\frac{\delta S_{\rm matter}}{\delta G^{MN}}.
\end{equation}

Let us consider the perturbation around the flat
Minkowski spacetime
$\eta_{MN}=\mbox{diag}(-, +, \cdots, +)$,
\begin{equation}
\label{perturbation_gravity}
G_{MN}(x)=\eta_{MN}+\kappa\widehat{h}_{MN}(x)+{\cal O}(\widehat{h}^2),
\end{equation}
where the coupling constant $\kappa=\sqrt{8\pi G_{\rm N}^{(D)}}$
has been introduced such that the kinetic term of
the graviton $\widehat{h}_{MN}(x)$ is canonically normalized.
Up to total derivatives, the action (\ref{eh}) can be written as
\begin{eqnarray}
S_{\rm EH}=\!\int\!\dd^D\!x
\bigg\{
&&
\frac{1}{2}
(\del_M\widehat{h}^M{}_N)(\del_L\widehat{h}^{LN})
-\frac{1}{2}(\del_M\widehat{h}_L{}^L)
(\del_N\widehat{h}^{MN})
	+\frac{1}{4}(\del_M\widehat{h}_L{}^L)
(\del^M\widehat{h}_K{}^K)-\frac{1}{4}(\del_M\widehat{h}_{NL})
(\del^M\widehat{h}^{NL}) \nonumber \\
	&&+\kappa\Big(\widehat{h}(\del \widehat{h})^2+\cdots\Big)
+{\cal O}(\kappa^2)\bigg\},
\label{eh_pert}
\end{eqnarray}
where the derivative coupling term,
which is proportional to $\kappa$, has been symbolically displayed.
By using the De Donder gauge
$0=\del_N\widehat{h}_M{}^N(x)-\frac{1}{2}\del_M\widehat{h}_N{}^N(x)$
to fix the general coordinate
transformation
$\delta\widehat{h}_{MN}(x)=\del_M\varepsilon_N(x)+\del_N\varepsilon_M(x)$,
we can obtain the graviton propagator\footnote{
As usual, we add the gauge fixing term to the action including
the Faddeev-Popov ghost $\eta_M(x)$
$$ {\cal L}_{\rm gauge-fixing}
=-\frac{1}{2}\Big(\del_N\widehat{h}_M{}^N
-\frac{1}{2}\del_M\widehat{h}_N{}^N\Big)^2
-\del_M\bar\eta_N\del^M\eta^N+{\cal O}(\widehat{h}^2).$$}
%
\begin{equation}
\label{propagator_graviton}
	\langle\widehat{h}_{MN}(-k)\widehat{h}_{KL}(k)\rangle
=-\frac{2i}{k^2}\Big(\eta_{MK}\eta_{NL}+\eta_{ML}\eta_{NK}
-\frac{2}{D-2}\eta_{MN}\eta_{KL}\Big).
 \end{equation}
The action of the matter field is also expanded as
 \begin{equation}
	S_{\rm matter}
=S_{\rm matter}[\psi, \eta_{MN}]
-\!\int\!\dd^D\!x\bigg\{\frac{\kappa}{2}T_{MN}\widehat{h}^{MN}
+{\cal O}(\kappa^2)\bigg\}.
 \end{equation}

Due to the coupling constant with a negative mass dimension,
conventional perturbative methods do not work due to bad behavior at UV.
Only in the low energies
\begin{equation}
\label{low_erergy_00}
\omega\ll \frac{1}{\kappa^{\frac{2}{D-2}}},
\end{equation}
the perturbation, in which the dimensionless coupling for
given characteristic energy scale $\omega$
is $\omega\kappa^{2/(D-2)}$, makes a sense.
In order to realize the low energy limit,
it is sometimes convenient to send the energy scale to infinity
or equivalently take a limit
\begin{equation}
\label{low_energy_01}
	\kappa\to0
 \end{equation}
with keeping the energies constant.
In this low energy regime, the interactions turn to be weaker and under control.

A similar well-known example suffering from the UV divergence
can be found in the electroweak theory.
The electroweak theory was originally described
by the four-Fermi interaction with
a negative mass dimension coupling constant.
This interaction is not renormalizable and breaks the perturbation
theory at sufficient high energies.
In this case, in high energy, we know new degrees of freedom come
into the theory, i.e.\ $W$ and $Z$ bosons which mediate
the contact interaction, and the UV divergence becomes soft.
Indeed, the theory turns to be well-defined in the sense of the renormalization.

String theory is one of the formulations
with well-defined consistent quantum gravity.
Essential departure from the conventional field theory of point particle
is to consider one-dimensional extended object i.e.\ string
as a fundamental variable.
Indeed, new (infinitely many) degrees of freedom
just like $W$ and $Z$ bosons can be introduced.
String world sheets which are formed by the string trajectories
could restrict their amplitudes through their topologies.
As a result, one can calculate consistent finite perturbative
quantum gravity amplitudes.

In string theories, there exist not only strings but also various
extended objects called $p$-brane which is a $p$-dimensional extended object.
These extended objects move in the bulk $D$-dimensional spacetime
and their trajectories become hypersurfaces embedded in the bulk.
From the causality point of view, the hypersurface should be timelike,
meaning that this should include a time direction.
In this notation, for example, a point particle is $0$-brane whose
trajectory describes the world line and a string is $1$-brane which sweeps out
the world sheet in the bulk spacetime.
The 2-brane is a ``membrane'' and the higher $p$-dimensional extended
object is called as ``$p$-brane''.
The free motion of the $p$-branes is described by the minimal action
principle for the length, the surface, and the volumes.
In the next subsections, we formulate these extended objects.
 \begin{figure}[htbb]
 \begin{minipage}{1.0\textwidth}
 \begin{center}
 \includegraphics[clip, width=8cm]{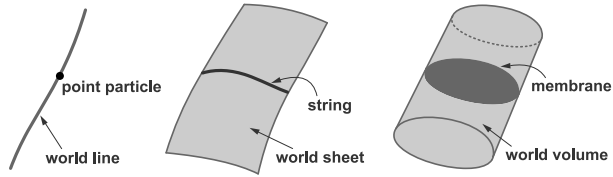}
 \caption{Extended objects}
 \label{extended_objects}
 \end{center}
 \end{minipage}
 \end{figure}

\subsection{Point particle}

The action of the particle with mass $m$ can be given
by the length of the ``timelike'' world line $\gamma$.
In order to obtain the dynamical equation of motion,
we need to introduce the target space coordinates
$X^M(\lambda)$ with $M=0, 1, \cdots,  D-1$ which are
fields on the world line parameterized by $\lambda$.
The coordinates $X^M(\lambda)$  provide the location of the particle
in the spacetime, i.e.\ the embedding coordinates of the world line into the bulk.
Then, the action is given by
\begin{equation}
S_0
=-m\times {\rm (length)}=-m\!\int_\gamma\!\dd s
=-m\!\int_\gamma\!
\sqrt{-\dd X^M\dd X^NG_{MN}(X)}
=-m\!\int_{\lambda_0}^{\lambda_1}
\!\!\!\dd\lambda\sqrt{-g_{\lambda\lambda}(\lambda)},
\label{world_line}
\end{equation}
where we have introduced
\begin{equation}
g_{\lambda\lambda}\big(X(\lambda)\big)
=\frac{\dd X^M}{\dd\lambda}\frac{\dd X^N}{\dd\lambda}G_{MN}(X)
\label{induced_worldline}
\end{equation}
which is the ``induced metric'' on the world line.
Apart from the manifest general covariance,
the action is invariant under the change $\lambda\to\lambda'(\lambda)$
with the boundary conditions $\lambda'(\lambda_0)=\lambda_0$ and
$\lambda'(\lambda_1)=\lambda_1$.

In general, there is no natural choice of the world line
parameterization.
One could fix this invariance in various ways.
One useful choice is the ``static'' gauge
\begin{equation}
\lambda=X^0 \ (\equiv t),
\label{static_gauge}
\end{equation}
which identifies the timelike coordinate of the world line with
the time in the spacetime.
In this gauge, with the flat Minkowski spacetime background
$G_{MN}(x)=\eta_{MN}={\rm diag}(-, +, \cdots, +)$
the action (\ref{world_line}) becomes
\begin{equation}
S_0=-m\!\int_{t_0}^{t_1}\!\!\!\dd t\sqrt{1-\vec{v}^2},
\label{to_nonrela}
\end{equation}
where $\vec{v}\equiv \dd\vec{X}/\dd t$.
If one goes to the nonrelativistic approximation
$|\vec{v}|\ll 1 (=c)$, up to the second order,
the action (\ref{to_nonrela}) gives the potential energy of the rest
mass $m$
and the usual  kinetic energy $m\vec{v}^2/2$ with proper sign.

Another useful gauge fixing is
\begin{equation}
\frac{\dd X^M}{\dd \tau}\frac{\dd X^N}{\dd\tau}G_{MN}
\equiv U^MU_M=-1,
\label{proper_time}
\end{equation}
where we identify $\tau$ as the proper time of the particle and
introduce the velocity $U^M\equiv\dd X^M/\dd\tau$.
In this gauge, the equation of motion for $X^M(\tau)$ could be
the timelike geodesic equation:
\begin{equation}
0=\frac{\dd^2 X^M}{\dd\tau^2}
+\Gamma^M{}_{NL}\frac{\dd X^N}{\dd\tau}\frac{\dd X^L}{\dd\tau}
=U^N\nabla_NU^M,
\label{geodesic}
\end{equation}
where $\Gamma^M{}_{NL}$ is the Christoffel symbol.
In flat Minkowski spacetime or local inertial frames,
where the Christoffel symbol vanishes,
the motion simply becomes free fall with the mass-shell
constraint (\ref{proper_time}).
Thus, the Christoffel symbol determines how the acceleration
could be modified from the straight line of the particle with the velocity.

\subsubsection{Newtonian limit}

We can consider Newton gravity by taking the Newtonian limit
where the weak and static gravitation field and the nonrelativistic velocity
$\dd\vec{X}/\dd\tau\ll\dd t/\dd\tau$ are required.
We expand the static metric ($0=\del_t G_{MN}(x)$)
around flat Minkowski spacetime
\begin{equation}
G_{MN}(x)=\eta_{MN}+{h}_{MN}(x)+{\cal O}({h}^2).
\end{equation}
Then, the geodesic equation (\ref{geodesic}) is reduced to be an
equation of motion under a gravitation potential,
\begin{equation}
 \frac{\dd^2\vec{X}}{\ \dd t^2}=-\vec{\nabla}V(x),
\end{equation}
where the proper time has been eliminated through the choice $\tau=t$.
The Newton potential $V(x)$ is given by the deviation of the
time-time
component of the metric
\begin{equation}
V(x)=-\frac{1}{2}{h}_{tt}(x).
\end{equation}
The Newton potential is determined by the Einstein equation
(\ref{einstein}), which becomes the Poisson equation in the Newtonian
limit,
\begin{equation}
\Delta V(x)=8\pi G_{\rm N}^{(D)}\frac{D-3}{D-2}\ \rho(x),
\end{equation}
where $\rho(x)=T_{tt}(x)$ is the rest energy which is dominant
in $T_{MN}(x)$  ($|T_{ij}|\ll |T_{tt}|$) and $\Delta$  is the Laplacian for
$(D-1)$-dimensional spatial coordinates.
For a point particle with a rest mass $M$  at the origin
i.e.\ $\rho(x)=M\delta^{D-1}(\vec{x})$, one can find the
potential\footnote{
In this normalization, we arrive at the usual form of 4D Newton gravity potential
$V(r)=-G_{\rm N}^{(4)}M/r$.
}
\begin{equation}
V(r)=-\frac{8\pi G_{\rm N}^{(D)}M}{(D-2)V_{S^{(D-2)}}}
\frac{1}{r^{D-3}},
\end{equation}
where $r=|\vec{x}|$ is the radial ``transverse''  to the point particle,
and $V_{S^{d}}$ is the volume of $d$-dimensional
unit sphere. 
Therefore, the presence of the rest point particle with a mass $M$
modifies the flat metric as
\begin{equation}
G_{tt}=-1+\frac{16\pi G_{\rm N}^{(D)}M}{(D-2)V_{S^{(D-2)}}}
\frac{1}{r^{D-3}}
+{\cal O}(1/r^{D-2}).
\label{mass_formula}
\end{equation}
This can be used to observe the mass in asymptotically ($r\to\infty$)
flat spacetime\footnote{
More precisely, we could define the mass in several formalisms
such as ADM mass or Komar mass.
}.
We define  the Schwarzschild radius $r_{\rm S}$ as
\begin{equation}
r_{\rm S}^{D-3}=\frac{16\pi G^{(D)}_{\rm N} M}{(D-2)V_{S^{(D-2)}}}
\label{schwarzschild_radius}
\end{equation}
which gives a characteristic length of spacetime.
This also implies that the backreaction measured by the deviation from the
flat metric can be observed at the scale of the
order $r_{\rm S}$.

This kind of deformation of the spacetime typically appears in black hole
geometries.
The most simplest example is the Schwarzschild black hole
which is a vacuum solution of the Einstein equation:
\begin{equation}
\dd s^2=-\Big(1-\frac{r^{D-3}_{\rm S}}{r^{D-3}}\Big)\ \dd t^2
+\frac{\dd r^2}{\Big(\displaystyle1-\frac{r^{D-3}_{\rm S}}{r^{D-3}}\Big)}
+r^2\dd\Omega^2_{D-2},
\label{schwarzschild_metric}
\end{equation}
where $r_{\rm S}$ is given in (\ref{schwarzschild_radius}) and
$M$ is the mass of Schwarzschild black hole.
This is the static spherical solution and asymptotically flat.
If one approaches to the small $r$ region, one faces the event horizon
located at $r=r_{\rm S}$.

Let us now discuss  the gravitational red shift.
In the curved spacetime, 
the emitted/observed frequencies of a signal in two different points
are different.
By using (\ref{proper_time}) with $X^M=(t, x^i)$,
one observes the relation between the proper time interval $\dd\tau$
and the coordinate time interval $\dd t$
as $\dd\tau=\sqrt{-G_{tt}(x)}\ \dd t$,
where we assumed that the clock is at rest with respect to the reference
frame $\dd x^i=0$.
Assuming that the gravitational field is stationary, we refer to the
coordinate time $t$ as the universal time.
Therefore, the time intervals in two points,
say the points $x_1$ and $x_2$ where the signal is emitted and observed,
respectively, are the same
$\Delta t_1=\Delta t_2$.
The emission frequency of the source defined by
$\omega_1\equiv 1/\Delta\tau_1$ is given as
$\omega_1=1/(\sqrt{-G_{tt}(x_1)}\Delta t_1)$.
Since the same is true for observer at $x_2$, one can conclude that
\begin{equation}
\omega_2
=\frac{\sqrt{-G_{tt}(x_1)}}{\sqrt{-G_{tt}(x_2)}}\ \omega_1.
\label{redshift}
\end{equation}
Applying this to the Schwarzschild metric (\ref{schwarzschild_metric})
at the points $r_1<r_2$,
one obtains the red shift $\omega_2<\omega_1$
because of $|G_{tt}(x_2)|>|G_{tt}(x_1)|$.
Moreover, if the source approaches the horizon $r_1\to r_{\rm S}$,
the observed frequency at $r_2$ tends to zero in the finite frequency $\omega_1$.
The horizon can be characterized as a surface of infinite
red shift.

\subsubsection{Electromagnetic interaction}

Now let us introduce the electromagnetic field in the bulk spacetime.
The action of the Maxwell gauge field $A_1(x)=A_M(x)\dd x^M$ reads as usual,
\begin{equation}
S_{\rm Maxwell}
=-\frac{1}{4}\!\int_{\cal M}\!\!\dd^Dx\sqrt{-G}F^{MN}F_{MN}
=-\frac{1}{2}\!\int_{\cal M}\!\!\dd^Dx\sqrt{-G}|F_2|^2
=-\frac{1}{2}\!\int_{\cal M}\!\!\!F_2\wedge* F_2,
\label{maxwell_1}
\end{equation}
where the field strength is given by
$
\frac{1}{2}F_{MN}\dd x^M\wedge\dd x^N=
F_2=\dd A_1
=\frac{1}{2}\Big(\del_M A_N-\del_N A_M\Big)\dd x^M\wedge\dd x^N.
$
The Maxwell field in the bulk could couple to the particle world line
as
\begin{equation}
S_{0\rm WZ}
=-e\!\int_\gamma\!\dd X^M\!A_M(X)
=-e\!\int_\gamma\!\dd\lambda\frac{\dd X^{M}}{\dd \lambda}A_M(X),
\label{wz_0}
\end{equation}
where $e$ is the charge of the particle and we refer to this sort of
term as Wess-Zumino term.
The gauge invariance of the action (\ref{wz_0}) under
the gauge transformation with a gauge parameter $\theta(x)$,
\begin{equation}
\delta A_M=\del_M \theta\, ,
\end{equation}
requires the world line not to be terminated if there are no sources.
The dynamics of the system is described by the actions  (\ref{world_line}),
(\ref{maxwell_1}), and (\ref{wz_0}).
In order to obtain the equation of motion for the Maxwell field
$A_M(x)$, it is convenient to use the identity
$$
A_M(X)=\!\int_{\cal M}\!\dd^Dx\delta^D(x-X(\lambda))A_M(x),
$$
in which the delta function localizes the Maxwell field $A_M(x)$
to the world line.
Varying the actions with respect to the Maxwell field $A_M(x)$,
we obtain
\begin{equation}
0=\nabla_MF^{MN}-J^N
\quad\Longleftrightarrow\quad
0=\dd*F_2-*J,
 \label{eom_maxwell_1}
\end{equation}
where the current $J^M(x)$ is defined by
\begin{equation}
J^M(x)
\equiv\frac{e}{\sqrt{-G}}
\!\int_\gamma\!\dd\lambda\frac{\dd X^M}{\dd\lambda}
\delta^D(x-X(\lambda)).
\end{equation}
By using the equation of motion (\ref{eom_maxwell_1}),
we arrive at the continuity equation
\begin{equation}
0=\nabla_M J^M \quad\Longleftrightarrow\quad 0=\dd*J.
\label{continuity_eq_1}
\end{equation}
The continuity equation can be also obtained by requiring the
gauge invariance of the action (\ref{wz_0}).

One can introduce the conserved charge through the continuity
equation (\ref{continuity_eq_1}),
\begin{equation}
Q
=\!\int_{V_t}\!\dd^{D-1}x\sqrt{-G}J^0
=\!\int_{V_t}\!*J,
\label{e_charge_1}
\end{equation}
where $V_t$ is the volume at some time $t$.
In the static gauge (\ref{static_gauge}),
it is easy to see that the charge $Q$ is nothing but $e$.
By using the equation of motion (\ref{eom_maxwell_1}),
we can rewrite the charge (\ref{e_charge_1}) in terms of the electric flux
\begin{equation}
Q=\!\int_{S^{(D-2)}}\!\!*F,
\end{equation}
where we have assumed that the constant time slice has the boundary with
topology $S^{(D-2)}$.

In the flat Minkowski spacetime with the static gauge,
the action (\ref{wz_0}) is reduced to be
 \begin{equation}
S_{0{\rm WZ}}
=-e\!\int_\gamma\!\dd t
\Big(A_0(t, X^i)+\frac{\dd X^i}{\dd t}A_i(t, X^i)
\Big).
\end{equation}
Together with the free part of the action (\ref{to_nonrela}),
we could obtain the equation of motion through the variation of $X^i$,
\begin{equation}
0=m\frac{\dd}{\dd t}
\bigg(\frac{v_i}{\sqrt{1-\vec{v}^2}}\bigg)
+e\Big(F_{i0}+F_{ij}v^j\Big).
\end{equation}
The second term on the right correctly represents the electromagnetic 
forces acting on the charged particle.

\subsubsection{Path-integral  in flat background spacetime}

We now quantize point particles in flat background spacetime.
We first consider the case where the particle trajectory is given by
the open path.
For precise quantization, we depart from the nonlinear
action (\ref{world_line}) and consider the following on-shell
equivalent action
\begin{equation}
\widetilde{S}_0[e, X]
=-\frac{1}{2}\!\int_{\lambda_0}^{\lambda_1}\!\!\!\dd\lambda
\sqrt{-\gamma_{\lambda\lambda}}
\Big(\gamma^{\lambda\lambda}\del_\lambda X^M\del_\lambda X^N\eta_{MN}
+m^2\Big)
=\frac{1}{2}\!\int_{\lambda_0}^{\lambda_1}\!\!\!\dd\lambda
\, e\Big(\frac{1}{e^2}\del_\lambda X^M\del_\lambda X_M-m^2\Big),
\label{world_line_1}
\end{equation}
where we have introduced
the world line metric (or the einbein)
$\gamma_{\lambda\lambda}(\lambda)$
(or $e(\lambda)=\sqrt{-\gamma_{\lambda\lambda}(\lambda)}$).
By using the equation of motion for the einbein $e(\lambda)$,
\begin{equation}
e^2=-\frac{1}{m^2}g_{\lambda\lambda}
=-\frac{1}{m^2}\del_\lambda X^M\del_\lambda X_M,
\end{equation}
we could be back to the original action (\ref{world_line}).

We now proceed to quantize the point particle through the path-integral
formulation with the classical action (\ref{world_line_1})~\cite{polyakov}.
Since there is a one dimensional diffeomorphism invariance of the action;
\begin{equation}
\begin{array}{rl}
&
\lambda\to\lambda'=\lambda'(\lambda), \quad
\mbox{with boundary conditions} \quad
\lambda'(\lambda_0)=\lambda_0, \ \
\lambda'(\lambda_1)=\lambda_1, \\
\mbox{and}
&
e(\lambda)\to e'(\lambda')
=\bigg(\displaystyle\frac{\dd\lambda}{\dd\lambda'}\bigg)
e(\lambda), \qquad 
X^M(\lambda)\to X'^M(\lambda')=X^M(\lambda),
\end{array}
\label{1d_diffeo}
\end{equation}
the path-integral should be evaluated over gauge inequivalent configurations.
Therefore, the transition amplitude is formally given by
\begin{equation}
\langle X_{\rm f}^M\!=\!X^M(\lambda=1)|
X_{\rm i}^M\!=\!X^M(\lambda=0)\rangle
=\!\int\!\frac{{\cal D}e\,{\cal D}X}{V_{\rm gauge}}\,
\eee^{\scriptstyle-\frac{1}{2}
\!\int_0^1\!\!\dd\lambda e\big(\frac{1}{e^2}
(\del_\lambda X^M)^2+m^2\big)},
\label{amplitude_particle}
\end{equation}
where, for convenience,
we have fixed the parameterization of the end points as $\lambda_0=0$ and
$\lambda_1=1$ and taken the Wick rotation $\lambda\to -i\lambda$ and
$X^0\to -iX^0$.
In (\ref{amplitude_particle}),
the formal gauge volume is denoted by $V_{\rm gauge}$ and
we put it inside the integral.
In order to define the integration measures and the gauge volume,
we first introduce the reparameterization (gauge) invariant norm
through the inner product of the tangent functional space,
\begin{equation}
||\delta e||^2_e
=\langle\delta e, \delta e\rangle_e
\equiv
\int_0^1\!\dd\lambda \frac{1}{e}(\delta e)^2,
\qquad
||\delta X||^2_e
=\langle\delta X, \delta X\rangle_e
\equiv
\int_0^1\!\dd\lambda\, e (\delta X)^2.
\label{inner_product}
\end{equation}
These norms give some heuristic guiding principle to find the
volume elements.
In finite dimensional case, the Riemann structures with
line element (\ref{riemann_metric}) induce the natural
volume elements $\sqrt{|G(x)|}\,\dd x^1\wedge\cdots\wedge\dd x^D$.
We may be able to read off ``functional metric'' through the line
element (\ref{inner_product}) and define the integration measure in the
similar manner.
What we should do first is to choose precise ``coordinates system'' in the
functional space.

Infinitesimal gauge variation of (\ref{1d_diffeo}) with
$\delta\lambda(\lambda)=\lambda'(\lambda)
-\lambda\equiv-\epsilon(\lambda)$ is given by
\begin{equation}
\delta_{\rm gauge} e=\del_\lambda(\epsilon e).
\label{infinitesimal_e}
\end{equation}
On the other hand,
there exists physical variation $\delta_{\perp}e(\lambda)$
which could not be achieved through the gauge
transformation (\ref{infinitesimal_e}).
The orthogonality in terms of the inner product defined by
(\ref{inner_product}) gives
\begin{equation}
0=\!\int_0^1\dd\lambda\frac{1}{e}
\big(\delta_{\rm gauge}e\big)\big(\delta_\perp e\big)
=-\!\int_0^1\!\dd\lambda\epsilon e
\del_\lambda\Big(\frac{1}{e}\delta_\perp e\Big),
\quad {\rm i.e.}\quad
\delta_\perp e=\delta c e, \quad {\rm with}\ \  \delta c: {\rm const.},
\label{scale_transformation_particle}
\end{equation}
where we have integrated by part and imposed the boundary conditions
$\epsilon(\lambda=0, 1)=0$.
Therefore, the general variation of the einbein is given by
the orthogonal sum of the reparameterization $\delta_{\rm gauge}e(\lambda)$
and the global scale transformation $\delta_\perp e(\lambda)$.
It is easy to see the scale transformation defined
in (\ref{scale_transformation_particle}) could be related to
the variation of the length of the world line $l$ which is gauge invariant
\begin{equation}
(0<) \ l=\!\int_0^1\!\dd\lambda e
\quad \Longrightarrow \quad
\delta l=\delta c \,l.
\label{length}
\end{equation}
Plugging the variation $\delta e=\delta_{\rm gauge}e+\delta_\perp e$
into the invariant norm (\ref{inner_product}),
we obtain
\begin{equation}
||\delta e||^2_e
=\!\int_0^1\!\dd\lambda\,
e^3\epsilon
\Big(-\frac{1}{e^2}\del_\lambda\frac{1}{e}\del_\lambda e\Big)\epsilon
+\frac{\delta l^2}{l},
\label{line_element_ein}
\end{equation}
where we have defined an invariant inner product of the gauge parameter
\begin{equation}
||\epsilon||^2_e\equiv\!\int_0^1\!\dd\lambda\, e^3 \epsilon^2.
\label{gauge_volume}
\end{equation}
By using ``one'' gauge degrees of freedom $\epsilon(\lambda)$,
we may fix the gauge as $e(\lambda)=l$ where the constant $l$ should be
the length of the world line defined in (\ref{length}).
The parameter $l$ is called as the modular parameter.
In order to estimate the volume forms through
(\ref{line_element_ein}) and (\ref{gauge_volume}),
we need to fix the topology of the world line,
which affects the boundary conditions of the function
$\epsilon(\lambda)$.

First, as before,
we discuss the case of open path with the boundary condition
$\epsilon(\lambda=0, 1)=0$.
It is convenient to use eigenfunctions of the real operator
$-\del_\lambda^2$ in (\ref{line_element_ein})
which are given by $\xi_n(\lambda)=\sqrt{2}\sin(n\pi\lambda)$,
$(n=1, 2, \cdots)$.
By using the linear combination
$\epsilon(\lambda)=\sum_{n=1}^\infty \delta a_n\xi_n(\lambda)$,
we can evaluate (\ref{line_element_ein}) and (\ref{gauge_volume})
as
\begin{equation}
||\delta e||^2_{e=l}
=
\sum_{n=1}^\infty\pi^2 l n^2 \delta a_n^2
+\frac{\delta l^2}{l},
\qquad \mbox{and}\qquad
||\epsilon||^2_{e=l}
=
\sum_{n=1}^\infty l^3\delta a_n^2.
\end{equation}
Thus, we could obtain the volume forms\footnote{
We here adopt the zeta function regularization for infinite products.
Since the zeta function $\zeta(s)$ can be analytically continued to a
meromorphic function of $s$ in the complex $s$-plane,
$\zeta(s)=\sum_{n=1}^\infty\frac{1}{n^s}$ and
$\zeta'(s)=-\sum_{n=1}^{\infty}\frac{\log n}{n^s}$
provide
$$
1+1+\cdots =\zeta(0)=-\frac{1}{2}, \qquad
1\cdot2\cdots=\eee^{-\zeta'(0)}=\sqrt{2\pi}.
$$
};
\begin{equation}
{\cal D}e
=
\Big(\prod_{n=1}^\infty\pi\sqrt{l}n\, \dd a_n\Big)
\frac{1}{\sqrt{l}}\dd l
=\sqrt{2}{\cal D}\epsilon\, \dd l,
\quad
\mbox{with}
\quad
{\cal D}\epsilon =l^{-3/4}\prod_{n=1}^\infty\dd a_n
\equiv V_{\rm gauge}.
\label{volume_mesure}
\end{equation}
We  decompose the functional degrees of freedom of the einbein
$e(\lambda)$ to those for the gauge ``coordinate'' $\epsilon(\lambda)$
and for the modular parameter (``coordinate'') $l$.
Since the gauge volumes will be canceled out 
in (\ref{amplitude_particle}), the amplitude
has the finite integration over $l$.

Next, we shall do path-integral for $X^M(\lambda)$,
decomposing the classical part and the fluctuation
part $\widetilde{X}^M(\lambda)$:
\begin{equation}
X^M(\lambda)=X_{\rm i}^M
+\big(X_{\rm f}^M-X_{\rm i}^M\big)\lambda
+\widetilde{X}^M(\lambda),
\label{decomposition_0}
\end{equation}
where the boundary conditions $X^M(0)=X_{\rm i}^M$ and
$X^M(1)=X_{\rm f}^M$ imply $\widetilde{X}^M(\lambda=0, 1)=0$.
Since the classical part fixed by the boundary conditions
does not represent any degrees of freedom for the integration, we consider only
the fluctuation part  in the path-integration.
Plugging (\ref{volume_mesure}) and (\ref{decomposition_0})
into (\ref{amplitude_particle}) and
using the boundary condition for $\widetilde{X}^M(\lambda)$,
we get
\begin{equation}
\langle X_{\rm f}^M|X_{\rm i}^M\rangle
\propto 
\!\int_0^\infty\!\!\!\dd l
\, \eee^{-\frac{1}{2}\big(\frac{1}{l}(X_{\rm f}-X_{\rm i})^2+lm^2\big)}
\!\int\!{\cal D}\widetilde{X}\,
\eee^{-\frac{1}{2}\!\int_0^1\!\dd\lambda
\widetilde{X}^M\big(-\frac{1}{l}\del_\lambda^2\widetilde{X}_M\big)}.
\end{equation}
Following the evaluation of the integration measure of the vielbein,
we expand the fluctuation as
$\widetilde{X}^M(\lambda)
=\sum_{n=1}^\infty\widetilde{x}_n^M\sqrt{2}\sin(n\pi\lambda)$.
Then the integration of the fluctuation can be performed as
\begin{equation}
\int\!{\cal D}\widetilde{X}\,
\eee^{-\frac{1}{2}\!\int_0^1\!\dd\lambda\widetilde{X}^M
\big(-\frac{1}{l}\del_\lambda^2\widetilde{X}_M\big)}
=\Big(\!\int\!
\Big(\prod_{n=1}^\infty\sqrt{l}\dd\widetilde{x}_n\Big)
\, \eee^{-\frac{1}{2}
\sum_{n=1}^\infty\frac{\pi^2}{l}n^2\widetilde{x}_n\widetilde{x}_n}\Big)^D
=\Big(\frac{\pi}{2}\Big)^{D/4}(2\pi l)^{-D/2}.
\end{equation}
Therefore, we could organize
\begin{eqnarray}
\langle X_{\rm f}^M|X_{\rm i}^M\rangle
&\propto&
\!\int_0^\infty\!\!\!\dd l\,
(2\pi l)^{-D/2}
\eee^{-\frac{1}{2}\big(\frac{1}{l}(X_{\rm f}-X_{\rm i})+lm^2\big)}
=\!\int_0^\infty\!\!\!\dd l\!\int\!\frac{\dd^Dp}{(2\pi)^D}
\, \eee^{ip(X_{\rm f}-X_{\rm i})-\frac{l}{2}(p^2+m^2)}
\nonumber
\\
&=&
\!\int\!\frac{\dd^Dp}{(2\pi)^D}
\, \eee^{ip(X_{\rm f}-X_{\rm i})}
\!\int_0^\infty\!\!\!\dd l\, \eee^{-l H}
=\!\int\!\frac{\dd^Dp}{(2\pi)^D}\frac{\eee^{ip(X_f-X_i)}}{H},
\label{feynman_prop}
\end{eqnarray}
which is nothing but Feynman propagator for the free relativistic scalar
field in $D$-dimensions.
In the expression (\ref{feynman_prop}), we have defined the Hamiltonian
$H=(p^2+m^2)/2$ so that the parameter $l$ corresponds to
Schwinger parameter.

Let us briefly consider the case of closed path i.e.\ one-loop vacuum
amplitude.
Due to the periodic boundary condition
$\epsilon(\lambda)=\epsilon(\lambda+1)$,
the eigenfunction may be given by
$\epsilon_n(\lambda)=\eee^{i2\pi n\lambda}$, $(n=0, \pm1, \cdots)$.
Comparing the open path,
the constant zero mode which corresponds to the translation in the
circle appears.
The same procedure gives the measures as
\begin{equation}
{\cal D}e=\Big(\prod_{n\ne0}\dd a_n\Big)\frac{\dd l}{l},
\qquad
{\cal D}\epsilon
=\prod_n\dd a_n
=\Big(\prod_{n\ne0}\dd a_n\Big) L
\equiv V_{\rm gauge},
\end{equation}
where we formally replaced the integration of the zero mode by $L$.
Then, the one-loop amplitude is given by
\begin{equation}
Z_{S^1}
\propto \!\int_0^\infty\!\frac{\dd l}{2l}
\, \eee^{-\frac{1}{2}lm^2}
\!\int\!{\cal D}X\,
\eee^{-\frac{1}{2}\!\int_0^1\!\dd\lambda
X^M\big(-\frac{1}{l}\del_\lambda^2X_M\big)}
=
\!\int_0^\infty\!\frac{\dd l}{2l}
\, \eee^{-\frac{1}{2}lm^2}
(2\pi l)^{-D/2}
=
\!\int\!\frac{\dd^Dp}{(2\pi)^D}
\!\int_0^\infty\!\frac{\dd l}{2l}\, \eee^{-lH}.
\label{particle_one_loop}
\end{equation}
It should be mentioned that
in any $D$-dimensional spacetime,
in UV regime where $l\to 0$, the amplitude diverges.
On the other hand,
in IR, the amplitude converges as long as $m^2>0$.
We will compare this result of particle one-loop amplitude
with string one-loop.

\subsection{$p$-brane}

So far we have discussed the motion of a particle in a general background
spacetime, the coupling to the external electromagnetic fields,
and the path-integral quantization.
One could generalize the world line description (\ref{world_line}),
(\ref{induced_worldline}), (\ref{wz_0}) and (\ref{amplitude_particle})
to that for a $p$-dimensional extended object.
The parameterization of this object is now given by $\sigma^m$
$(m=0, 1, \cdots, p)$ and the embedding is given through the target
spacetime (bulk) coordinates
$X^M(\sigma)$.
The action may be proportional to the volume swept by the $p$-brane.
The invariant world volume, which is a natural generalization of the world
line (\ref{world_line}), is given by
\begin{equation}
S_p
=-T_p\times {\rm (volume)}=-T_p\!\int_{\Sigma}\!\dd^{1+p}\sigma
\sqrt{-\det{g_{mn}}(\sigma)},
\label{world_volume}
\end{equation}
where the induced metric on the world volume is now
\begin{equation}
g_{mn}\big(X(\sigma)\big)=\del_mX^M\del_nX^NG_{MN}(X)\, .
\label{induced_metric}
\end{equation}
The determinant inside the square root is for the indices of
the world volume labeling $m$ and $n$.
$T_p$ is the tension of the $p$-brane whose mass dimension
is ${\rm mass}^{1+p}$.

Let us analyze the interaction between $p$-brane and other spacetime fields.
Here we consider a $(p+1)$-form gauge potential $A_{p+1}(x)$,
which is a natural generalization of the usual Maxwell field
$A_1(x)=A_M(x)\dd x^M$:
\begin{equation}
A_{p+1}(x)\equiv\frac{1}{(p+1)!}A_{M_1\cdots M_{p+1}}(x)
\dd x^{M_1}\wedge\cdots\wedge\dd x^{M_{p+1}},
\end{equation}
where $A_{M_1\cdots M_{p+1}}(x)$ is a totally antisymmetric tensor
field.
A $(p+2)$-form field strength
$F_{M_1\cdots M_{p+2}}(x)$ can be defined by
$
F_{p+2}(x)\equiv
\dd A_{p+1}(x)$.
The field strength is invariant under the  gauge transformation
\begin{equation}
\delta A_{p+1}(x)=\dd \theta_p(x),
\end{equation}
where $\theta_p(x)$ is a $p$-form gauge parameter.
The kinetic term of the $(p+1)$-form field,
which is the natural analog of (\ref{maxwell_1}),  is given by
\begin{equation}
S_{(p+1){\rm-form}}
=-\frac{1}{2}\!\int_{\cal M}\! F_{(p+2)}\wedge * F_{(p+2)}
=-\frac{1}{2}\!\int_{\cal M}\!
\dd^{D}x\sqrt{-G}|F_{p+2}|^2.
\label{p+1-form_f}
\end{equation}

We could introduce the minimal coupling between the $(p+1)$-form
gauge field and the $p$-brane,
which is the extended version of (\ref{wz_0}),
\begin{eqnarray}
S_{p\rm WZ}
&=&
-q_p\!\int_{\Sigma}\! A_{p+1}
=-q_p\!\int_{\Sigma}
\frac{1}{(p+1)!}A_{M_1\cdots M_{p+1}}(X)\dd X^{M_1}\wedge\cdots\wedge
\dd X^{M_{p+1}}
\nonumber
\\
&=&
-q_p\!\int_{\Sigma}
\frac{1}{(p+1)!}A_{M_1\cdots M_{p+1}}(X)
(\del_{m_1}X^{M_1})\cdots(\del_{m_{p+1}}X^{M_{p+1}})
\dd\sigma^{m_1}\wedge\cdots\wedge\dd\sigma^{m_{p+1}},
\label{pwz}
\end{eqnarray}
where $q_p$ is the charge of the $p$-form field and the support of
the integration is the $(p+1)$-dimensional world volume.
The gauge invariance tells us that the $p$-brane should not have the
boundary.
However if the boundary is attached to other branes,
one could define the charge.

The equation of motion for the gauge field is
\begin{equation}
0=\nabla_M F^{MN_1\cdots N_{p+1}}
-J^{N_1\cdots N_{p+1}}
\quad\Longleftrightarrow\quad
0=\dd*F_{p+2}-*J_{p+1},
\label{maxwell_p}
\end{equation}
where
\begin{equation}
J^{M_1\cdots M_{p+1}}(x)
=\frac{q_p}{\sqrt{-G}}
\!\int_{\Sigma}\!
\dd\sigma^{m_1}\wedge\cdots\wedge\dd\sigma^{m_{p+1}}
(\del_{m_1}X^{M_1})\cdots(\del_{m_{p+1}}X^{M_{p+1}})
\delta^D(x-X(\sigma)).
\end{equation}
The continuity equation
\begin{equation}
0=\nabla_MJ^{MN_1\cdots N_p}
\quad\Longleftrightarrow\quad
0=d*J_{p+1}
\end{equation}
implies a $p$-brane charge $q_p$
\begin{equation}
q_p=\!\int_{B^{(D-p-1)}}\!\!\!* J_{p+1}
=\!\int_{S^{(D-p-2)}}\!\!\! *F_{p+2},
\label{q_charge_00}
\end{equation}
where  $\del B^{D-p-1}=S^{D-p-2}$ and we have used
the equation of motion (\ref{maxwell_p}) and the Stokes' theorem.

Once the field strength $F_{p+2}$ (electric) is defined,
one can always consider its
dual $\widetilde{F}_{D-p-2}$ (magnetic) via Hodge dual operation,
$$
*F_{p+2}\equiv\widetilde{F}_{D-p-2}
=\dd\widetilde{A}_{D-p-3}.
$$
The dual gauge field $\widetilde{A}_{D-p-3}(x)$ may couple to the
$(D-p-4)$-brane, so that
the $p$-brane and the $(D-p-4)$-brane are dual to each other.
It has been shown that
the charges should satisfy the Dirac quantization condition~\cite{dirac_q}
\begin{equation}
q_p q_{D-p-4}=2\pi n, \qquad n\in Z.
\label{dirac_q}
\end{equation}
%

\subsection{String}

We here consider the case $p=1$, i.e.\ string.
Comparing with the other extended objects,
string with 2D world sheet is special.
In flat background spacetime, one could quantize the string
itself and analyze physics in the target spacetime through the 2D world
sheet dynamics by using well-developed CFT and 2D quantum gravity.

String theory appeared originally through attempts to
explain the hadron resonance spectrum of the strong interaction.
The Veneziano scattering amplitude~\cite{venetiano} described the dynamics of
relativistic open string.
However, experiments in those days  turned out to support the gauge theory
i.e. QCD instead of string.
On the other hand, relativistic closed strings have a massless
spin 2 particle in their spectrum, which precisely corresponds to
the graviton.
The close string theory turned to be the theory of the
gravity~\cite{ssy}.
This is the initial place for string theory as a unified theory.

Possible shapes of this one-dimensional object are line and loop
which are referred as an open string and a closed string,
respectively\footnote{
We  consider only orientable world sheets.
}.
The length of string depends on the physics under consideration.
When the string theory is describing quantum gravity,
the typical size of string is the Planck length
$l_{\rm P}$.
Much below the string scale, there is not enough resolution to
distinguish string to point particle.
However,
it contains various oscillation modes which correspond to various
quantum numbers.
The energies and the polarizations of the oscillation modes are related
to the masses and the spins of the corresponding elementary
particles.

\subsubsection{String world sheet}

For strings the action (\ref{world_volume})
is nothing but the Nambu-Goto action for the area of the world sheet
parameterized by the coordinates $\sigma^m=(\tau, \sigma)$ with
$0\le\sigma\le l$ for the spatial coordinate:
\begin{equation}
S_1[X]
=-\frac{1}{2\pi\alpha'}
\!\int_{\Sigma}\!\dd^2\sigma\sqrt{-\det g_{mn}(\sigma)}.
\label{nambu_goto}
\end{equation}
The string length $l_{\rm s}$ is defined by
$l_{\rm s}^2=\alpha'$,
which is the only dimensionful external parameter in string
theories\footnote{
In 2D, such a dimensionful coupling constant is not irrelevant but
marginal.
The world sheet theory does not suffer from UV divergences.
}.
Closed string provides the world sheet without boundaries,
while open string sweeps out world sheet with boundaries.
\begin{figure}[htbb]
\begin{minipage}{1.0\textwidth}
\begin{center}
\includegraphics[clip, width=4cm]{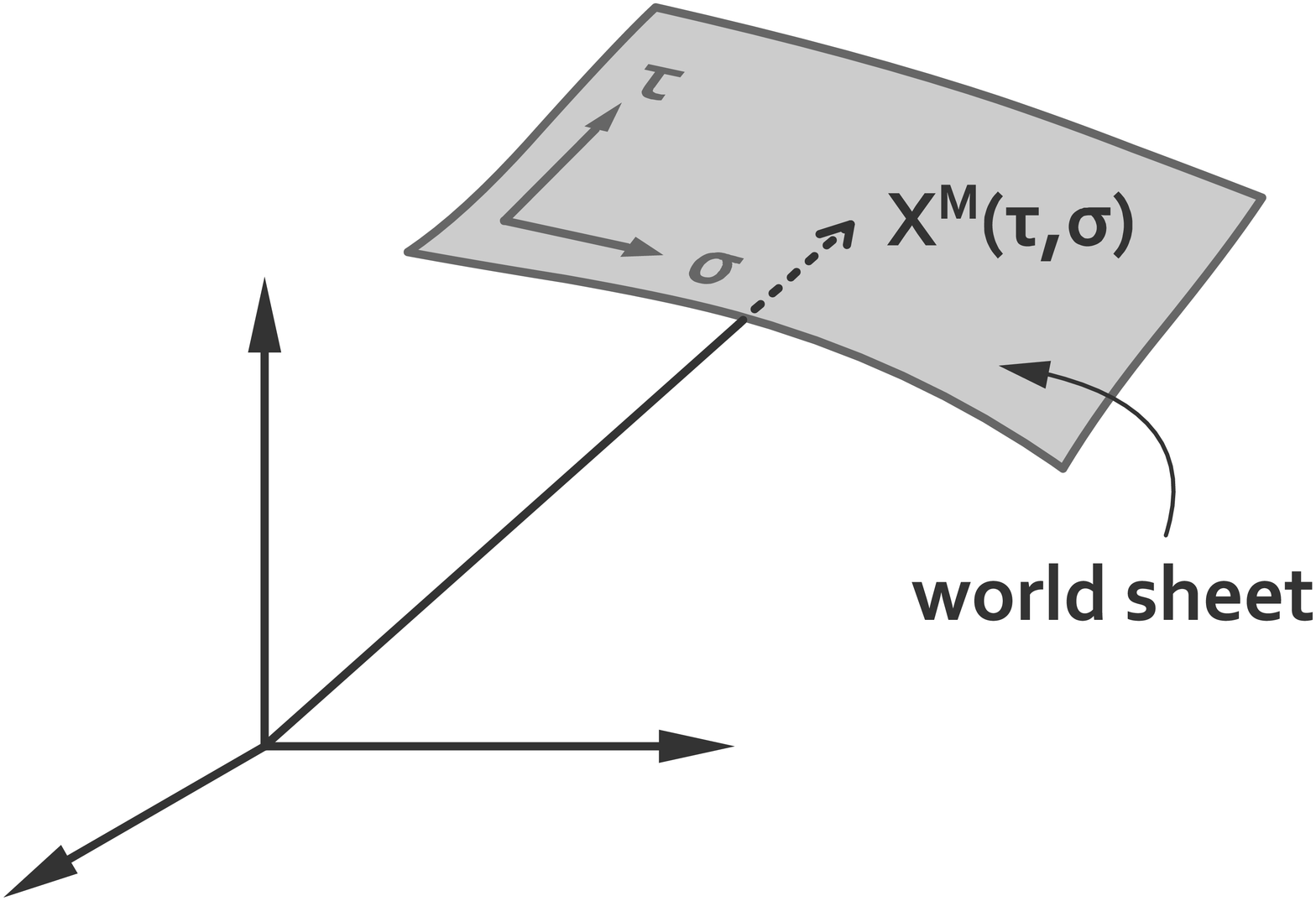}
\qquad\qquad
\includegraphics[clip, width=6.5cm]{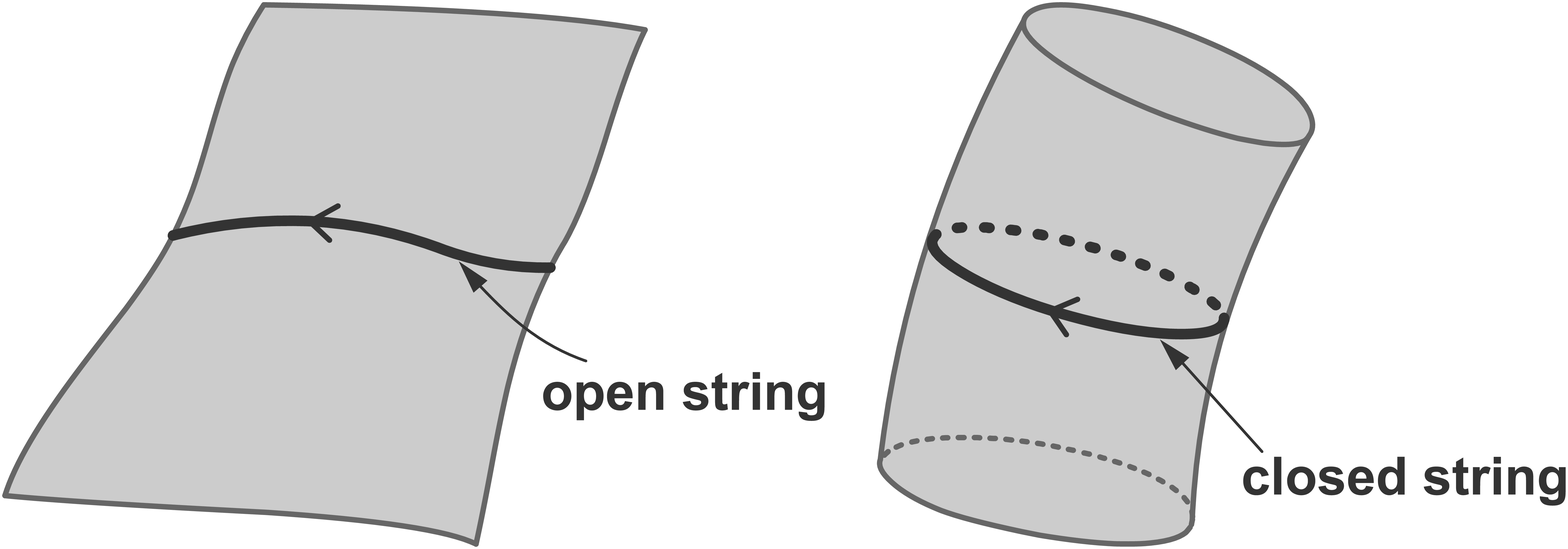}
\caption{String world sheets}
\label{world_sheet}
\end{center}
\end{minipage}
\end{figure}

In quantum theory, it is not convenient to use the action
(\ref{nambu_goto}) mainly because of its high nonlinearity of the square
root.
As we discussed in the case of particles,
we move on to the Polyakov action which provides on-shell equivalent
description  of the original Nambu-Goto action:
\begin{equation}
\widetilde{S}_1[X, \gamma]
=-\frac{1}{4\pi\alpha'}
\!\int_\Sigma\!\dd^2\sigma
\sqrt{-\gamma}\gamma^{mn}\del_mX^M\del_nX^NG_{MN}(X),
\label{polyakov}
\end{equation}
where the auxiliary field $\gamma_{mn}(\sigma)$, which can be understood as
the metric on the world sheet, has been introduced\footnote{
The metric $\gamma_{mn}(\sigma)$ is conceptually different from
the induced metric $g_{mn}(\sigma)$.
}.
The action (\ref{polyakov}) now acquires an additional local symmetry,
i.e.\ Weyl symmetry (local scale symmetry)
$\gamma_{mn}(\sigma)\to\eee^{2\lambda(\sigma)}\gamma_{mn}(\sigma)$,
which plays a central role in string theories.
As we will see, this Weyl rescaling may be used to
reshape awkward diagrams to standard forms which are easy to deal with.
Together with the general coordinate transformations,
we could fix the 2D world sheet metric to the conformal gauge
$\gamma_{mn}(\sigma)=\eta_{mn}$, at least locally. 
In a usual world sheet with nontrivial topologies,
the conformal gauge cannot be imposed globally. 
Note that the conformal gauge cannot fix the gauges completely.
There exists residual symmetry which is conformal symmetry.

The action (\ref{polyakov}) is the  nonlinear sigma model
of 2D scalar fields $X^M(\sigma)$ in the curved background spacetime
whose metric is given by $G_{MN}(X)$.
We will analyze the world sheet action (\ref{polyakov}) by using
2D quantum field theory.
From the point of view of renormalizability,
the action (\ref{polyakov}) is not complete.
We will later find other pieces to complete the full nonlinear sigma model action.

By using the equation of motion of $\gamma_{mn}(\sigma)$,
which is the energy momentum tensor in the 2D world sheet,
\begin{equation}
0=T_{mn}
\equiv\frac{2}{\sqrt{-\gamma}}
\frac{\delta\widetilde{S}_1}{\delta\gamma^{mn}}
=-\frac{1}{2\pi\alpha'}
\Big(g_{mn}-\frac{1}{2}\gamma_{mn}\gamma^{pl}g_{pl}\Big),
\label{em_world_sheet}
\end{equation}
and
eliminating the auxiliary field $\gamma_{mn}(\sigma)$
through the algebraic equation above,
one could be back to the original (\ref{nambu_goto}).
The energy momentum tensor is traceless $0=\gamma^{mn}T_{mn}$ which reflects
that the theory is classically scale invariant.

The variation of the action (\ref{polyakov}) with respect to
$X^M(\sigma)$ reads
\begin{eqnarray}
\delta\widetilde{S}_1[X, \gamma]
&=&
\frac{1}{2\pi\alpha'}
\!\int_{\Sigma}\!
\dd^2\sigma\sqrt{-\gamma}\Big(\nabla^2X^M
+\gamma^{mn}\del_mX^L\del_nX^P\Gamma^M{}_{LP}\Big)G_{MN}\delta X^N
\nonumber
\\
&&
-
\frac{1}{2\pi\alpha'}
\!\int_{\del\Sigma}\!
(\dd\Sigma)^m\del_mX^MG_{MN}\delta X^N.
\end{eqnarray}
The first line gives the equation of motion for $X^M(\sigma)$, and
the second line is the boundary term.
For closed string, the boundary term simply vanishes, while
that is not the case for the open string.
We need to impose boundary conditions at the endpoints of the open string.
These are Neumann boundary condition and Dirichlet boundary
condition, respectively:
\begin{equation}
n^m\del_m X^M\Big|_{\del\Sigma}=0,
\qquad
X^M\Big|_{\del\Sigma}=c^M,
\label{neumann_dirichlet}
\end{equation}
where $n^m$ is the unit normal vector to the boundary and
$c^M$ is the constant which describes the position of fixed endpoints
of the open string.
It should be noted that these boundary conditions preserve the local
scale symmetry in the world sheet.
Since the momentum conservation breaks down at the string endpoints
with Dirichlet boundary conditions,
 the string should be attached to another extended
object, which should be dynamical.
As we will see, the Dirichlet boundary condition could not be imposed in the
string perturbation in the vacuum, but in a solitonic background.

The boundary condition (\ref{neumann_dirichlet}) can be imposed
in each spacetime direction and  string endpoint, independently.
For simplicity, we impose the same boundary conditions on each direction;
Neumann boundary conditions for directions
$(M=m=0, 1, \cdots, p)$ and Dirichlet boundary conditions for the others
$(M=a=p+1, \cdots, D-1)$.
In this case, the string endpoints make a single hypersurface in the target
spacetime.
This hypersurface can be interpreted as the world volume of
$p$-dimensional object.
This is the Dirichlet $p$-brane, in short D$p$-brane.
\begin{figure}[htbb]
\begin{minipage}{1.0\textwidth}
\begin{center}
\includegraphics[clip, width=3.2cm]{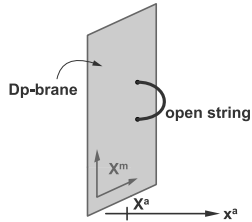}
\caption{D$p$-brane}
\label{dp-brane}
\end{center}
\end{minipage}
\end{figure}

Let us consider the partition function in the path-integral formulation:
\begin{equation}
Z=\!\int\!\frac{{\cal D}X{\cal D}\gamma}{V_{\rm gauge}}
\, \eee^{-\widetilde{S}_1[X, \gamma]}.
\end{equation}
As we have demonstrated in the case of particles,
we have to sum up all possible gauge inequivalent physical fluctuations
on the world sheet,
where the gauge degrees of freedom are those for diffeomorphism and Weyl
rescaling.
Physically, this summation corresponds to that for all possible
string excitations which are equivalent to the particle states.

\begin{figure}[htbb]
\begin{minipage}{1.0\textwidth}
\begin{center}
\includegraphics[clip, width=6cm]{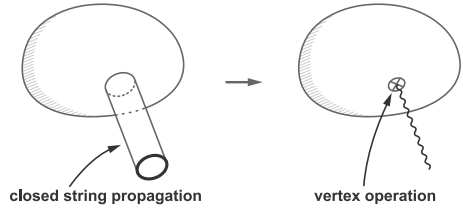}
\caption{String interaction and vertex operator}
\label{vertex_operator}
\end{center}
\end{minipage}
\end{figure}

We now perturb the bulk metric from the flat Minkowski spacetime,
\begin{equation}
G_{MN}(X)=\eta_{MN}+h_{MN}(X).
\end{equation}
Apart from the kinetic term in the flat
Minkowski background, the action (\ref{polyakov}) produces the
interaction term between string and the external field
which is called the vertex operator (external source term),
\begin{equation}
V_{h}\equiv\frac{1}{4\pi\alpha'}\!\int_\Sigma\!
\dd^2\sigma\sqrt{-\gamma}\gamma^{mn}\del_mX^M\del_nX^Nh_{MN}(X).
\label{vertex_gravity}
\end{equation}
In the path-integral formulation,
this corresponds to
the local operator insertion to the world sheet
\begin{equation}
Z=\!\int\!\frac{{\cal D}X{\cal D}\gamma}{V_{\rm gauge}}
\, \eee^{-\widetilde{S}_1|_{\rm free}}
\Big(1-V_h+\frac{1}{2!}V_h^2+\cdots\Big).
\end{equation}
This can be understood from the following world sheet point of view.
Consider first a closed string propagating in bulk and interacting
with a string world sheet (left side of Fig.\ref{vertex_operator}).
With the Weyl rescaling, the string propagator may be replaced by
the point on the world sheet.
This injection is nothing but the insertion of the vertex operator.
The vertex operator corresponds to emission/absorption of a string in
a particular mass eigenstate,
that is one of the various modes which exist in initial string states.
If one takes the perturbation as the plane wave i.e.\
\begin{equation}
h_{MN}(X)=\zeta_{MN}\ \eee^{ik\cdot X},
\label{plane_wave}
\end{equation}
and imposes the local scale invariance even in  quantum theory,
the operator is restricted as
\begin{equation}
k^2=0, \qquad \zeta_{MN}k^N=\zeta_{MN}k^M=0.\label{condi}
\end{equation}
These massless and transverse polarization conditions imply that
the vertex operator (\ref{vertex_gravity})
describes a emission/absorption of graviton.
Instead of this $S$-matrix calculation,
we could consider the general source $G_{MN}(X)$.
As we will see in the following sections,
again requiring the Weyl invariance of the world sheet,
the spacetime metric $G_{MN}(X)$ turns out to follow the Einstein equation.

\subsubsection{Strings in general background}

We now introduce another renormalizable term which respects the Weyl symmetry
in the world sheet:
\begin{equation}
S_{\Phi_0}
=\Phi_0
\bigg(\frac{1}{4\pi}\!\int_\Sigma\!\dd^2\sigma\sqrt{-\gamma}R(\gamma)
+\frac{1}{2\pi}\!\int_{\del\Sigma}\!\dd sK\bigg),  \quad
\Phi_0 :{\rm const.},
\label{dilaton_0}
\end{equation}
where $R(\gamma)$ is the scalar curvature of $\gamma_{mn}(\sigma)$  and
$K$ is the extrinsic (geodesic) curvature of the boundary.
This is the Einstein-Hilbert action in 2D spacetime with boundary.
One could freely add this to the Polyakov action (\ref{polyakov})
without changing the equation of motion for the world
sheet metric $\gamma_{mn}(\sigma)$ (\ref{em_world_sheet}).
Indeed, Riemann-Roch theorem tells that
the parenthesis in (\ref{dilaton_0}) gives topological invariant of the
world sheet
(Euler number $\chi$),
\begin{equation}
\frac{1}{4\pi}\!\int_\Sigma\!\dd^2\sigma\sqrt{-\gamma}R(\gamma)
+\frac{1}{2\pi}\!\int_{\del\Sigma}\!\dd sK=\chi=2-2g-b,
\end{equation}
where $g$ and $b$ are the number of handles and  boundaries
of the Riemann surface, respectively.

\begin{figure}[htbb]
\begin{minipage}{1.0\textwidth}
\begin{center}
\includegraphics[clip, width=6cm]{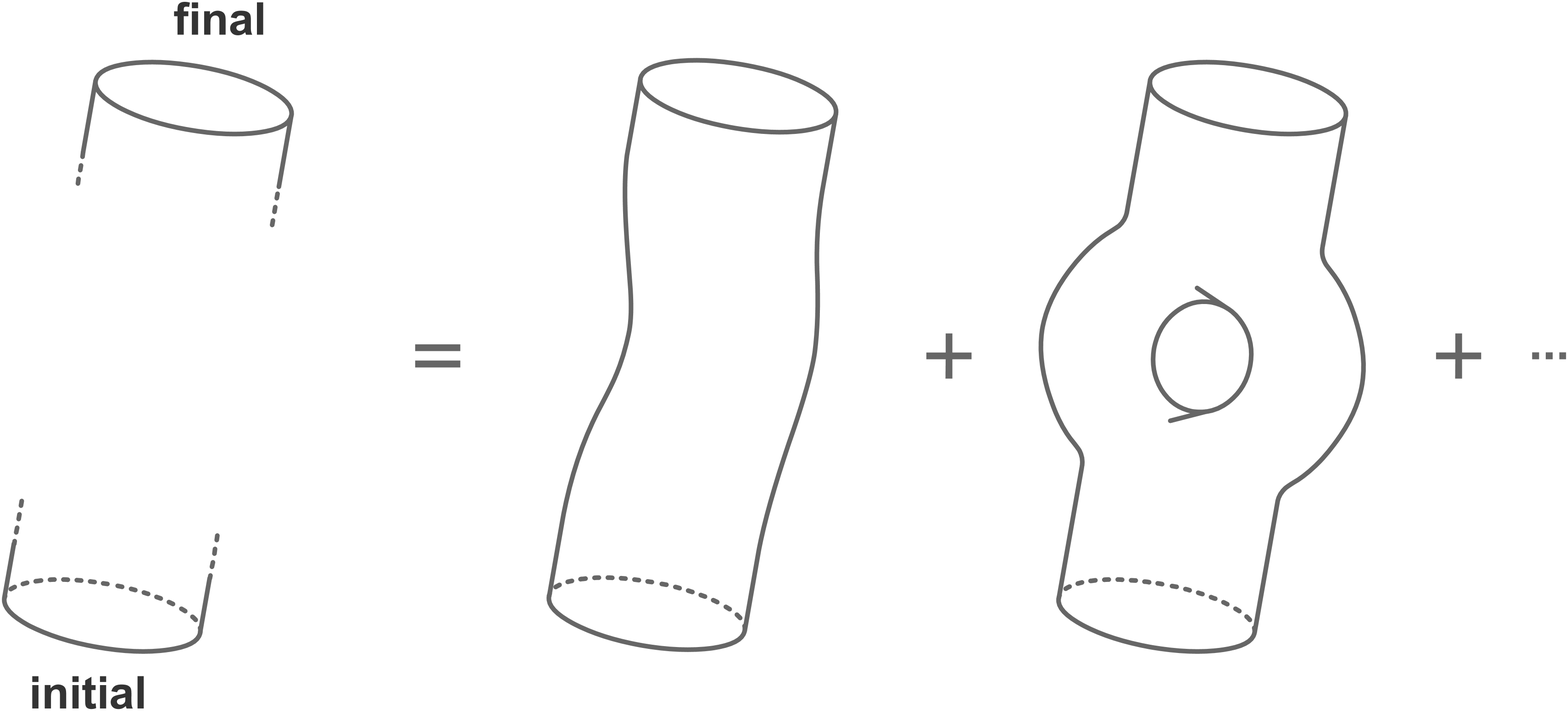}
\qquad\qquad
\includegraphics[clip, width=6cm]{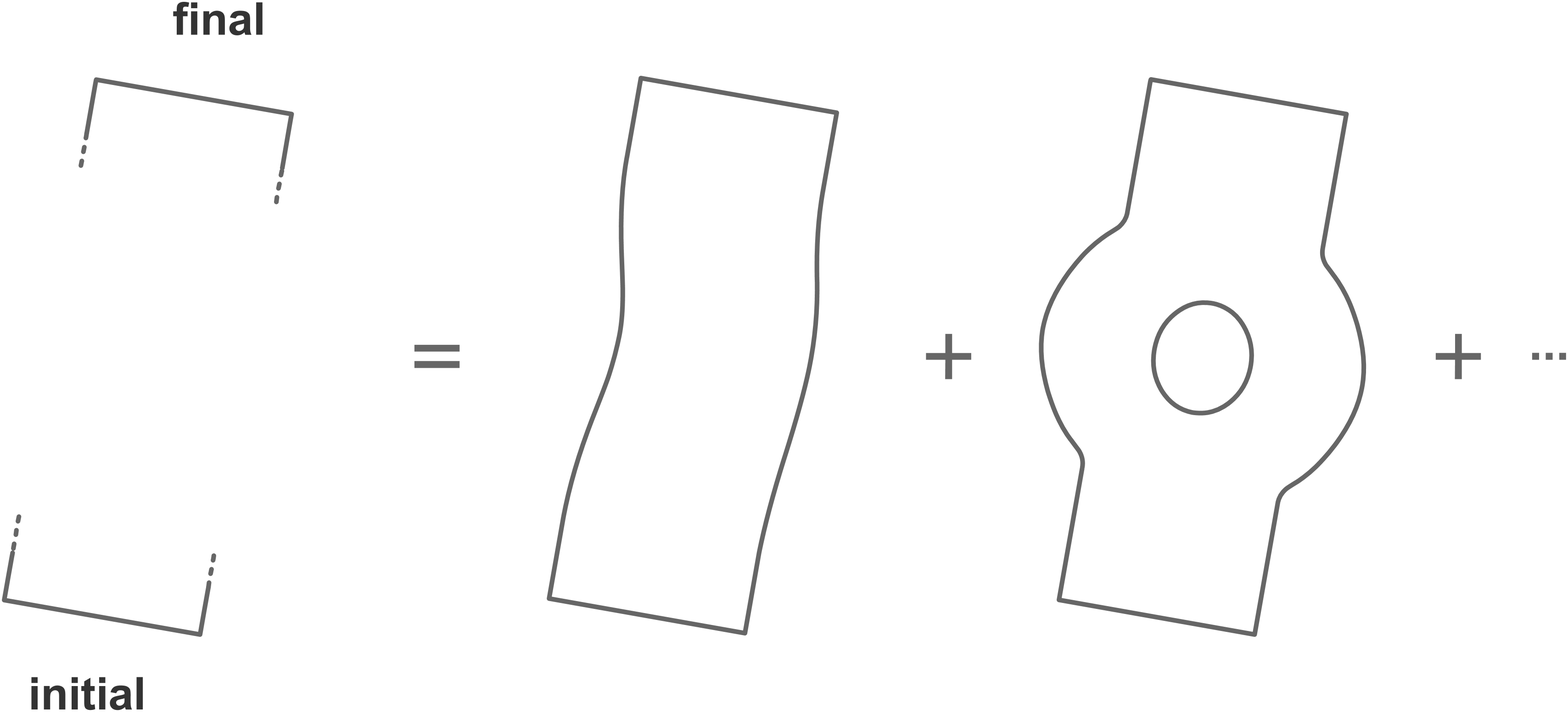}
\caption{Genus expansions for closed string (left) and open string
 (right)}
\label{genus_expansions}
\end{center}
\end{minipage}
\end{figure}

This world sheet topology is important in string interactions.
Indeed, in the path-integral formulation,
the summations should include  the world
sheet topologies.
For point particles, the path-integral  sums over all paths
connecting the initial and the final states.
For strings, one could naturally generalize this to summing over all
world sheets, i.e.\ all embedding $X^M(\sigma)$ and
all world sheet metric $\gamma_{mn}(\sigma)$,
connecting the initial and the final strings (see
Fig.\ref{genus_expansions}).

The string partition function may be modified by adding $S_{\Phi_0}$:
\begin{equation}
Z=\!\int\!\frac{{\cal D}X{\cal D}\gamma}{V_{\rm gauge}}
\, \eee^{-\widetilde{S}_1-S_{\Phi_0}}
=
\sum_{T} (\eee^{\Phi_0})^{-\chi(T)}
\!\int_{\Sigma_T}\!\!\frac{{\cal D}X{\cal D}\gamma}{V_{\rm gauge}} \
\eee^{-\widetilde{S}_1},
\label{partition_1}
\end{equation}
where the summation over metric can be decomposed with fixed
topology $T$.
The factor $(\eee^{\Phi_0})^{-\chi(T)}$ gives the relative weight of
different topologies.
For a closed string case,
if a handle is added in the world sheet, one gets a factor
$\eee^{2\Phi_0}$ through the Euler number.
On the other hand, adding the handle can be regarded as the process
for emitting and reabsorbing a closed string, i.e.\ two events of string
interactions.
Therefore one could identify a closed string coupling $g_{\rm closed}$ as
$g_{\rm closed}=\eee^{\Phi_0}$. 
Similar consideration could be applied for open string world sheets with
handles and boundaries.
As a result, open string coupling $g_{\rm open}$ is the square root of the
closed string coupling.
We then define the string coupling constant $g_{\rm s}$ as
\begin{equation}
g_{\rm s}\equiv\eee^{\Phi_0} =g_{\rm closed}=g^2_{\rm open}.
\end{equation}
For $g_{\rm s}\ll1$, the summation over topologies
defines a perturbative expansion, the genus expansion.
The genus is the quantum string loop in the target
spacetime, analogous to  particle loops in quantum field theory.
The dimensionless constant $\Phi_0$ can be understood as
the vacuum expectation value of the dilaton field,
one of the massless spectra in string theories.
In string theories, there are no dimensionless coupling constants, though
the dimensionless coupling constants may be provided dynamically 
through the vacuum expectation values
of dynamical fields, most probably nonperturbatively.
So one may generalize the action (\ref{dilaton_0}) to
\begin{equation}
S_{\Phi}=\frac{1}{4\pi}\!\int_\Sigma\!\dd^2\sigma\sqrt{-\gamma}
R(\gamma)\Phi(X)
+\frac{1}{2\pi}\!\int_{\del\Sigma}\!\dd sK\Phi(X),
\label{dilaton}
\end{equation}
and define the constant $\Phi_0$ as the asymptotic value of the dilaton
field $\Phi_0=\Phi(X=\infty)$.
Setting a fluctuation part of the dilaton as
$\widetilde{\Phi}(X)=\Phi(X)-\Phi_0$,
we define the effective dynamical coupling
$g_{\rm s}^{\rm eff}(X)=\eee^{\Phi(X)}
=g_{\rm s}\eee^{\widetilde{\Phi}(X)}$.
The perturbation now makes  sense for $g_{\rm s}^{\rm eff}(X)\ll1$.
Due to the generic dilaton $\Phi(X)$, the action (\ref{dilaton})
itself looses Weyl invariance.
However, the symmetry could be restored through the quantum
effects coming from other Weyl invariant actions.

One could find another action  which preserves the Weyl invariance
in the string world sheet.
Since the string is a one-dimensional extended object,
it naturally couples to a two-form external gauge potential $B_2(X)$
through the action (\ref{pwz}),
\begin{equation}
S_{\rm 1WZ}
=\frac{1}{4\pi\alpha'}
\!\int_\Sigma\! B_2
=\frac{1}{4\pi\alpha'}
\!\int_\Sigma\!\dd^2\sigma\sqrt{-\gamma}\epsilon^{mn}
\del_mX^M\del_nX^NB_{MN}(X),
\label{1wz}
\end{equation}
where $\epsilon^{mn}(\sigma)$ is the Levi-Civita tensor in the 2D world sheet.
We here just follow the convention of the charge $q_1=1/(4\pi\alpha')$.
Since the action (\ref{1wz}) does not depend on the 2D metric,
that is Weyl invariant automatically, it
does not change the equation (\ref{em_world_sheet}).
As we discussed in the previous section,
the gauge invariance requires that the world sheet should not have any boundary,
i.e.\ this interaction is only for closed strings.
For open string world sheet, we need to perform the gauge transformation on the
gauge field coupled to the boundary.

\subsubsection{String equations and low energy effective theory}

Now, we have the world sheet actions (\ref{polyakov}), (\ref{dilaton}),
and (\ref{1wz}), which may respect Weyl symmetry.
From the target spacetime point of view,
these describe the string in the external background fields $G_{MN}(X)$,
$B_{MN}(X)$, and $\Phi(X)$, which are the lightest modes in string
theories\footnote{
For completeness, we provide an action for background tachyon
field $T(X)$ which couples to 2D cosmological constant term,
$$
S_{\sigma}^T
=\frac{1}{4\pi\alpha'}\!\int\!\dd^2\sigma\sqrt{\gamma}T(X).
$$
This term leads to an effective action which may be added
to (\ref{effective_action_bc}),
$$
S^{\rm eff}_{\rm tachyon}
=\frac{1}{2\kappa_0^2}\!\int_{\cal M}\!\!
\dd^Dx\sqrt{-G}\ \eee^{-2\Phi}
\Big\{-(\del_M T)^2+\frac{4}{\alpha'}T^2
\Big\}.
$$
However, since in superstring theories the tachyon could be projected out,
we here simply neglect this effective action.
}:
\begin{equation}
S_{\sigma}
=\frac{1}{4\pi\alpha'}
\!\int_\Sigma\!\dd^2\sigma\sqrt{\gamma}
\Big\{
\gamma^{mn}\del_mX^M\del_nX^NG_{MN}(X)
+i\epsilon^{mn}\del_mX^M\del_nX^NB_{MN}(X)
+\alpha'R(\gamma)\Phi(X)\Big\}\, .
\label{sigma_model}
\end{equation}
Here we adopted the Euclidean signature and  considered only
the closed string world sheet.
In general, it is difficult to exactly solve  this interacting world sheet theory
in a general background on which strings propagate.
However, if the background is weakly curved, we could treat the
world sheet interactions perturbatively.
In this case, in order to analyze the world sheet sigma model (\ref{sigma_model}),
one could use the background field method.
We first expand the coordinates $X^M(\sigma)$ around a classical solution
$X^M_0$,
\begin{equation}
X^M(\sigma)= X^M_0+\sqrt{\alpha'}Y^M(\sigma),
\label{expansion}
\end{equation}
where $Y^M(\sigma)$ is chosen to be dimensionless.
One can expand the action (\ref{sigma_model}) and perform
2D world sheet field theory for the fluctuations $Y^M(\sigma)$.
Expansions of the external fields $G_{MN}(X)$,
$B_{MM}(X)$, and $\Phi(X)$ around $X^M_0$
generate infinitely many spacetime derivative
terms which can be regarded as couplings to the fields $Y^M(\sigma)$.
Since this derivative expansion can be taken as an expansion in the
powers of
$
(\sqrt{\alpha'}/r_{\rm S})=(l_{\rm s}/r_{\rm S})
$
with
the characteristic length scale $r_{\rm S}=(\del G/\del X)^{-1}$
of the target spacetime,
in the perturbation theory, we should impose
\begin{equation}
\frac{\sqrt{\alpha'}}{r_{\rm S}}\ll1.
\label{alphap_expansion}
\end{equation}
Therefore, these expansions are valid when the string length
$\sqrt{\alpha'}=l_{\rm s}$  is too small to feel the curvature
of the target spacetime.
The string could be regarded as the point particle here.
In this case, 
as in the conventional quantum field theory, starting from the free theory,
we compute deviations order by order with the expansion parameter
$(\sqrt{\alpha'}/r_{\rm S})$.

In the quantized string theories, we would like to keep the world sheet
Weyl invariant.
This puts some restrictions on the target spacetime.
One can calculate the trace of energy momentum tensor in one-loop for the
sigma model and the tree level of the string
perturbation,
\begin{equation}
\langle T_m{}^m\rangle
=-\frac{1}{2\alpha'}\beta^G_{MN}\gamma^{mn}\del_mX^M\del_nX^N
-\frac{i}{2\alpha'}\beta^B_{MN}\epsilon^{mn}\del_mX^M\del_nX^N
-\frac{1}{2}\beta^\Phi R,
\end{equation}
and require that this should vanish.
The beta functionals can be calculated as
\begin{eqnarray}
\beta^G_{MN}
&=&
\alpha'\Big(R_{MN}+2\nabla_M\nabla_N\Phi
-\frac{1}{4}H_{MKL}H_N{}^{KL}\Big)+{\cal O}(\alpha'^2),
\nonumber
\\
\beta^B_{MN}
&=&
\alpha'\Big(-\frac{1}{2}\nabla^KH_{KMN}+(\del^K\Phi) H_{KMN}\Big)
+{\cal O}(\alpha'^2),
\\
\beta^\Phi
&=&
\frac{(D-D_{\rm c})}{6}
+\alpha'\Big(-\frac{1}{2}\nabla^2\Phi
+(\del_M\Phi)^2-\frac{1}{24}H_{KMN}H^{KMN}\Big)
+{\cal O}(\alpha'^2),
\nonumber
\end{eqnarray}
where $H_{MNL}$ is the field strength of two-form gauge field $B_{MN}$,
and $D_{\rm c}$ is a contribution from Weyl anomaly, which is 26 for
the bosonic string.
Equations $\beta^G_{MN}=\beta^B_{MN}=\beta^\Phi=0$
can be given as the equation of motion of the following
action, which is the low energy effective action
\begin{equation}
S^{\rm eff}_{\rm closed}
=\frac{1}{2\kappa_0^2}
\!\int_{\cal M}\!\!\dd^Dx\sqrt{-G}
\ \eee^{-2\Phi}
\bigg\{
R+4(\del_M\Phi)^2
-\frac{1}{12}H_{MNL}H^{MNL}
-\frac{2(D-D_{\rm c})}{3\alpha'}+{\cal O}(\alpha')
\bigg\},
\label{effective_action_bc}
\end{equation}
where the coupling constant $\kappa_0$ will be determined later.
The factor $\eee^{-2\Phi}$ indicates a tree
level contribution from the string perturbation.
It should be mentioned that
the effective action (\ref{effective_action_bc}) can be perfectly
reproduced a priori through a different way via
the string $S$-matrix computation.
The solutions of the equations of motion describe
backgrounds for string theory where strings could be consistently
quantized in the lowest order in $\alpha'$ and the string coupling
constant.

One solution of the equations of motion is $R_{MN}=B_{MN}=0$,
$\Phi=\Phi^0$ (const.) and $D=D_{\rm c}=26$ which defines
the critical dimensions of bosonic string.
In the next subsection, we discuss the quantization of the
bosonic string.

So far we have discussed the equation of motion for the closed string
from the viewpoint of the world sheet nonlinear sigma model.
Let us consider the same thing for open string in the closed string
background where the closed string excitations are needed for consistent
open string interactions.
Since the endpoint of open string can be charged like a point particle,
the spacetime gauge field naturally couples to the 1D boundary
of the open string world sheet:
\begin{equation}
S_{\rm boundary}
=\!\int_{\del\Sigma}\!A_1
=\!\int_{\del\Sigma}\!\dd\Sigma^m\del_mX^MA_M(X).
\label{dbi_0}
\end{equation}
Here we assume that the Neumann boundary conditions are imposed in all
directions.
Together with the sigma model action (\ref{sigma_model})
and  the boundary term in (\ref{dilaton}),
the beta functional for the coupling $A_M$ could be evaluated.
The equation of motion  obtained by imposing the vanishing of
the anomaly could be derived from the low energy effective action
which is known as the Dirac-Born-Infeld action \cite{dbi_0},
\begin{equation}
S^{\rm eff}_{\rm open}
=-T_{D-1}\!\int_{\cal M}\!\!\dd^Dx \ \eee^{-\Phi}
\sqrt{-\det(G_{MN}+B_{MN}+2\pi\alpha' F_{MN})}.
\end{equation}
%

\subsubsection{Quantization of  bosonic string in flat background}

We start with the Polyakov action in the flat background,
\begin{equation}
\widetilde{S}_1[X, \gamma]
=-\frac{1}{4\pi\alpha'}\!\int\!\dd^2\sigma
\sqrt{-\gamma}\gamma^{mn}\del_mX^M\del_nX^N\eta_{MN}.
\label{polyakov_flat}
\end{equation}
With the conformal gauge for 2D world-sheet
$\gamma_{mn}=\eta_{mn}=\mbox{diag}(-, +)$
the action (\ref{polyakov_flat}) reduces to
\begin{equation}
 S_{\rm B}=\frac{1}{\pi\alpha'}\!\int\!\dd^2\sigma
\del_+X^M\del_-X_M,
\label{polyakov_light_cone}
\end{equation}
where we have used light-cone coordinates on the world-sheet
$\sigma^\pm\equiv\tau\pm\sigma$.
The equation of motion of string coordinates turns to be free wave
equation,
\begin{equation}
0=(-\del_\tau^2+\del_\sigma^2)X^M=\del_+\del_-X^M.
\label{eom_string}
\end{equation}
When one imposes the conformal gauge, 
the equation of motion for $\gamma_{mn}$ becomes a constraint, 
\begin{equation}
0=T_{mn}, \quad \mbox{i.e.} \quad
0=\del_+X^M\del_+X_M=
\del_- X^M\del_- X_M.
\label{emtensor}
\end{equation}
This constraint is the first class constraint which could 
generate gauge symmetry. 
This is nothing but conformal symmetry.
The wave equation (\ref{eom_string}) can be solved
by the sum of left and right movers
which travel in the opposite directions along the string:
\begin{equation}
X^M(\tau, \sigma)
=X^M_{\rm L}(\sigma^+)+X^M_{\rm R}(\sigma^-).
\end{equation}

For closed string, imposing the periodic boundary
condition $X^M(\tau, \sigma)=X^M(\tau, \sigma+l)$, one can get
\begin{subequations}
\begin{eqnarray}
X^M_{\rm L}(\sigma^+)
&=&
\frac{1}{2}x^M+\frac{\pi\alpha'}{l}p^M\sigma^+
+i\sqrt{\frac{\alpha'}{2}}
\sum_{n\ne0}
\frac{{\alpha}^M_n}{n}\eee^{-i\frac{2\pi}{l} n\sigma^+},
\\
X^M_{\rm R}(\sigma^-)
&=&
\frac{1}{2}x^M+\frac{\pi\alpha'}{l}p^M\sigma^-
+i\sqrt{\frac{\alpha'}{2}}
\sum_{n\ne0}
\frac{\widetilde{\alpha}^M_n}{n}\eee^{-i\frac{2\pi}{l}n\sigma^-},
\end{eqnarray}
\end{subequations}
where the zero modes, $x^M$ and $p^M$,
express the position and momentum of center of mass, respectively.
Defining the canonical momentum
$\Pi^M(\tau, \sigma)=\delta \widetilde{S}_1/\delta (\del_\tau X_M)$,
the Hamiltonian is given by
\begin{equation}
H=\frac{1}{2\pi\alpha'}\!\int_0^{l}\!\dd\sigma
\big((\del_+X)^2+(\del_-X)^2\big)
=\frac{\pi}{l}
\Big(\alpha'p^2+\sum_{n\ne0}\big(\alpha_{-n}^M\alpha_{n M}
+\widetilde{\alpha}_{-n}^M\widetilde{\alpha}_{n M}\big)\Big).
\end{equation}

For open string, depending on the boundary conditions
(\ref{neumann_dirichlet})
i.e.\ Neumann (N) and Dirichlet (D),
left and right modes are related by
$\widetilde\alpha^M_n=\pm\alpha^M_n$ i.e.\ $\sin/\cos$ waves.
We list below the possible mode expansions and their Hamiltonians:

\vspace*{1mm}
\noindent
{\bf (NN)} ($\del_\sigma X^M(\tau, \sigma=0)
=\del_\sigma X^M(\tau, \sigma=l)=0$)
\begin{subequations}
\begin{eqnarray}
X^M(\tau, \sigma)
&=&
x^M+\frac{2\pi\alpha'}{l}p^M\tau+i\sqrt{2\alpha'}\sum_{n\ne0}
\frac{\alpha^M_n}{n}\, \eee^{-i\frac{\pi}{l}n\tau}
\cos\Big(\frac{n\pi\sigma}{l}\Big),
\label{boson_nn}
\\
H
&=&
\frac{\pi}{2l}\Big(2\alpha'p^2+\sum_{n\ne0}\alpha^M_{-n}\alpha_{Mn}\Big).
\end{eqnarray}
\end{subequations}
{\bf (DD)} ($X^M(\tau, \sigma=0)=x^M_0$ and
$X^M(\tau, \sigma=l)=x^M_1$)
\begin{subequations}
\begin{eqnarray}
X^M(\tau, \sigma)
&=&
x^M_0+\frac{x^M_1-x^M_0}{l}\sigma
+\sqrt{2\alpha'}\sum_{n\ne0}
\frac{\alpha^M_n}{n}\, \eee^{-i\frac{\pi}{l}n\tau}
\sin\Big(\frac{n\pi\sigma}{l}\Big),
\label{boson_dd}
\\
H
&=&
\frac{1}{4\pi\alpha'l}\big(x_1^M-x_0^M\big)^2
+\frac{\pi}{2l}\sum_{n\ne0}\alpha^M_{-n}\alpha_{Mn}.
\end{eqnarray}
\end{subequations}
{\bf (ND)} ($\del_\sigma X^M(\tau, \sigma=0)=0$ and
$X^M(\tau, \sigma=l)=x^M_0$)
\begin{subequations}
\begin{eqnarray}
X^M(\tau, \sigma)
&=&
x^M_0+i\sqrt{2\alpha'}\sum_{n\in Z+\frac{1}{2}}\frac{\alpha^M_n}{n}
\eee^{-i\frac{\pi}{l}n\tau}\cos\Big(\frac{n\pi\sigma}{l}\Big),
\label{nd}
\\
H
&=&
\frac{\pi}{2l}\!\!\sum_{n\in Z+1/2}\!\!\!\alpha^{M}_{-n}\alpha_{Mn}.
\end{eqnarray}
\end{subequations}
The momentum operator appears only in the (NN) directions.

For canonical quantization,
we replace the Poisson bracket $\{X^M(\tau, \sigma), \Pi^N(\tau,
\sigma')\}_{\rm PB}
=\eta^{MN}\delta(\sigma-\sigma')$ by the commutator.
In terms of the mode expansions, we find the following commutation
relations:
\begin{equation}
[x^M, \ p^N]=i\eta^{MN}, \quad
[\alpha_m^M, \ \alpha_n^N]=
[\widetilde{\alpha}_m^M, \ \widetilde{\alpha}_n^N]
=m\delta_{m+n}\eta^{MN},
\quad
[\alpha^M_m, \ \widetilde{\alpha}^N_n]=0.
\label{commutation_relation}
\end{equation}
By using the modes (\ref{commutation_relation}),
we could construct Fock space,
defining the ground state $|0; k\rangle$ as
$\hat{p}^M|0; k\rangle=k^M|0; k\rangle$ and
$\alpha^M_m|0; k\rangle=\widetilde{\alpha}^M_m|0; k\rangle=0$
for $m>0$.
In order to obtain physical Hilbert space, we need to impose the
physical condition governed by (\ref{emtensor}).
This condition indeed removes such as a negative norm state
following from $[\alpha^0_m, \ {\alpha^0_m}^\dagger]=-1$
with $(a^M_m, \ {a^M_m}^\dagger)\equiv
(\alpha^M_m/\sqrt{m}, \ \alpha^M_{-m}/\sqrt{m})$.
We first define the Virasoro generators via the Fourier 
transformation of the energy momentum tensors:
\begin{equation}
L_m\equiv-\frac{l}{4\pi^2}\!\int_0^l\!\dd\sigma\, T_{--}
\eee^{-i\frac{2\pi}{l}m\sigma}
=\frac{1}{2}\sum_n\alpha_{m-n}^M\alpha_{n M},
\quad
\widetilde{L}_m
\equiv-\frac{l}{4\pi^2}\!\int_0^l\!\dd\sigma\, T_{++}
\eee^{i\frac{2\pi}{l}m\sigma}
=\frac{1}{2}\sum_n\widetilde{\alpha}^M_{m-n}\widetilde{\alpha}_{n M},
\label{virasoro}
\end{equation}
where we denote
$\alpha_0^M=\widetilde{\alpha}_0^M=\sqrt{\alpha'/2}\, p^M$ for
closed string and $\alpha_0^M=\sqrt{2\alpha'}p^M$ for open string.
The Virasoro generators $L_0$ and $\widetilde{L}_0$ are related to
the Hamiltonian
\begin{equation}
H=\frac{2\pi}{l}(L_0+\widetilde{L}_0), \quad
\mbox{(for closed string)} \qquad
H=\frac{\pi}{l}L_0,  \quad \mbox{(for open string)}.
\end{equation}
As usual in the canonical quantization,
one should take the normal ordering of operators
where all lowering operators are placed to the right.
In order to construct  the Virasoro generators (\ref{virasoro})
as  quantum operators,
$L_0$ and $\widetilde{L}_0$ should be treated with some care.
We define $L_0$ and $\widetilde{L}_0$ with a common constant $a$
which will be determined later:
\begin{equation}
L_0 \longrightarrow L_0-a=\frac{1}{2}\alpha^2_0
+\sum_{n=1}^\infty\alpha_{-n}^M\alpha_{n M}-a,
\quad
\widetilde{L}_0\longrightarrow
\widetilde{L}_0-a=\frac{1}{2}\widetilde{\alpha}^2_0+
\sum_{n=1}^\infty\widetilde{\alpha}^M_{-n}\widetilde{\alpha}_{nM}
-a.
\end{equation}
Then, we can define the Hilbert space which satisfies the condition
\begin{equation}
(L_m-a\delta_m)|\mbox{phys}\rangle=
(\widetilde{L}_{m}-a\delta_m)|\mbox{phys}\rangle=0, \quad m\ge0,
\quad \mbox{and}\quad
 (L_0-\widetilde{L}_0)|\mbox{phys}\rangle=0.
\label{virasoro_condition}
\end{equation}
Due to the hermiticity of the Virasoro generators 
$L_m^\dagger=L_{-m}$, we only need to impose the physical state condition
(\ref{virasoro_condition}) for $m\ge0$.
As we will see, 
the conditions (\ref{virasoro_condition}) for $L_0$ and $\widetilde{L}_0$
yield the mass relation. 
The Virasoro operators $L_{-m}(m>0)$ act on physical states and 
create the highest representation of the Virasoro algebra. 
In 26D, these states can be identified with the gauge degrees of 
freedom (null states).

Let us discuss the normal ordering constant $a$.
In practical computations,
the normal ordering constant is related to the following relation,
\begin{equation}
\frac{1}{2}\sum_n\alpha_{-n}\alpha_n
=\frac{1}{2}\sum_{n}:\alpha_{-n}\alpha_n:
+\frac{1}{2}\sum_{n=1}^\infty n, \quad
\mbox{with} \quad
[\alpha_m, \ \alpha_n]=m\delta_{m+n}.
\label{integer}
\end{equation}
Indeed the constant $a$ for closed string is formally given by
\begin{equation}
a_{\rm closed}=(D-2)\times
\Big(-\frac{1}{2}\sum_{n=1}^\infty n\Big).
\label{aclosed}
\end{equation}
It is clear in the light-cone gauge that
the coefficient $(D-2)$ can be interpreted as the number of physical transverse
modes of a string  in $D$ dimensions.
As we have used before, we could here also adopt the zeta function
regularization; $\sum_{n=1}^\infty n=\zeta(-1)=-1/12$.

We here consider open strings in $D$ dimensions which
consist of $n$ (ND) directions and $(D-n)$ (NN) and (DD)
directions.
Regarding the normal ordering,
(NN) and (DD) directions have the same structure as  (\ref{integer})
because of the ``integer'' modes, i.e.\ for each direction,
\begin{equation}
a_{\rm NN/DD}=\frac{1}{24}.
\label{nndd}
\end{equation}
However, for (ND) directions (\ref{nd}),
those contain half-integer (more correctly half-odd integer) modes.
The same discussion that led to (\ref{nndd}) gives
the constant $a$ for one (ND) string as\footnote{
We use $\zeta(s, q)=\sum_{n=1}^\infty(n+q)^{-s}$
with analytic continuation
$\zeta(-1, q)=-(6q^2-6q+1)/12$.
}
\begin{equation}
a_{\rm ND}=-\frac{1}{2}\sum_{n=1}^\infty\Big(n+\frac{1}{2}\Big)
=-\frac{1}{48}.
\label{ndnd}
\end{equation}
Therefore, the open string has the normal ordering constant
\begin{equation}
a_{\rm open}
=
\frac{1}{24}\big(\#({\rm NN})+\#({\rm DD})\big)
-\frac{1}{48}\big(\#({\rm ND})\big)
=\frac{D-2}{24}-\frac{n}{16}.
\end{equation}
The mass $M^2$ of closed  and open string excitations  with
$n$ (ND) directions
can be estimated through the relation
(\ref{virasoro_condition})
\begin{subequations}
\begin{eqnarray}
M^2_{\rm closed}
&\equiv&
-p^2=\frac{2}{\alpha'}\big(N_{\rm closed}
+\widetilde{N}_{\rm closed}-2a_{\rm closed}\big),
\quad \mbox{with} \quad N_{\rm closed}=\widetilde{N}_{\rm closed},
\label{closed_mass}
\\
M^2_{\rm open}
&\equiv&
-p^2=\frac{1}{\alpha'}\big(N_{\rm open}-a_{\rm open})
+\frac{(x_0-x_1)^2}{(2\pi\alpha')^2}.
\label{open_mass}
\end{eqnarray}
\end{subequations}
The number operators $N$ are given by
\begin{equation}
N_{\rm closed}
=
\sum_{n=1}^\infty \alpha^M_{-n}\alpha_{Mn},
\quad
\widetilde{N}_{\rm closed}
=\sum_{n=1}^\infty\widetilde{\alpha}^M_{-n}\widetilde{\alpha}_{Mn},
\quad
N_{\rm open}
=
\sum_{n=1}^\infty \alpha^i_{-n}\alpha_{i n}
+\!\!\!\!\!\sum_{n\in Z+\frac{1}{2}>0}\!\!\!\!\!\alpha^a_{-n}\alpha_{a
n},
\end{equation}
where $i$ and $a$ run through (NN), (DD) directions and (ND) directions,
respectively.
In 26D, the normal ordering constants may be $a_{\rm closed}=1$
and $a_{\rm open}=1-n/16$.
We can observe the tension of (ND) directions contribute to
the open string mass spectrum (\ref{open_mass}).

Now, we could construct the Hilbert space.
For closed strings,
the mass-shell relation (\ref{closed_mass}) tells that
the ground state $|0; k\rangle$ is a tachyonic scalar.
We could construct the following massless states as
the first excited states,
\begin{center}
\begin{tabular}{rc}
$\alpha_{-1}^{\{ M}\widetilde{\alpha}_{-1}^{N\}}
-\frac{1}{D}\eta^{MN}(\alpha_{-1}\cdot\widetilde{\alpha}_{-1})|0;k\rangle$ :
&
symmetric tensor $G_{MN}(x)$
\\
$\alpha_{-1}^{[M}\widetilde{\alpha}_{-1}^{N]}|0; k\rangle$ :
&
antisymmetric tensor $B_{MN}(x)$
\\
$\frac{1}{D}
(\alpha_{-1}\cdot\widetilde{\alpha}_{-1})|0; k\rangle$ :
&
scalar $\Phi(x)$
\end{tabular}
\end{center}
Virasoro constraints produce the condition (\ref{condi}) for gravitons.

Let us move on to open strings.
For simplicity, we here discuss the case with no (ND)
directions, i.e.\
$a_{\rm open}=1$.
Moreover, we consider a particular open string in Fig.\ref{dp-brane},
i.e.\  $x_1^a=x_0^a$.
As in the closed strings, the ground state is tachyonic.
The first excited states are given by
\begin{center}
\begin{tabular}{rc}
$\alpha_{-1}^m|0; k\rangle$ \quad and \quad
$\alpha_{-1}^a|0; k\rangle$:
&
vector $A_{m}(\sigma)$  \quad and \quad scalar $\phi_a(\sigma)$
\end{tabular}
\end{center}
where $m$ and $a$ denote the (NN) and (DD) directions, respectively.
From (\ref{open_mass}), these are massless states.
The momentum $k$  depends only on the direction for Neumann boundary
conditions.
The scalar fields $\phi_a(\sigma)$ correspond to the position of D-brane.
More about D-branes will be  given in later.

Both for closed and open strings, in addition to these massless modes,
infinitely many massive states are measured
by the inverse of string length $1/\sqrt{\alpha'}$.
Sending $\alpha'\to0$, the massive modes may be decoupled.

\subsubsection{Closed string one-loop}

Let us briefly discuss
closed string vacuum amplitudes.
As we discussed,  in general, string Feynman diagram can be classified by the genus
expansions.
The first nontrivial contribution to the partition function is
string one-loop, i.e.\ torus.
One could calculate this by essentially the same way done in the case of particles
in closed path (\ref{particle_one_loop}).

\begin{figure}[htbb]
\begin{minipage}{1.0\textwidth}
\begin{center}
\includegraphics[width=10cm]{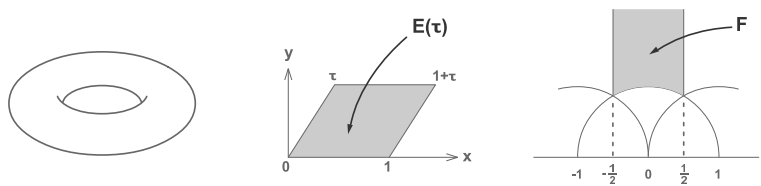}
\caption{Torus, complex structure, and fundamental domain}
\label{torus}
\end{center}
\end{minipage}
\end{figure}

Regarding a torus world-sheet, we first specify physical deformation
of the torus which corresponds to the length of the world-line $l$ for
particle trajectory i.e.\ $S^1$ for closed path.
Since a torus has two non-contractible cycles $S^1\times S^1$,
there may exist two parameters which characterize the torus.
This can be represented by a complex parameter $\tau=\tau_1+i\tau_2$
in the complex plane called moduli parameter, which remains after suitable gauge
fixing for the diffeomorphism and Weyl rescaling and should be integrated
in the path-integral.
A torus can be constructed by a parallelogram $E(\tau)$ with identifications of
two sides.
There exists the ``large'' coordinate transformations
$\tau'=(a\tau+b)/(c\tau +d)$ with $ad-bc=1$,
$(a, b, c, d \in Z)$ generated by $\tau'=\tau+1$
and $\tau'=-1/\tau$ which could not be absorbed by diffeomorphism/Weyl
rescaling.
This deformation is called as moduli.
It turns out that topologically inequivalent torus with $\tau$ can be defined
in so-called the fundamental domain $F$ which is usually taken to be,
\begin{equation}
F=\Big\{\tau_2 > 0, \quad -\frac{1}{2}< \tau_1 \le \frac{1}{2},
\quad |\tau|\ge 1\Big\}\, .
\end{equation}
The path-integration should be performed over the fundamental domain $F$.
By doing the diffeomorphism and the Weyl rescaling on the world sheet,
we could bring the world-sheet metric to be constant which is
\begin{equation}
\gamma_{ab}
=
\left(
\begin{array}{cc}
1 & \tau_1 \\
\tau_1 & |\tau|^2
\end{array}
\right)
\quad \Longleftarrow \quad
\dd s^2=|\dd z|^2= \gamma_{ab}\dd x^a\dd x^b,
\quad \mbox{with} \quad z=1\cdot x^1 +\tau x^2,~ 0\le x^1,~ x^2 \le 1\, ,
\end{equation}
where $z$ specifies the point on the torus with coordinates $(x^1, x^2)$.
This corresponds to fixing the gauge as $e=l$ in the particle case.
Using the preparations above, one could evaluate the path-integral
in the essentially same way done in the case of particles.
Choosing suitable ``eigenfunctions'' with periodic boundary conditions,
we could obtain
\begin{equation}
Z_{\rm torus}=
\!\int\!\frac{{\cal D}X{\cal D}\gamma}{V_{\rm gauge}}
\, \eee^{-\widetilde{S}_1[X, \gamma]}
\propto
\!\int_F\frac{\dd^2\tau}{4\tau_2}(4\pi^2\alpha'\tau_2)^{-13}
|\eta(\tau)|^{-48},
\label{string_one_loop}
\end{equation}
where $\eta(\tau)=q^{1/24}\prod_{n=1}^\infty (1-q^n)$ with
$q=\eee^{i2\pi\tau}$ is the
Dedekind's eta function.

Since the string partition function should be the same for equivalent
tori, it should be invariant under the modular transformations.
One can indeed check this with the result in (\ref{string_one_loop}).
The modular invariance is an important concept in string theory.
We could require this to construct consistent string theories.
One of the examples is the GSO projection in Type II superstring
theories.

It is interesting to compare the results between the case of particle
(\ref{particle_one_loop})
and string (\ref{string_one_loop}).
The UV divergence which has been observed in the case of particle
with $l\to 0$
does not exist in the string case where $\tau_2\sim l$ is bounded below
in the fundamental domain.
The geometrical restriction of torus topology brings natural
cut-off in string theory.

On the other hand, for IR region i.e.\ large $\tau_2\sim l$,
let us compare the result for the one-loop amplitude
(\ref{string_one_loop}) after performing $\tau_1$ integration
and that for the particle
(\ref{particle_one_loop}) in $D=26$,
\begin{equation}
Z_{\rm torus}
\sim\!\int\!\frac{\dd l}{l}\frac{\eee^{2l}}{l^{13}}
\quad
\Longleftrightarrow
\quad
\int\!\frac{\dd l}{l}\frac{\eee^{-\frac{1}{2}lm^2}}{l^{26/2}}
=Z_{S^1}\, .
\end{equation}
This indicates that there exists the tachyon with mass $m^2=-4$.
This is the lightest mode of bosonic string which could be first observed
in this low energy limit.
This IR divergence could be cured by introducing supersymmetry,
and thus we could observe the finiteness of string theories.
It is believed that the finiteness exists in the higher loops.

\subsection{Superstring}

By definition, in bosonic string theory so far discussed,
there are no fermionic fields in spacetime/world sheet.
Moreover in the bosonic theory, the ground state is tachyonic\footnote{
In field theories, the presence of tachyon means that one chooses
the wrong vacuum.
This may  be also the case in string theories~\cite{tachyon}.
}.
These are resolved if the supersymmetry is introduced.
As a result, requiring the tachyon  and the anomaly free,
five superstring theories (Type I, Type IIA, Type IIB, Heterotic with gauge groups
$E_8\times E_8$ and $SO(32)$)
can be consistently formulated in 10D target
spacetime~\cite{anomaly_free}.
We only consider Type II theories in this paper.
These possess ${\cal N}=2$ supersymmetry:
Type IIA has (1,1) supersymmetry with non chiral spinors while Type IIB
is the chiral theory with (2, 0) supersymmetry.
Type II superstrings are for closed string theories.
However, these may provide a seat for D-brane,
so that open string could participate and play an interesting role.

\subsubsection{Quantization of superstring in flat background}

Let us enlarge the world-sheet symmetry to 2D local supersymmetry.
First of all, we need to introduce superpartners of the bosonic fields
$(X^M(\sigma), \gamma_{mn}(\sigma))$ in the Polyakov
action (\ref{polyakov_flat}),
i.e.\
$D$ fermionic ``matter'' $\psi^M(\sigma)$ and gravitino
$\chi_m(\sigma)$.
This approach leads to Neveu-Schwarz-Ramond superstring\cite{nsr}
\footnote{
The Polyakov action is invariant under the global Poincar\'e
transformation.
One can enlarge this symmetries to global invariance of the super
Poincar\'e group.
This leads to an another formulation of superstring so-called
Green-Schwarz superstring~\cite{gs}.
}.
The resultant supersymmetric action can be found in \cite{dzbdvh}.
By using diffeomorphism, Weyl, and their fermionic counterparts,
one could take the superconformal gauge i.e.
$\gamma_{mn}(\sigma)=\eta_{mn}$ and $\chi_m(\sigma)=0$.
The bosonic part becomes the same as (\ref{polyakov_light_cone}),
while the fermionic part may be given by
\begin{equation}
S_{\rm F}
=-\frac{1}{2\pi}\!\int\!\dd^2\sigma\,
i\big(\bar{\psi}^M\rho^m\del_m\psi_M\big)
=\frac{i}{\pi}\!\int\!\dd^2\sigma\,
\big(\psi_+^M\del_-\psi_{+M}+\psi_-^M\del_+\psi_{-M}\big),
\label{fermion}
\end{equation}
where
we have taken 2D gamma matrix as $\rho^0=i\sigma^2$ and
$\rho^1=\sigma^1$.
The fermionic matters $\psi^M=(\psi_+^M \ \psi_-^M)^{\rm T}$
are Majorana-Weyl fermion whose two components are real Grassmann
fields in this representation.
The field variation of (\ref{fermion}) gives
\begin{equation}
\delta S_{\rm F}
=\frac{2i}{\pi}\!\int\!\dd^2\sigma
\big(\delta\psi_+^M\del_-\psi_{+M}
+\delta\psi_-^M\del_+\psi_{-M}\big)
+\frac{i}{2\pi}\!\int\!\dd\tau
\big(
\delta\psi_+\psi_+-\delta\psi_-\psi_-\big)
\big|_{\sigma=0}^{\sigma=l}.
\label{boundary_term_0}
\end{equation}
The first part gives equations of motion
\begin{equation}
0=\del_+\psi_-=\del_-\psi_+,
\end{equation}
while the second part reads boundary terms which should vanish
with appropriate boundary conditions.
For closed strings, these should be done by imposing the
periodic/antiperiodic boundary conditions
$\psi^M_\pm(l)=\eta\psi^M_\pm(0)$ with $\eta^2=1$.
For open strings, we need to impose conditions between
left/right movers at the boundaries:
$\psi^M_\pm(0)=\psi^M_\mp(0)$ and
$\psi^M_\pm(l)=\eta \psi^M_\mp(l)$.

Let us next consider residual supersymmetries
after taking the superconformal gauge.
Under transformations
\begin{subequations}
\begin{eqnarray}
 \delta X^M
&=&
2i\alpha'\bar\epsilon\psi^M
=2i\alpha'(\epsilon_+\psi_-^M-\epsilon_-\psi_+^M),
\\
\delta\psi^M
&=&
(\rho^m\epsilon)\del_mX^M, \qquad \mbox{i.e.} \qquad
\delta\psi_\pm^M=\pm2\epsilon_\mp\del_\pm X^M,
\end{eqnarray}
\end{subequations}
the total action (\ref{polyakov_light_cone}) and
(\ref{fermion}) transforms as
\begin{eqnarray}
 \delta\big(S_B+S_F\big)
&=&
\frac{4i}{\pi}\!\int\!
\dd^2\sigma
\big(
(\del_+\epsilon_+)\psi_-^M\del_-X_M
-(\del_-\epsilon_-)\psi_+^M\del_+X_M
\big)
\nonumber
\\
&&
-\frac{i}{\pi}\!\int\!
\dd\tau
\big(
\epsilon_+\psi_-^M\del_+X_M
+\epsilon_-\psi_+^M\del_-X_M
\big)\big|_{\sigma=0}^{\sigma=l},
\label{boundary_term_1}
\end{eqnarray}
where $\epsilon(\sigma)=(\epsilon_+(\sigma), \epsilon_-(\sigma))^{\rm T}$ 
are infinitesimal supersymmetric parameters.
Therefore, residual symmetries are realized by holomorphic functions
$\epsilon_\pm(\sigma)=\epsilon_\pm(\sigma^\mp)$ with suitable
boundary conditions which make boundary terms vanish. 
Together with the bosonic part, residual symmetry becomes 
superconformal symmetry.

Now let us discuss the mode expansions
 taking the boundary
terms in (\ref{boundary_term_0}) and (\ref{boundary_term_1}) into account.
For closed fermionic strings, we obtain the following
solutions,
\begin{eqnarray}
\psi_\pm^M(\tau, \sigma)
=\sqrt{\frac{2\pi}{l}}\sum_{n\in Z} d^M_n\,
\eee^{-i\frac{2\pi}{l}n\sigma_\pm} \qquad \mbox{and} \qquad
\psi_\pm^M(\tau, \sigma)
=\sqrt{\frac{2\pi}{l}}\!\!\sum_{n\in Z+\frac{1}{2}}\!\!
b^M_n\, \eee^{-i\frac{2\pi}{l}n\sigma_\pm}
\label{closed_rns}
\end{eqnarray}
The first one which satisfies the periodic boundary condition is
called Ramond sector,
while the second one that satisfies anti-periodic boundary condition is
called Neveu-Schwarz sector.
Therefore, in closed strings which contain independent left/right
movers, there exist four sectors
i.e.\ Ramond-Ramond, (Neveu-Schwarz)-(Neveu-Schwarz),
(Neveu-Schwarz)-Ramond and Ramond-(Neveu-Schwarz) sectors.
The consistency at string loop level requires that all four sectors should be included.

For open strings, we have options for choosing the
Neumann/Dirichlet boundary conditions for bosonic strings
(\ref{boson_nn}), (\ref{boson_dd}) and (\ref{nd}).
Since supersymmetries relate bosonic/fermionic strings,
we could also take various directions of fermionic strings such as
(NN), (ND), and (DD) directions.
Defining Ramond and Neveu-Schwarz sector in (NN) directions as

\noindent
{\bf (NN)}
\begin{subequations}
\begin{equation}
(\mbox{R}) \qquad
\psi_\pm^M(\tau, \sigma)
=
\sqrt{\frac{\pi}{l}}\sum_{n\in Z}d^M_n\,
\eee^{-i\frac{\pi}{l}n\sigma_\pm} \qquad
(\mbox{NS}) \quad
\psi_\pm^M(\tau, \sigma)
=
\sqrt{\frac{\pi}{l}}\!\!\sum_{n\in Z+\frac{1}{2}}\!\!b^M_n\,
\eee^{-i\frac{\pi}{l}n\sigma_\pm}  \quad
\end{equation}
which could fix boundary conditions of the supersymmetric
parameters,
we can sequentially find mode expansions for each sectors
in another directions:

\noindent
{\bf (DD)}
\begin{equation}
(\mbox{R}) \qquad
\psi_\pm^M(\tau, \sigma)
=
\pm\sqrt{\frac{\pi}{l}}\sum_{n\in Z}d^M_n\,
\eee^{-i\frac{\pi}{l}n\sigma_\pm} \qquad
(\mbox{NS}) \quad
\psi_\pm^M(\tau, \sigma)
=
\pm\sqrt{\frac{\pi}{l}}\!\!\sum_{n\in Z+\frac{1}{2}}\!\!b^M_n\,
\eee^{-i\frac{\pi}{l}n\sigma_\pm}  
\end{equation}
\noindent
{\bf (ND)}
\begin{equation}
(\mbox{R}) \qquad
\psi_\pm^M(\tau, \sigma)
=
\sqrt{\frac{\pi}{l}}\!\!\sum_{n\in Z+\frac{1}{2}}\!\!d^M_n\,
\eee^{-i\frac{\pi}{l}n\sigma_\pm} \qquad
(\mbox{NS}) \quad
\psi_\pm^M(\tau, \sigma)
=
\sqrt{\frac{\pi}{l}}\sum_{n\in Z}b^M_n\,
\eee^{-i\frac{\pi}{l}n\sigma_\pm}
\end{equation}
\end{subequations}
It should be noted that
in the (ND) directions the Ramond and the Neveu-Schwarz sectors
turn to have half-integer and integer mode numbers, respectively.
As in the closed string, Hilbert space should contain both  R and NS sectors.

We now quantize the fermionic string canonically, given
the anti-commutation relation:
\begin{equation}
\{d^M_m, \ d^N_n\}=\{b^M_m, \ b^N_n\}=\delta_{m+n}\eta^{MN}.
\label{anti-commutators}
\end{equation}

The ground state of NS sector is defined
by $\alpha_m^M|0\rangle_{\rm NS}=b^M_n|0\rangle_{\rm NS}=0$ $(m>0,
n>0)$,
and similar conditions should be satisfied for
$\widetilde{\alpha}^M_m$ and $\widetilde{b}^M_n$ in closed string.
As the definition above, the ground state $|0\rangle_{\rm NS}$ is unique
and describes the spacetime scalar.
We could build up the Fock space on the ground state by multiplying the
oscillator modes $\alpha^M_{-m}$ and $b^M_{-r}$ $(m>0, r>0)$.

For R sector, non-zero modes $d^M_n$ $(n>0)$ play a similar role,
$\alpha^M_m|0\rangle_{\rm R}=d^M_n|0\rangle_{\rm R}=0$.
However, we observe  zero-mode parts of the anti-commutation relations
in (\ref{anti-commutators}) follow the Clifford algebra
with replacements $d^M_0=\Gamma^M/\sqrt{2}$,
\begin{equation}
\{d^M_0, \ d^N_0\}=\eta^{MN}.
\end{equation}
This indicates the Ramond ground states represent the Clifford algebra
in $D$-dimensional spacetime,
i.e.\ spacetime spinor.
The world-sheet supersymmetry has generated spacetime fermion.

As we did in the bosonic string,
we would like to construct mass relations.
In order to fix the normal ordering constant,
we use the zeta function regularization.
The result is given by
\begin{equation}
a_{\rm integer\ mode}=-\frac{1}{24},
\qquad
a_{\rm half-integer \ mode}=+\frac{1}{48},
\end{equation}
which would be compared with (\ref{nndd}) and (\ref{ndnd}).
By using these constants, we could estimate the mass relation
\begin{subequations}
\begin{eqnarray}
M^2_{\rm open}
&=&\frac{1}{\alpha'}\big(N_{X}+ N_\psi^{\rm R}\big)
+\frac{(x_0-x_1)^2}{(2\pi\alpha')^2}, \qquad
\mbox{(R)}
\label{open_r}
\\
M^2_{\rm open}
&=&
 \frac{1}{\alpha'}\big(N_X+N_\psi^{\rm NS}\big)
+\frac{(x_0-x_1)^2}{(2\pi\alpha')^2}
-\frac{1}{\alpha'}\Big(\frac{1}{2}-\frac{n}{8}\Big),
\qquad \mbox{(NS)}\, ,
\label{open_ns}
\end{eqnarray}
\end{subequations}
where the newly added number operator of the fermionic part
is given by $N_\psi^{\rm R}=\sum_{n>0}nd_{-n}^Md_{nM}$
and $N_\psi^{\rm NS}=\sum_{n>0}nb_{-n}^Mb_{nM}$.
$M$ runs through (NN), (DD), (ND) directions and
$n$ should be understood as integer/half-integer depending on the
directions.
The mass spectrum does depend on the number of (ND) directions $n$.

We need to impose so-called GSO projection.
We first introduce an operator $F$ which counts the number of
oscillators $F_{\rm NS}=\sum_{r>0}b_{-r}b_{r}$ and
$F_{\rm R}=\sum_{r>0}d_{-r}d_r$ for each sectors.
We now define the operator $G=(-)^{F_{\rm NS}+1}$ and
$G=\Gamma^{11}(-)^{F_{\rm R}}$ with the 10D chirality operator
$\Gamma^{11}$.
Then GSO projection implies $G|\mbox{phys}\rangle=|\mbox{phys}\rangle$.
Through this projection, in the NS sector, states with only odd numbers of
oscillators could survive in the Hilbert space.
This removes tachyonic ground state and keeps the first excited states
which correspond to the massless gauge (scalar) fields.
In the R sector, we could project states with either  even or odd
numbers of oscillator excitations depending on the chirality of the
spinor ground state.
The GSO projection provides spacetime supersymmetry and also ensures
the modular invariance.
Depending on keeping/removing states, we obtain different theories.
Type IIA/IIB is just in this category.

\subsubsection{Supergravity}

Superstring contains a spacetime supersymmetry and also includes gravity.
Therefore, the low energy effective action could be that of supergravity.
Bosonic contents of the supergravity multiplets are
a graviton and its family ($G_{MN}(x)$, $B_{MN}(x)$, $\Phi(x)$) as in the
common sector referred to as NSNS sector,
and $p$-form gauge fields $C_p(x)$ with $p=1, 3$
for type IIA and
with $p=0, 2, 4$ for type IIB in RR sector.
Other sectors in closed strings i.e.\ RNS and NSR provide
superpartners of those fields.
In this paper we
only display these bosonic contents in the supergravity theories.
Note that in RR sectors in Type II string theories, 
physical states are identified not with gauge potentials 
but with their field strength. 
In other words, there exist no states with RR charges 
in string perturbation theories.
As we will discuss, RR charge can appear through D-branes 
in nonperturbative string theory.

The common NSNS sector of IIA/IIB supergravity action
is the same with (\ref{effective_action_bc}) with
\begin{equation}
D=D_{\rm c}=10.
\end{equation}
The actions for $p$-form gauge fields in RR sector
are given by
\begin{subequations}
\begin{eqnarray}
S^{\rm IIA}_{\rm RR}
&=&
-\frac{1}{4\kappa_0^2}
\!\int_{\cal M}\!\!\dd^{10}x
\sqrt{-G}
\bigg(
|F_2|^2+|\widetilde{F}_4|^2\bigg)
-\frac{1}{4\kappa_0^2}
\!\int_{\cal M}\!\!
B_2\wedge F_4\wedge F_4,
\label{IIa_sugra}
\\
S^{\rm IIB}_{RR}
&=&
-\frac{1}{4\kappa_0^2}
\!\int_{\cal M}\!\!\dd^{10}x
\sqrt{-G}
\bigg(|F_1|^2+|\widetilde{F}_3|^2
+\frac{1}{2}|\widetilde{F}_5|^2\bigg)
-\frac{1}{4\kappa_0^2}\!\int_{\cal M}\!\!C_4\wedge H_3\wedge F_3,
\label{IIb_sugra}
\end{eqnarray}
\end{subequations}
with
$$
\widetilde{F}_4=F_4-C_1\wedge H_3,
\qquad
\widetilde{F}_3=F_3-C_0\wedge H_3,
\qquad
\widetilde{F}_5=F_5-\frac{1}{2}C_2\wedge H_3+\frac{1}{2}B_2\wedge F_3\, .
$$
Here we define the field strength of so-called RR $p$-form gauge
fields as $F_{p+1}(x)=\dd C_p(x)$ and denote $H_3(x)=\dd B_2(x)$ for the
Kalb-Ramond two-form field.
We need to impose the self-dual condition $*\widetilde{F}_5(x)=F_5(x)$
at the level of the equations of motion for IIB.
The massless closed superstring spectrum does not only contain the
graviton and its family but also RR $p$-forms.
Apart from the Chern-Simons terms which are the second parts of
the actions (\ref{IIa_sugra}) and (\ref{IIb_sugra}),
schematically, we write the low energy effective action for type II
superstring theories as
\begin{equation}
S_{{\rm NSNS-RR} p}
=\frac{1}{2\kappa_0^2g_{\rm s}^2}
\!\int_{\cal M}\!\!\dd^{10}x\sqrt{-G}
\bigg\{\eee^{-2\widetilde{\Phi}}
\Big(R+4(\del_M\widetilde{\Phi})^2
-\frac{1}{12}H_{MNL}H^{MNL}\Big)
-\frac{g_{\rm s}^2}{2}|F_{p+2}|^2\bigg\},
\label{sp_0}
\end{equation}
where we have divided $\Phi(x)=\Phi_0+\widetilde{\Phi}(x)$ as before.
The action (\ref{sp_0}) is defined in the string frame which
is constructed from the string perturbation theory.
In order to obtain the canonical form of the Einstein-Hilbert action
we could rescale the string frame metric $G_{MN}(x)$ to that of the
Einstein frame $G_{MN}^{\rm E}(x)$ through the Weyl
rescaling, 
\begin{equation}
G_{MN}^{\rm E}(x)=\eee^{-\widetilde{\Phi}(x)/2}G_{MN}(x),
\end{equation}
so that
\begin{equation}
S_{{\rm NSNS-RR} p}
=
\frac{1}{2\kappa_0^2g_{\rm s}^2}
\!\int_{\cal M}\!\!\dd^{10}x\sqrt{-G^{\rm E}}
\bigg\{R(G^{\rm E})
-\frac{1}{2}(\del_M\widetilde{\Phi})^2
-\frac{1}{12}\eee^{-\widetilde{\Phi}}H_{MNL}H^{MNL}
-\frac{g_{\rm s}^2}{2}\ \eee^{\frac{3-p}{2}\widetilde{\Phi}}
|F_{p+2}|^2\bigg\}.
\label{spe}
\end{equation}
Here the kinetic term of the dilaton field also becomes its
canonical form.
The canonical form for Einstein-Hilbert action is
used in this paper to estimate the masses of the objects.
By using this form, we define the 10D Newton constant $G_{\rm N}^{(10)}$
as
\begin{equation}
16\pi G_{\rm N}^{(10)}=2\kappa_0^2g_{\rm s}^2\equiv2\kappa^2.
\label{10d_newton}
\end{equation}
In the dimensional analysis,
the gravitational coupling constant can be estimated in terms of the
string length $\sqrt{\alpha'}$,
\begin{equation}
\kappa\propto\alpha'^2.
\end{equation}
%

\subsubsection{RR $p$-brane solutions}

In supergravity theories, there exist various classes of exact
solutions for equations of motion derived from 
the supergravity action~\cite{p-brane}.
Let us consider $p$-brane solitonic solutions which couple to $(p+1)$-form
gauge fields.
We are interested in the classical solutions which contain minimal
subset of the supergravity field contents, i.e.\ the graviton, the
dilaton, and $(p+1)$-from gauge fields.
Setting the Kalb-Ramond field $B_{MN}(x)=0$,  we consider the following
truncated action:
\begin{equation}
\widetilde{S}_{{\rm NSNS-RR} p}
=\frac{1}{2\kappa_0^2g_{\rm s}^2}
\!\int_{\cal M}\!\!\dd^{10}x\sqrt{-G}
\bigg\{\eee^{-2\widetilde{\Phi}}
\Big(R+4(\del_M\widetilde{\Phi})^2
\Big)
-\frac{g_{\rm s}^2}{2}|F_{p+2}|^2\bigg\}.
\label{sp}
\end{equation}

We would like to find a solution like a static charged point particle.
The ``point'' would be replaced to the $p$-dimensional flat object,  and
the transverse directions to the object might have spherical symmetry
where the object looks charged ``point''.
In addition, in order to isolate the extended object,
we need to impose asymptotically flat condition to the transverse
directions.
With these ansatz, the solution has the following form in $p<7$ in the
string frame \cite{bpbrane};
\begin{eqnarray}
\dd s^2
&=&
\frac{1}{\sqrt{H_p(r)}}
\Big(-f(r)\dd t^2+\sum_{i=1}^p\dd x^i{}^2\Big)
+\sqrt{H_p(r)}\Big(\frac{\dd r^2}{f(r)}+r^2\dd\Omega^2_{8-p}\Big),
\nonumber
\\
\eee^{\widetilde{\Phi}}
&=&
H_p^{(3-p)/4}(r),
\label{non_extremal}
\\
F_{p+2}
&=&
\frac{1}{g_{\rm s}}\sqrt{1+\frac{r_{\rm H}^{7-p}}{L^{7-p}}}
\ \dd t\wedge\dd x^1\wedge\cdots\wedge\dd x^p\wedge\dd(H_p^{-1}(r)),
\quad
\Big(\mbox{or} \quad F_{8-p}=*F_{p+2}\Big),
\nonumber
\end{eqnarray}
with
\begin{equation}
H_p(r)=1+\frac{L^{7-p}}{r^{7-p}},
\qquad
f(r)=1-\frac{r_{\rm H}^{7-p}}{r^{7-p}}.
\label{harmonic}
\end{equation}
We have parameterized the $(p+1)$-dimensional world volume
coordinates as $(t, x^1, \cdots, x^p)$ and the transverse
$(9-p)$-dimensional space by
the polar coordinate system with
radial coordinate $r\, \big(=\sqrt{(x^{p+1})^2+\cdots+(x^9)^2}\, \, \big)$.
The functions $H_p(r)$ and $f(r)$ are harmonic functions of the transverse
coordinates, meaning $0=(\del_{p+1}^2+\cdots+\del_9^2)H_p(r)$ and
the same for $f(r)$.
They are normalized such that the metric becomes
asymptotically  flat in the spatial infinity.
In the case of $p=3$, we need to add the Hodge dual of the expression
$F_5$ in (\ref{non_extremal}) to
satisfy the self-dual relation $*F_5=F_5$.

Two integration constants $L$ and $r_{\rm H}$ could be related
with physical parameters i.e.\ charge and mass.
Using the expression (\ref{q_charge_00}) and taking
the normalization of the action (\ref{sp}) into account,
we can define charge $Q_p$ in the covariant way
through the Gauss' law which is an integration of
flux over the transverse sphere $S^{8-p}$ (cf. (\ref{q_charge_00})),
\begin{equation}
Q_p\equiv\frac{1}{2\kappa_0^2}\!\int_{S^{8-p}}\!\!\!*F_{p+2}
=\frac{V_{S^{8-p}}}{2\kappa_0^2g_{\rm s}}(7-p)L^{\frac{7-p}{2}}
\sqrt{L^{7-p}+r_{\rm H}^{7-p}},
\label{charge_bp}
\end{equation}
where $V_{S^{8-p}}$ is the volume of unit sphere $S^{8-p}$.
The same result can be obtained in the Einstein frame, although the
definition of charge should be modified due to the dilaton factor for
the gauge fields in (\ref{spe}).
It is easy to see that at $r\to\infty$
the electric field behaves like a point
charge in the transverse space, which is the generalized Coulomb
potential,
$$
F_{p+2}=\frac{2\kappa_0^2g_{\rm s}Q_p}{V_{S^{8-p}}}\frac{1}{r^{8-p}}
\ \big(\dd t\wedge\dd x^1\wedge\cdots\wedge\dd x^p\wedge\dd r\big)+\cdots.
$$
Since the solution is asymptotically flat in the transverse space,
we could define a mass $M_p$.
As mentioned before for the point particle,
the mass could be read off from the deviation from the flat
metric in the asymptotic region (\ref{mass_formula}).
In order to proceed, we need to take two things into account.
First, the procedure only works for the action (\ref{eh})
i.e.\ in the canonical Einstein frame.
Second, our extended object has infinite volume in its longitudinal
directions.
We first regularize this volume and rewrite the action into the
canonical Einstein-Hilbert action for
point particle source
in the $(1+(9-p))$-dimensional transverse spacetime,
$$
\widetilde{S}_{{\rm NSNS-RR}p}
=\frac{V_p}{2\kappa_0^2g_{\rm s}^2}\!\int\!\dd^{10-p}x
\sqrt{-\hat{G}}\Big( R(\hat{G}) +\cdots\Big),
$$
where $V_p$ is the $p$-dimensional spatial volume of the $p$-brane and
we have eliminated the dilaton factor by using Weyl rescaling formula.
The resultant metric is
\begin{equation}
\hat{G}_{tt}=-1
+\bigg(\frac{7-p}{8-p}L^{7-p}+r_{\rm H}^{7-p}\bigg)\frac{1}{r^{7-p}}
+\cdots.
\label{gtt_asympt}
\end{equation}
Comparing with (\ref{mass_formula}),
we find
\begin{equation}
\frac{M_p}{V_p}=\frac{V_{S^{8-p}}}{2\kappa_0^2g_{\rm s}^2}
\Big(
(7-p)L^{7-p}+(8-p)r_{\rm H}^{7-p}
\Big).
\label{tension_bp}
\end{equation}
This $M_p/V_p$ can be interpreted as the tension
which is a natural analog of the mass of the $p$-brane.
The Schwarzschild radius $r_{\rm S}$ can be easily estimated through
(\ref{gtt_asympt}),
\begin{equation}
r_{\rm S}^{7-p}=\frac{7-p}{8-p}L^{7-p}+r_{\rm H}^{7-p}.
\end{equation}

In analogy with the charged black hole solution or Reissner-Nordstr\"om black
hole solution, the solution (\ref{non_extremal}) is known as a ``black'' $p$-brane
solution. 
In particular, this has two horizons which are related with $r=0$ and
$r_{\rm H}$.
To avoid the naked singularities, physical solutions may be restricted
in the parameter region $r_{\rm H}\ge0$.
This is reflected in terms of the tension and charge obtained in
(\ref{tension_bp}) and (\ref{charge_bp}) as
\begin{equation}
g_{\rm s}\frac{M_p}{V_p}\ge Q_p,
\label{bps_bound}
\end{equation}
which is known as the BPS bound.
The extremal case $r_{\rm H}=0$ is called BPS configurations which
correspond to the ground state.
The BPS state implies that  (tension)$=$(charge).
The situation can be understood in terms of supersymmetry.
In fact, the configuration given in (\ref{non_extremal}) preserves a half of
32 real supersymmetry transformations in type II supergravity
theories (1/2 BPS states).
In the theories controlled by supersymmetric algebra having the central
charges, the relation (\ref{bps_bound}) describes the mass of a
state and its central charge  which is the RR $p$-brane charge in our case.
If the inequality is saturated,
the unitary representation of the algebra belongs to a special class i.e.\
the short representation.
The relation for the BPS state is a consequence of the supersymmetry
algebra which is protected from quantum corrections.
Therefore, this minimum energy states are completely stable.

The case with $p=3$ is special since singularities totally disappear.
The dilaton field is constant.
Therefore, the black RR 3-brane solution describes
a smooth solitonic object in the type IIB supergravity theory.
All the other brane solutions contain singularities.

Let us look at the black RR 3-brane solution in the case of $p=3$ in
(\ref{non_extremal}),
\begin{equation}
\dd s^2
=\frac{1}{\sqrt{\displaystyle1+\frac{L^4}{r^4}}}
\bigg\{
-\Big(1-\frac{r_{\rm H}^4}{r^4}\Big)\dd t^2+\sum_{i=1}^3\dd x^{i2}
\bigg\}
+\sqrt{1+\frac{L^4}{r^4}}
\bigg\{\frac{\dd r^2}{\Big(\displaystyle1-\frac{r_{\rm H}^4}{r^4}\Big)}
+r^2\dd\Omega^2_5
\bigg\}.
\label{ads_s0}
\end{equation}
The outer horizon appears at $r=r_{\rm H}$.
We further consider the extremal case $r_{\rm H}=0$:
\begin{equation}
\dd s^2
=\frac{1}{\sqrt{\displaystyle1+\frac{L^4}{r^4}}}
\bigg\{
-\dd t^2+\sum_{i=1}^3\dd x^{i2}
\bigg\}
+\sqrt{1+\frac{L^4}{r^4}}
\bigg\{\dd r^2
+r^2\dd\Omega^2_5
\bigg\}.
\label{ads_00}
\end{equation}

\begin{figure}[htbb]
\begin{minipage}{1.0\textwidth}
\begin{center}
\includegraphics[width=5cm]{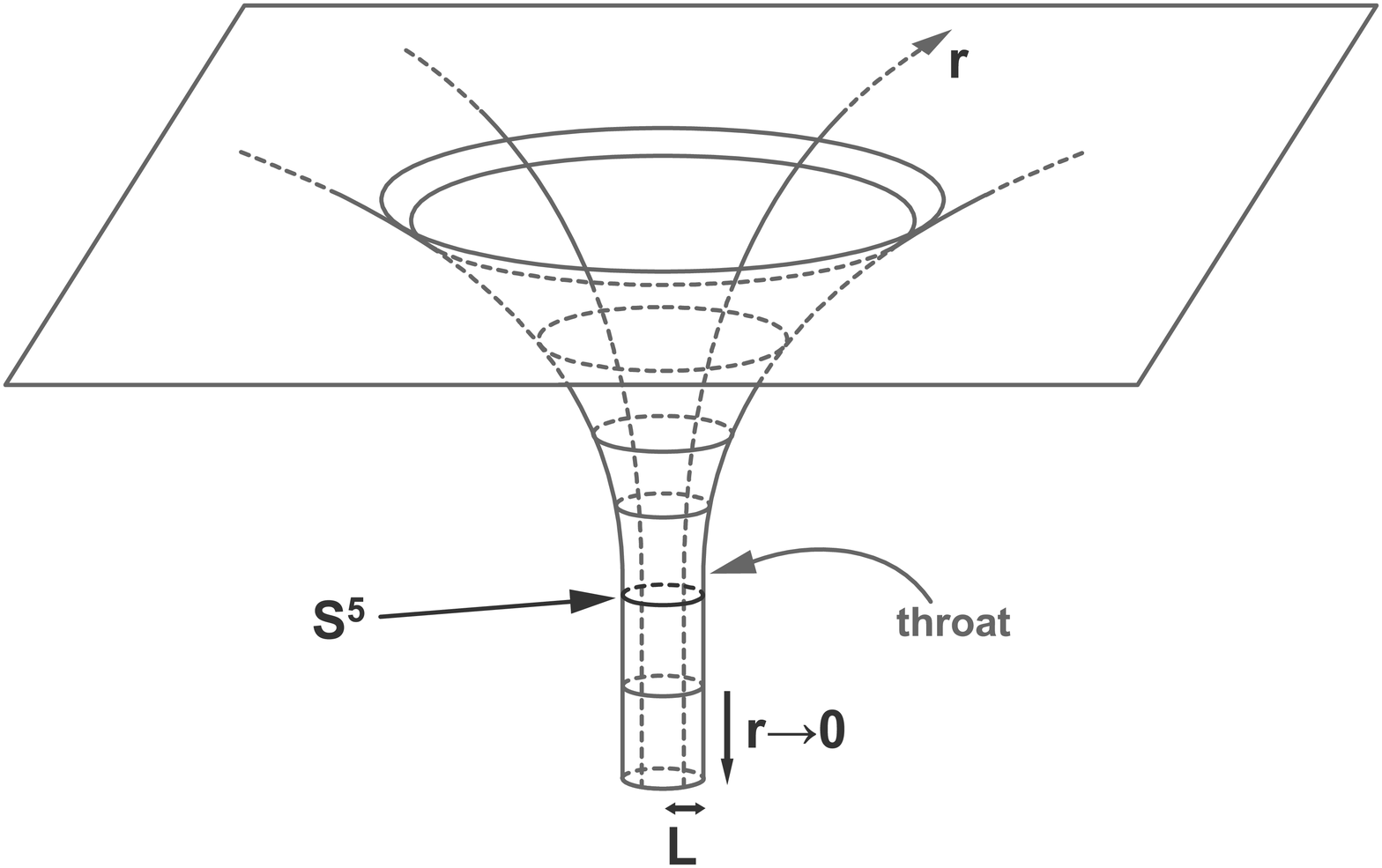}
\caption{Extremal $p$-brane}
\label{ads_embedding}
\end{center}
\end{minipage}
\end{figure}

We divide the spacetime in the metric (\ref{ads_00})
into two regimes $r>L$ and $r<L$.
In the region $r\gg L$ the metric (\ref{ads_00}) describes
the 10D flat Minkowski spacetime.
In the other limit $r\ll L$ where the harmonic function can
be approximately $H_3(r)=1+L^4/r^4\to L^4/r^4$,
we obtain the metric
\begin{equation}
\dd s^2=\underbrace{\frac{r^2}{L^2}
\bigg\{
-\dd t^2+\sum_{i=1}^3\dd x^{i2}
\bigg\}
+\frac{L^2}{r^2}\ \dd r^2}_{\scriptsize\mbox{AdS$_5$}}
+
\underbrace{L^2\dd\Omega^2_5}_{S^5}.
\label{ads_0}
\end{equation}
The metric has been divided into two parts, i.e.\  AdS$_5\times S^5$
with the same radius $L$.
We refer  to (\ref{ads_0}) as the near horizon geometry
where $r=0$ corresponds to (Killing) horizon.
The geometry (\ref{ads_00}) could be visualized as the
embedding surface in Fig.\ref{ads_embedding}.
Two asymptotic regions are separated by an infinitely long
``throat'' with the constant radius.

The black RR 3-brane solution is a solitonic solution which interpolates two
vacua, i.e.\ flat Minkowski space at infinity and the AdS$_5\times S^5$ in
the near horizon region.
It should be mentioned
that these asymptotic vacua preserve all 32 supersymmetries in
type IIB supergravity although the full solution breaks half of
them \cite{gunaydin_marcus}.

\subsection{D-brane}

D$p$-brane has been introduced through the boundary
conditions of open string (\ref{neumann_dirichlet}) and
defined by the $(p+1)$-dimensional hypersurface
where the movement of open string endpoints
are restricted.
This rigid hypersurface turns out to be a dynamical object through the
open string ending on it.
Moreover, closed string could also interact with D-brane through the
open string interactions.

\begin{figure}[htbb]
\begin{minipage}{1.0\textwidth}
\begin{center}
\includegraphics[clip, width=5.5cm]{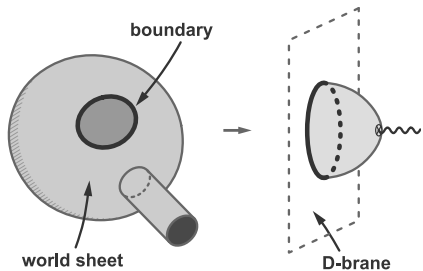}
\caption{Interactions around D-brane}
\label{interaction_dpbrane}
\end{center}
\end{minipage}
\end{figure}

Let us consider the D$p$-brane from the viewpoint  of the string
world sheet.
Suppose there is a closed string world sheet.
One can consider the interaction with gravity through the vertex
operator insertions.
Now let us consider the case of the world sheet with boundaries.
With the Dirichlet boundary conditions imposed in suitable directions,
D-brane could appear at the boundary.
Typical world sheet is given by the left side of Fig.\ref{interaction_dpbrane}.
After appropriate conformal mapping, one can smoothly deform the world
sheet to the right.
This can be interpreted as the interaction between the D-brane and
closed string including graviton.
One can estimate the strength of the interaction through the world
sheet topology.
Since the topology of the world sheet is a disc,
in terms of the string perturbation (\ref{partition_1}),
the weight of the path-integral is given by $1/g_{\rm s}$.
Since the presence of D-brane can be measured by its
tension, we could estimate the tension of D$p$-brane as
\begin{equation}
T_p \propto \frac{1}{g_{\rm s}}.
\end{equation}

As we have discussed, RR $p$-branes appear as solitonic solutions
in supergravity theory. 
These have $p$-form charges, and the extremal solutions are
BPS configurations. 
Polchinski identified this RR $p$-brane as the D$p$-brane and
showed the D-brane is a necessary object for consistency of the interactions
between open and closed superstrings \cite{Polchinski:1995mt}.

\subsubsection{D$p$-brane effective action}

We could generalize the action (\ref{dbi_0}) to D$p$-brane effective action
by simply replacing\footnote{
One of the proper treatments is given via T-duality transformation.
}
$ A_M(x)\to (A_m(\sigma), \ \phi_a(\sigma))$, 
where $m$ is for the world volume coordinate and $a$ is the 
direction for the Dirichlet boundary condition.
The scalar $\phi_a(\sigma)$ corresponds to the collective coordinates
of D$p$-brane.
In this case, we also need to replace the bulk field $G_{MN}(X)$ and
$B_{MN}(X)$ by ``induced'' quantities $g_{mn}(\sigma)$
and $B_{mn}(\sigma)$ defined before,
\begin{equation}
S_{\rm DBI}
=-T_p\!\int_{\Sigma}\!\dd^{1+p}\sigma\ \eee^{-\widetilde{\Phi}}
\sqrt{-\det(g_{mn}+B_{mn}+2\pi\alpha'F_{mn})}.
\label{dbi}
\end{equation}
Here we fix the D$p$-brane tension as\footnote{
We use the normalization $T_1\equiv1/(2\pi\alpha'g_{\rm s})=T/g_{\rm s}$
for D1-brane. The recursive relation $T_p=2\pi\sqrt{\alpha'}T_{p+1}$
follows from T-duality.
}
\begin{equation}
T_p=\frac{1}{(2\pi)^p\alpha'^{\frac{p+1}{2}}g_{\rm s}}.
\label{tension_p}
\end{equation}
This is the generalization of Nambu-Goto type action
(\ref{world_volume}).
The embedding coordinates $X^M(\sigma)$
defined through the induced metric $g_{mn}(\sigma)$
(\ref{induced_metric}) represent the shape of the D$p$-brane in the
target spacetime.
We divide the embedding functions to
longitudinal and transverse parts to D$p$-brane
$X^M(\sigma)=(X^m(\sigma),  X^a(\sigma)\equiv
2\pi\alpha'\phi^a(\sigma))$,
and take the static gauge for the world volume reparameterization
invariance $X^m(\sigma)=\sigma^m$.

Let us perturb the DBI action (\ref{dbi}) around the flat NSNS background
in the Einstein frame (\ref{spe}),
\begin{equation}
G_{MN}^{\rm E}(x)=\eta_{MN}+\kappa\widehat{h}_{MN}(x), \qquad
B_{MN}(x)=0+\kappa \widehat{b}_{MN}(x), \qquad
\widetilde{\Phi}(x)=0+\kappa\widehat{\Phi}(x)\, .
\label{perturbation_nsns}
\end{equation}
 On this background, we do expand the DBI action for small $\alpha'F_{mn}$.
Similar to (\ref{perturbation_gravity}),
the gravitational coupling constant $\kappa$ defined in
(\ref{10d_newton}) appears such that
the kinetic terms of the graviton and its family members are correctly
normalized.
Then, up to the linear order of the perturbation, we find
\begin{eqnarray}
S_{\rm DBI}
=-T_p(2\pi\alpha')^2\!\int_{\Sigma}\!\dd^{1+p}\sigma
\bigg\{
&&
\frac{1}{(2\pi\alpha')^2}
+\frac{\kappa}{(2\pi\alpha')^2}\bigg(\frac{p-3}{4}\widehat{\Phi}
+\frac{1}{2}\widehat{h}_m{}^m
\bigg)
+\bigg(\frac{1}{4}F_{mn}F^{mn}
+\frac{1}{2}\del_m\phi_a\del^m\phi^a\bigg)
\nonumber
\\
&&
+\frac{\kappa}{2\pi\alpha'}
\Big(\widehat{h}_{ma}\del^m\phi^a+\frac{1}{2}\widehat{b}_{mn}F^{mn}\Big)
\nonumber
\\
&&
+\kappa
\bigg(\widehat{\Phi}\Big(\frac{p-7}{16}F_{mn}F^{mn}
+\frac{p-3}{8}\del_m\phi_a\del^m\phi^a\Big)
\nonumber
\\
&&
\hspace*{8mm}
+\frac{1}{2}\widehat{h}_{mn}T^{mn}
+\frac{1}{2}\widehat{h}_{ab}\del_m\phi^a\del^m\phi^b
\bigg)
+\cdots
\bigg\},
\label{dbi_exp}
\end{eqnarray}
with
\begin{equation}
T_{mn}
=-F_{ml}F_n{}^l+\frac{1}{4}\eta_{mn}F_{kl}F^{kl}
-\del_m\phi_a\del_n\phi^a+\frac{1}{2}\eta_{mn}\del_l\phi_a\del^l\phi^a.
\label{emt_dbi}
\end{equation}
The first part corresponds to the tree level vacuum energy, and
the second part gives ``source'' terms of the dilaton 
and the graviton that are from
massless closed string modes\footnote{
The source term for the Kalb-Ramond field $B_{mn}(x)$ may not be given
by the DBI action but the string world sheet (\ref{1wz}).
}.
The third term provides the action for Yang-Mills and scalar fields
living in the world volume.
Comparing with the canonical normalization of the kinetic term of the
vector field, we can relate $p+1$-dimensional Yang-Mills coupling constant
to the tension (\ref{tension_p}),
\begin{equation}
g_{\rm YM}^2=(2\pi)^{p-2}\alpha'^{\frac{p-3}{2}}g_{\rm s}.
\label{coupling_0}
\end{equation}
$T_{mn}(x)$ defined in (\ref{emt_dbi})
can be understood as the energy-momentum tensor
in the world volume gauge theory.

We have considered the DBI action (\ref{dbi}) as the low energy
effective action of D-brane which couples to the NSNS background,
the graviton, the dilaton, and the Kalb-Ramond field.
As we discussed, the D$p$-brane naturally couples to the $(p+1)$-form
gauge field.
The natural candidates for this gauge potentials are
RR-gauge potentials in the RR sector.
Their low energy couplings are given by the Wess-Zumino term,
\begin{equation}
S_{\rm WZ}
=-\mu_p\sum_p
\!\int_{\Sigma_{p+1}}\!\!\!\!
C_{p+1}\wedge \eee^{B+2\pi\alpha'F}
=-\mu_p
\!\int_{\Sigma_{p+1}}\!\!\!\!
\Big(C_{p+1}+C_{p-1}\wedge (B+2\pi\alpha'F)
+\cdots\Big),
\label{wz_p}
\end{equation}
where the integration should be taken on the D$p$-brane world
volume. The $p$-form gauge field $C_p(\sigma)$ and Kalb-Ramond two-form
field $B(\sigma)$ have to
be understood as the induced quantities of the world volume.
The first term is a natural coupling given by (\ref{pwz}).
In addition, there exist nontrivial couplings with lower dimensional
RR forms.
For instance, for D3-brane, the WZ term (\ref{wz_p})
provides the term
$$
-\mu_3(2\pi\alpha')^2 \!\int\!C_0 F\wedge F
=\mu_3(2\pi\alpha')^2\!\int\!\dd^4\sigma
\sqrt{-g}\Big(\frac{C_0}{4}\epsilon^{mnkl}F_{mn}F_{kl}\Big),
$$
where the scalar field $C_0(\sigma)$ corresponds to the axion
which may fix the theta angle.
We will also use the WZ term to construct baryons in the holographic
QCD.

Through the open string fluctuations we obtained the
Yang-Mills system in $(1+p)$ dimensions
as the low energy effective action of D$p$-brane.
Instead of a single D-brane,
we now consider a stack of $N$ parallel D-branes which are placed on top of each
other.
There are now $N^2$ different types of open strings, where each type
of string starts from a particular one of the $N$ branes
and ends on a particular one\footnote{
Open strings which we discuss have an orientation.
A string starting from D-brane $i$ and ending on $j$
is different from the one starting from $j$ and ending on $i$.
}
.
As we have discussed, the endpoint of string  is charged under $U(1)$ gauge
field.
We assume that the low-energy theory describing such a stack of branes
has basically the same form with the theory
we have discussed, but  the fields are promoted to
$N\times N$ matrices i.e.\ $A_m\to (A_m)^i{}_j$ and
$\phi_a\to(\phi_a)^i{}_j$.
Therefore, the gauge theory becomes $U(N)$ gauge theory.
The diagonal components of the scalar fields $(\phi_a)^i{}_i$
correspond to the transverse fluctuations
of the $m^{\rm th}$ individual branes.
Although the direct way to obtain the effective action of such a system is
to generalize the BI action (\ref{dbi}) for nonabelian case,
it is not easy to write down this explicitly.
Alternatively,  under the small field strength approximation,
we generalize the Yang-Mills part of the action (\ref{dbi_exp})  as
\begin{equation}
S^{\rm eff}_{{\rm D}p{\rm -brane}}=-\frac{1}{g_{\rm YM}^2}
\!\int_{\Sigma}\!\dd^{1+p}\sigma\ {\rm Tr}
\bigg\{\frac{1}{4}F_{mn}F^{mn}
+\frac{1}{2}D_m\phi_aD^m\phi^a
-\frac{1}{4}\sum_{a\ne b}[\phi^a,  \phi^b]^2\bigg\},
\label{dp_ym}
\end{equation}
where we have  defined the covariant derivative as
$ D_m\phi^a=\del_m\phi^a+i[A_m, \phi^a]$. 
The structures of the commutator could be again understood from T-duality
perspective.
We obtained $U(N)$ Yang-Mills (+Higgs) theory in $(1+p)$-dimensional world volume
from the $N\times $D$p$-branes effective action.
The overall $U(1)$ factor of $U(N)$ corresponds to the overall position of the
branes and may decouple from the $SU(N)$ gauge dynamics.

Let us discuss Higgs mechanism in D-brane picture \cite{w_1995}.
As usual, by using gauge transformation, we could find
the potential minimum in the configuration
\begin{equation}
\phi^a={\rm diag}(v^a_1, \cdots, v^a_N).
\label{vev}
\end{equation}
The diagonal component $v^a_m$ represents the location of the
 $m^{\rm th}$ D-brane in the transverse directions.
$N$ D-branes sit on the same position, $U(N)$ symmetry holds.
If all of the D-branes are separated,  the VEV (\ref{vev}) breaks the gauge
symmetry $U(N)\to U(1)^N$ so that we could expect $W$-bosons to get mass.
The essential mechanism can be easily understood in $U(2)$ case. For simplicity
we consider a D-brane that has VEV only in one transverse direction $X^D$;
$\phi^D={\rm diag}(v_1, v_2)$. 
The gauge field $A_m$ can be expressed as
$$
A_m
=
\left(
\begin{tabular}{cc}
$A^1_m$ & $W_m$ \\
$W^*_m$ & $A^2_m$
\end{tabular}
\right).
$$
Through the covariant derivative in the action (\ref{dp_ym})
$W$-boson obtains a mass term
\begin{equation}
(v_1-v_2)^2|W_m|^2
=\frac{1}{(2\pi\alpha')^2}(X^D_1-X^D_2)^2|W_m|^2
=\frac{1}{(2\pi\alpha')^2}({\rm string \ length})^2|W_m|^2,
\label{w-boson_mass}
\end{equation}
while two gauge field $A_m^{(1, 2)}$ remain massless.
As we observed in (\ref{open_mass}),
the mass can be provided as the mass of open string
which is stretched between two D-branes.
\begin{figure}[htbb]
\begin{minipage}{1.0\textwidth}
\begin{center}
\includegraphics[clip, width=3.5cm]{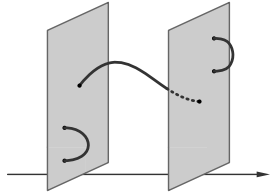}
\caption{Stringy Higgs mechanism}
\label{stringy_higgs}
\end{center}
\end{minipage}
\end{figure}
%

\subsubsection{D$p$-branes and RR $p$-brane solitons in type II supergravity}

In order to compare the D$p$-brane with RR $p$-brane,  
we need to know the charge $\mu_p$ in (\ref{wz_p})
carried by a single D$p$-brane.
In analogy with point charges,
we  consider the force between D-branes.

\begin{figure}[htbb]
\begin{minipage}{1.0\textwidth}
\begin{center}
\includegraphics[clip, width=9cm]{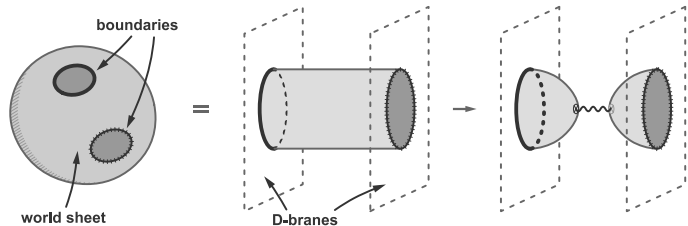}
\caption{Interaction between D$p$-branes}
\label{interaction_dbranes}
\end{center}
\end{minipage}
\end{figure}

We put two parallel flat D-branes with distance $|\vec{Y}|$.
The total force can be estimated through the exchanges of closed string
states (See Fig.\ref{interaction_dbranes}).
We calculate this amplitude by two ways and compare the results.
The first one is that from pure perturbative string theory point of view;
the diagram can be visualized as a cylinder where
closed strings propagate between two D-branes.
However, in the world sheet point of view,
we can interpret this as an open string
one loop amplitude (cf.\ (\ref{particle_one_loop})),
\begin{eqnarray}
{\cal A}_{\rm 1-loop}
&=&-\frac{1}{2}\log{\rm Det}({\rm propagator})^{-1}
=\frac{1}{2}\!\int_0^\infty\!\frac{\dd t}{t}{\rm Tr}\ \eee^{-t L_0}
\sim
2i\pi(1-1)(4\pi^2\alpha')^{3-p}V_{1+p}G_{9-p}(|\vec{Y}|),
\nonumber
\\
&=&
{\cal A}_{\rm NSNS}+{\cal A}_{\rm RR},
\label{one_loop}
\end{eqnarray}
where $V_{1+p}$ is the volume of $p$-brane and $G_d(x)$ is a
$d$-dimensional massless Green's function\footnote{
For later references, we give the formula for the Fourier transformation,
\begin{equation}
\int\frac{\dd^dk}{(2\pi)^d}
\ \eee^{ikx} k^n
=\frac{2^n}{\pi^{d/2}}\frac{\Gamma\big(\frac{d+n}{2}\big)}
{\Gamma\big(-\frac{n}{2}\big)}\frac{1}{|x|^{d+n}}.
\label{fourier_trans_formula}
\end{equation}
}
$$
G_{d}(x)=\!\int\!\frac{\dd^dk}{(2\pi)^d}\frac{\eee^{ikx}}{k^2}
=\frac{\Gamma\big((d-2)/2\big)}{4\pi^{d/2}|x|^{d-2}}.
$$
In (\ref{one_loop})
we have extracted the contributions from long range interactive massless modes
in closed string.
Here, ${\cal A}_{\rm NSNS}= 2i\pi(4\pi^2\alpha')^{3-p}V_{1+p}G_{9-p}(|\vec{Y}|),~
{\cal A}_{\rm RR}=- {\cal A}_{\rm NSNS}$.
The absence of the string coupling $g_{\rm s}$ in (\ref{one_loop})
reflects that the amplitude has been computed on the world sheet with
the cylinder topology.
So we observe that the force between D$p$-branes totally vanishes.
This can be interpreted as the cancellation between
the attractive forces due to graviton and dilaton and the
repulsive forces from the RR gauge fields.
In (\ref{one_loop}), $+1$ and $-1$ arise from gravity sector
and RR gauge sector, respectively.
Resulting stable configuration in fact indicates that D-branes are BPS
states.

Next, we calculate the same amplitude from the
effective theory point of view.
The ingredients here are the NSNS graviton sector and RR sector
everywhere,  the gauge fields on D$p$-brane, and the
scalar fields $\phi_a(\sigma)$.
Since we are interested in the flat D$p$-brane configuration,
the embedding functions $X^M(\sigma)$ have a trivial meaning that
the scalar fields $\phi_a(\sigma)$ are constants which just specify
the location of the D$p$-branes.
We perturb the NSNS-RR background fields through (\ref{perturbation_nsns})
and $C_{p+1}(x)=0+\kappa_0\widehat{C}_{p+1}(x)$,
and observe responses of D$p$-branes.
The relevant source/linear coupling terms can be found in the
DBI action (\ref{dbi_exp}) and the WZ term (\ref{wz_p}),
\begin{equation}
S^{\rm source}_{\rm DBI}
=-T_p\kappa\!\int_{\Sigma_{p+1}}\!\!\!\!\dd^{1+p}\sigma
\Big(\frac{p-3}{4}\widehat{\Phi}
+\frac{1}{2}\widehat{h}_m{}^m\Big),
\quad
S^{\rm source}_{\rm WZ}
=-\mu_p\kappa_0\!\int_{\Sigma_{p+1}}\!\!\!\!\widehat{C}_{p+1}
=-\mu_p\kappa_0
\!\int_{\Sigma_{p+1}}\!\!\!\!\dd^{1+p}\sigma\ \widehat{C}_{01\cdots p}.
\end{equation}
As we have already mentioned,
the antisymmetric Kalb-Ramond field $\widehat{b}_{mn}(\sigma)$ and
also the gauge field strength $F_{mn}(\sigma)$ do not couple linearly.
In the WZ term, a  component survived is only
$\widehat{C}_{01\cdots p}(x)$ in the flat embedding.
The bulk propagators which couple to the D-brane can be read off from
the action (\ref{spe}) under the perturbation
(\ref{perturbation_nsns})\footnote{
For the graviton sector, we refer to the propagator (\ref{propagator_graviton}).
},
\begin{eqnarray}
\langle \widehat{\Phi}(-k)\widehat{\Phi}(k)\rangle
&=&
-\frac{2i}{k^2},
\qquad
\langle\widehat{C}_{01\cdots p}(-k)\widehat{C}_{01\cdots p}(k)\rangle
=
\frac{2i}{k^2}.
\nonumber
\\
\langle\widehat{h}_{MN}(-k)\widehat{h}_{KL}(k)\rangle
&=&
-\frac{2i}{k^2}
\Big(\eta_{MK}\eta_{NL}+\eta_{ML}\eta_{NK}-\frac{1}{4}\eta_{MN}\eta_{KL}
\Big),
\end{eqnarray}
We are ready to compute the tree amplitudes for each sector.
The result is as follows for RR gauge field
sector and NSNS gravity sectors, respectively,
\begin{eqnarray}
{\cal A}_{\rm RR}
&=&
(i\mu_p\kappa_0)^2V_{1+p}\!\int\!\frac{\dd^{9-p}k}{(2\pi)^{9-k}}
\frac{2i}{k^2}\ \eee^{i\vec{k}\vec{Y}}
=-2i\mu_p^2\kappa_0^2V_{1+p}G_{9-p}(|\vec{Y}|).
\\
{\cal A}_{\rm NSNS}
&=&
(iT_{p}\kappa)^2V_{1+p}\!\int\!\frac{\dd^{9-p}k}{(2\pi)^{9-p}}
(-\frac{2i}{k^2})
\bigg\{\bigg(\frac{p-3}{4}\bigg)^2+\bigg(\frac{1}{2}\bigg)^2
\bigg(2\eta_m{}^n\eta_n{}^m
-\frac{1}{4}\eta_m{}^m\eta_n{}^n
\bigg)\bigg\}\ \eee^{i\vec{k}\vec{Y}}
\nonumber
\\
&=&
2i\kappa_0^2g_{\rm s}^2T_p^2V_{1+p}G_{9-p}(|\vec{Y}|).
\end{eqnarray}
Comparing the previous result (\ref{one_loop}),
we can obtain relations
$
\mu_p^2\kappa_0^2=\pi(4\pi^2\alpha')^{3-p}$ and
$\kappa_0^2g_{\rm s}^2T_p^2=\pi(4\pi^2\alpha')^{3-p}$,
and the BPS condition
\begin{equation}
\mu_p=g_{\rm s}T_p=\frac{1}{(2\pi)^p\alpha'^{\frac{p+1}{2}}},
\label{rr_charge}
\end{equation}
where for the last equality we have used (\ref{tension_p}).
The condition (\ref{rr_charge}) could be  a counter part of
the equality in (\ref{bps_bound}).
Moreover we can fix the 10D Newton constant in (\ref{10d_newton})
in terms of the parameter in string theory,
\begin{equation}
G_{\rm N}^{(10)}=\kappa_0^2g_{\rm s}^2/(8\pi)
=8\pi^6g_{\rm s}^2\alpha'^4
\equiv\kappa^2/(8\pi),
\qquad {\rm i.e.} \quad \kappa=2\pi^{3/2}g_{\rm s}(2\pi\alpha')^2.
\label{newton_c}
\end{equation}
We can also show that the charge is quantized.
Comparing the normalization factors of the kinetic terms between
(\ref{p+1-form_f}) and (\ref{sp}),
we can conclude the Dirac quantization condition (\ref{dirac_q})
is satisfied with $n=1$,
\begin{equation}
\widetilde{\mu}_p\widetilde{\mu}_{6-p}=2\pi,
\end{equation}
where we have rescaled the charge (\ref{rr_charge}) as
$\widetilde{\mu}_p\equiv\sqrt{2}\kappa_0\mu_p$.

We have observed that
D$p$-brane and the extremal RR $p$-brane share the same properties.
Both of them correspond to BPS states and possess RR charges.
This suggests that the D$p$-brane and the extremal RR $p$-brane could be identified.
Under the identification of charges (\ref{charge_bp})
and (\ref{rr_charge}),
\begin{equation}
 Q_p=N\times\mu_p, \qquad
\mbox{i.e.} \qquad
\int_{S^{8-p}}\!\!\!\!*F_{p+2}=\frac{N}{(2\pi\sqrt{\alpha'})^{p-7}},
\label{identification}
\end{equation}
where $N$ is the number
of coincident D$p$-branes\footnote{
As BPS configuration indicates,
in general, there also exist multi-center solutions in the supergravity
which correspond to parallel branes separated by some distance.
},
the parameter $L$ corresponding to the Schwarzschild radius
can be fixed as
\begin{equation}
L^{7-p}=
\frac{(2\pi)^{7-p}\alpha'^{\frac{7-p}{2}}Ng_{\rm s}}{(7-p)V_{S^{8-p}}}.
\label{factor_l}
\end{equation}

However, the D$p$-brane and RR $p$-brane pictures
are reliable in different regimes.
If one introduces $N\times$ D$p$-branes in string perturbation theory,
the effective string coupling is raised to be $N\times g_{\rm s}$
according to the open string loop counting (\ref{partition_1}).
Therefore, in the string perturbation theory, the condition
\begin{equation}
\lambda\equiv
Ng_{\rm s}\ll1
\label{weak}
\end{equation}
should be required.
In the bulk spacetime point of view,
the condition (\ref{weak}) is equivalent to
\begin{equation}
r_{\rm S}\ll\sqrt{\alpha'},
\label{oposit}
\end{equation}
where $r_{\rm S}=\Big(\frac{7-p}{8-p}\Big)^{1/(7-p)}L$ is the
Schwarzschild radius.
The gravitational scale is much smaller than the string scale,
so that the physics with the length scale $r_{\rm S}$ is not affected by
strings.
The gravitational backreaction of D-brane are arbitrary small,
and D-brane does not modify the geometry in the perturbative regime.
This can be also understood in the Einstein equation (\ref{einstein})
by computing the energy momentum tensor ($\sim$ tension) from D$p$-brane:
\begin{equation}
R_{tt}-\frac{1}{2}G_{tt}R=8\pi G^{(10)}_{\rm N}T_{tt}
\propto g_{\rm s}^2\times\bigg(N\times\frac{1}{g_{\rm s}}\bigg)
\propto\lambda.
\label{backreaction}
\end{equation}
Therefore, the D-brane and its open string modes are defined in the flat
spacetime.

On the other hand, RR $p$-branes are nontrivial solutions of type II supergravity
actions which are the low energy effective action of the closed string.
As we discussed,
the low energy effective action itself can be obtained
under the weak curvature approximation measured by the string unit
(\ref{alphap_expansion}),  i.e.\
\begin{equation}
r_{\rm S}\gg\sqrt{\alpha'},
\label{no_stringy}
\end{equation}
which means the Schwarzschild radius is now macroscopic.
The condition (\ref{no_stringy}) is equivalent to
\begin{equation}
\lambda\gg 1.
\label{strong_lambda}
\end{equation}
According to (\ref{backreaction}), in this case, D-brane would
gravitationally collapse to the RR $p$-brane soliton.
It should be noted that the number $N$ should be large
in order to keep the condition (\ref{strong_lambda}) under the string
perturbation $g_{\rm s}\ll1$.

These observations suggest the ``duality'' between two pictures
i.e.\ D$p$-brane and RR $p$-brane soliton pictures,  in the different
limits of the parameter $\lambda$.

\section{AdS/CFT correspondence}

The AdS/CFT correspondence is a (gravity theory in AdS spacetime)
/(conformal field theory) duality
in string theory \cite{Maldacena:1997re}
(for review see \cite{adscft_review}).
This correspondence is based on two ideas.
One is holography which relates theories living in spaces with different
dimensionality.
Inspired by the ``area low'' of black hole entropies,  
the holography is introduced as the statement:
quantum gravity in some region may be described in terms of a non
gravitational theory living on its boundary \cite{ths}.
Another important idea is that the nonabelian gauge theory might have
a dual description in terms of strings \cite{th}, i.e.\
the 't Hooft large $N$ expansion.
These two can be realized with precise meaning through the AdS/CFT
correspondence.
Despite the name, the correspondence can be naturally extended to
non AdS spacetime and non-conformal gauge field theories.

\subsection{Maldacena conjecture}

\begin{figure}[htbb]
\begin{minipage}{1.0\textwidth}
\begin{center}
\includegraphics[clip, width=11cm]{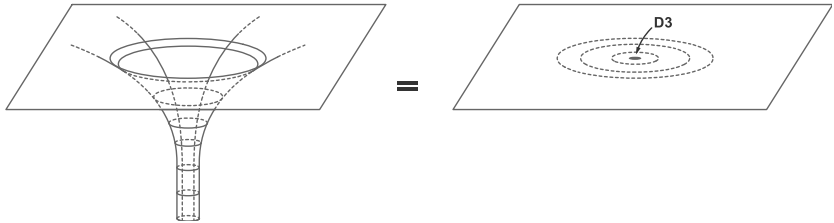}
\caption{RR $p$-brane and D$p$-brane}
\label{pbrane_dbrane}
\end{center}
\end{minipage}
\end{figure}

In the last section,
the D$p$-branes and the RR $p$-brane solutions in supergravity theories
describe the same BPS states.
Pictorial identification between D3-brane and extremal RR-3-brane
is given by Fig.\ref{pbrane_dbrane}.
By looking into the detail of the two descriptions
and taking certain limits, we can obtain a new insight: the AdS/CFT correspondence.
As the original and the concrete example,
we start with $N_{\rm c}$ coincident D3-branes
in type IIB superstring theory in 10D.

Let us first consider $N_{\rm c}$ D3-branes in string perturbation
theory where 
there exist open strings ending on these D3-branes and also closed strings
in the bulk.
At the low energy
\begin{equation}
\omega\ll1/{\sqrt{\alpha'}}\propto 1/\kappa^{1/4},
\label{low_energy_string}
\end{equation}
we may obtain the low energy effective action
\begin{equation}
S=S_{\rm bulk}+S_{\rm D3-branes}+S_{\rm interactions}.
\end{equation}
The bulk action is that of 10D supergravity which describes
the massless closed string excitation modes (\ref{spe}), and 
some derivative corrections.
The D-brane actions are given by (\ref{dbi}), whose $\alpha'$ expansion
is (\ref{dbi_exp}),  and (\ref{wz_p}).
As we mentioned, the D-brane effective action contains the Yang-Mills
gauge theory with coupling (\ref{coupling_0}), i.e.\ for D3-branes,
\begin{equation}
g_{\rm YM}^2=2\pi g_{\rm s}.
\label{d3_ym}
\end{equation}
In the case of the $N_{\rm c}$ D3-branes, its 4D world volume gauge theory is
${\cal N}=4$ $SU(N_{\rm c})$ supersymmetric
Yang-Mills theory whose bosonic part is the same as
(\ref{dp_ym})\footnote{
D-brane preserves 16 supercharge as the BPS states.
In 4D these supercharges can be decomposed into four Weyl spinors.
Thus the low energy effective action of D3-brane should have ${\cal N}=4$
supersymmetries.}.
The interaction part $S_{\rm interactions}$ may be generated
by integrating out the higher energy modes of string excitations.
These  are typically proportional to the positive power of the
gravitation constant $\kappa$.

Let us now focus on the 4D super Yang-Mills theory.
First of all, as we have observed in (\ref{dbi_exp}),
in general, the super Yang-Mills couples to the bulk modes including graviton.
If we are interested in the super Yang-Mills gauge theory itself,
we need to decouple these bulk gravity modes.
This could be achieved by taking the limit; the gravitational constant
$\kappa\to0$ while keeping the Yang-Mills coupling constant $g_{\rm
YM}$ finite.
Looking at the couplings (\ref{10d_newton}) and
(\ref{d3_ym}),
this is equivalent to the limit $\alpha'\to 0$ with fixed $g_{\rm s}$.
In this limit with fixed energies
all of the interactions vanish except for BPS
condition. 
For the gravity part, the action simply follows (\ref{eh_pert}) in 10D.   
Therefore,  the total system falls into the two independent
ones, i.e.\ 4D pure ${\cal N}=4$ $SU(N_{\rm c})$ super Yang-Mills
on the D3-brane and free type IIB
supergravity in the 10D bulk.
This is so-called decoupling limit.
Schematically,
\begin{eqnarray}
&&
\mbox{(D3-branes in flat 10D Minkowski spacetime)}
\nonumber
\\
&&
\hspace*{10mm}
\mbox{------ (low energy limit)} \longrightarrow
\left\{
\begin{tabular}{l}
\mbox{(${\cal N}=4$ super Yang-Mills in 4D world volume)}
\\
\mbox{(decoupled 10D supergravity on flat spacetime)}
\end{tabular}
\right.
\label{d-brane_picture}
\end{eqnarray}

It is worth to mention  the energy scale in the gauge
theory side described by open strings~\cite{Maldacena:1997re}.
To take the limit $\alpha'\to0$,
in general, we need to fix the energy scale.
Below the string energy scale,
the natural scale may be the mass of
$W$-boson (\ref{w-boson_mass}),
\begin{equation}
\omega_W\equiv \frac{r}{\alpha'},
\label{low_energy_00}
\end{equation}
where $r$ is the distance between D-branes\footnote{
This can be realized as the following situation;
we put the stack of $(N_{\rm c}-1)$ D3-branes at the origin
and introduce the probe single D3-brane at the position $r$.
By definition, the probe does not do anything to the others.
The decoupling limit implies that we could only observe the stack of
D3-branes at the origin.
}.
Since the low energy dynamics of the gauge theory are described by the
open strings attached to the D-branes,
it would be natural to take the Higgs mechanism as a typical source of the energy
scale.
Keeping the energy (\ref{low_energy_00}) finite and taking the
limit $\alpha'\to0$ requires the scaling behavior
$r\propto\alpha'\to0$. 
This indicates that an observer should be located far from the
D3-branes. 
The low energy limit (\ref{low_energy_string})
directly suggests
\begin{equation}
r\ll\sqrt{\alpha'},
\label{near_horizon_region}
\end{equation}
which will be realized as the near horizon limit (\ref{ads_0})
in the dual geometry.

We shall consider the same $N_{\rm c}$ D3-branes system
from the point of view of
type IIB superstring on the solitonic background (\ref{ads_00})
with
\begin{equation}
L^4=4\pi\alpha'^2 N_{\rm c}g_{\rm s},
\label{d3_l}
\end{equation}
which comes from the identification (\ref{factor_l}).
Since the geometry possesses the asymptotic
flat region,  it is natural to see physics from that region as we 
eventually did in the previous perturbative string picture. 
In the background (\ref{ads_00}),
due to the gravitational red shift (\ref{redshift}),
the observer in the asymptotic region will measure an energy $\omega_\infty$
which is smaller than the original energy $\omega_r$ emitted from the
position $r$,
\begin{eqnarray}
\omega_\infty
&=&
\sqrt{-G_{tt}(r)}\ \omega_r
=\Big(1+\frac{L^4}{r^4}\Big)^{-\frac{1}{4}} \omega_r.
\end{eqnarray}
For $r>L$,
the low energy excitations are
those of free type IIB supergravity.
In the region $r<L$, the red shift factor becomes more important,
especially in the near horizon $r\ll L$,
the energy is measured as
\begin{equation}
\omega_\infty\sim\frac{r}{L}\omega_r\, .
\end{equation}
This implies that the ``finite'' energy modes near horizon ($r\ll L$) may be
observed as the low energy modes in the asymptotic region.
Indeed string excitation
modes $\omega\sim 1/\sqrt{\alpha'}$ could be observed as
\begin{equation}
\omega_\infty\sim\frac{r}{L}\omega_r
\propto\frac{r}{\sqrt{\alpha'}}\frac{1}{\sqrt{\alpha'}}
=\frac{r}{\alpha'}\, ,
\label{red_shifted_low_energy}
\end{equation}
where we have used the relation (\ref{d3_l}).
This could be the energy scale which should be fixed during  the
limit $\alpha'\to0$ which we have discussed as the decoupling limit in the
gauge theory.
Indeed the fixed energy scale (\ref{red_shifted_low_energy}) is
the same with (\ref{low_energy_00}).
The low energy decoupling limit (\ref{near_horizon_region}) also
supports the near horizon region i.e.\ $r\ll\sqrt{\alpha'}\sim L$.

As a result, low energy excitations at infinity
can be categorized into two types, i.e.\
the low energy excitations far from the branes and
finite energy excitations near horizon.

It is important to see that these excitations could be decoupled each other.
As a typical experiment at the asymptotic region,
we consider the graviton scattering with the  excitations
on the D3-branes in the low energy.
The characteristic behavior of the absorption cross section of
the D3-branes can be found with the incident energy
$\omega$ as~\cite{klebanov_97}
\begin{equation}
\sigma_{\rm D3}\propto\omega^3 L^8
\propto\omega^3\alpha'^4N_{\rm c}^2g_{\rm s}^2,
\end{equation}
where the degrees of freedom for the gauge field $N^2_{\rm c}$
show up.
This indicates  that, in low energies $\omega\ll1/\sqrt{\alpha'}$
with fixed $g_{\rm s}$ and $N_{\rm c}$,
D3-branes do not interact with supergravity excitations in the
asymptotic region.
In other words,
since the low energy long wave length $1/\omega\gg\sqrt{\alpha'}$
cannot enter  the small throat $L\sim\sqrt{\alpha}$,
they could not interact each other.
Therefore, we conclude that in the solitonic background picture,
the excitation modes measured in the low energy
could be well approximated by two isolated parts,
i.e.\ those for the 10D bulk free supergravity and for the type IIB
superstring in the near horizon region.

As we did in the perturbative string picture, we shall take the low energy limit
$\alpha'\to0$ with fixed energies and fixed dimensionless quantities.  
Keeping the energy (\ref{red_shifted_low_energy}) fixed,
in this energy scale, the metric (\ref{ads_00}) becomes
the near horizon metric
\begin{equation}
\frac{\dd s^2}{\alpha'}
=\frac{U^2}{\sqrt{4\pi N_{\rm c}g_{\rm s}}}
\Big(-\dd t^2+\sum_{i=1}^3\dd x^{i 2}\Big)
+\frac{\sqrt{4\pi N_{\rm c}g_{\rm s}}}{U^2}\dd U^2
+\sqrt{4\pi N_{\rm c}g_{\rm s}}\dd\Omega^2_5,
\label{ads_10}
\end{equation}
where we have defined the coordinate $U=r/\alpha'$.

Another natural energy scale has been considered by the
probe of the field in supergravity to the stack of
D3-branes~\cite{uvir}.
The typical energy scale of the supergravity field in (\ref{ads_10})
becomes $u=r/L^2(\propto r/\alpha')$.
In this energy scale,  the near horizon metric becomes
\begin{equation}
\frac{\dd s^2}{L^2}=
u^2\Big(-\dd t^2+\sum_{i=1}^3\dd x^{i2}
\Big)
+\frac{1}{u^2}\dd u^2
+\dd\Omega^2_5.
\label{ads_11}
\end{equation}
These metrics (\ref{ads_10}) and (\ref{ads_11}) describe
AdS$_5\times S^5$ geometry.
Therefore schematically,
\begin{eqnarray}
&&
\mbox{(D3-branes as the solitonic background)}
\nonumber
\\
&&
\hspace*{10mm}
\mbox{------ (low energy limit)} \longrightarrow
\left\{
\begin{tabular}{l}
\mbox{(excitations in} {\mbox AdS}$_5\times S^5$
{\rm region)}
\\
\mbox{(decoupled 10D supergravity on flat spacetime)}
\end{tabular}
\right.
\label{rr-brane_picture}
\end{eqnarray}

Comparing these two pictures in the low energy
(\ref{d-brane_picture}) and (\ref{rr-brane_picture}),
Maldacena conjectured that
\begin{equation}
\mbox{(${\cal N}=4$ super Yang-Mills in 4D)}
\sim
\mbox{(type IIB string theory on AdS$_5\times S^5$)}
\label{adscft}
\end{equation}
We discuss this conjecture more precisely in the next subsection.

\subsection{Correspondence}

\subsubsection{Symmetries}

In the field theory side, i.e.\ 4D ${\cal N}=4$ super Yang-Mills theory,
field contents in the adjoint
representation
are massless gauge fields, 6 real massless scalars, and
four massless Weyl fermions as gaugino.
Under $SU(4)$ $R$-symmetry transformation,
which is a global symmetry for 4D ${\cal N}=4$
supersymmetry with 16 supercharges,
the scalars and fermions transform as ${\bm 6}$ and ${\bm 4}$,
respectively.
It can be shown that the beta function vanishes at least 3 loops and
believed to be vanished in all orders of the perturbation.
This feature indicates that the 4D ${\cal N}=4$ super Yang-Mills theory
is a conformal field theory.

Let us discuss the relation between the conformal symmetry in the
gauge theory and the isometry (symmetries of the metric) of AdS spacetime.
The conformal group in $D$ dimensional spacetime is
the group of reparameterizations which preserve the spacetime
metric up to a local scale factor,
\begin{equation}
g_{\mu\nu}(x)\to g_{\mu\nu}'(x)=\eee^{\Lambda(x)} g_{\mu\nu}(x).
\end{equation}
The conformal transformations can be generated
by the infinitesimal coordinate transformations
$x^\mu\to x^{\mu'}=x^\mu+\varepsilon^\mu(x)$
which satisfy
\begin{equation}
\delta g_{\mu\nu}(x)
=g'_{\mu\nu}(x)-g_{\mu\nu}(x)
=\nabla_\mu\varepsilon_\nu+\nabla_\nu\varepsilon_\mu
=\Lambda(x)g_{\mu\nu}(x).
\end{equation}
Taking the trace of the both sides,
one can eliminate the  factor $\Lambda(x)$
and obtain the conformal Killing equation
\begin{equation}
\nabla_\mu\varepsilon_\nu+\nabla_\nu\varepsilon_\mu
=\frac{2}{D}g_{\mu\nu}(g^{\rho\sigma}\nabla_\rho\varepsilon_\sigma).
\label{conformal_killing}
\end{equation}
In flat spacetime ($D\ge 3$),
one can solve the equation (\ref{conformal_killing}) and obtain
the generator $\varepsilon^\mu(x)$ of the form:
\begin{equation}
\varepsilon^\mu(x)
=a^\mu+\omega^\mu{}_\nu x^\nu+\lambda x^\mu
+2(c_\lambda x^\lambda)x^\mu-c^\mu (x_\lambda x^\lambda).
\end{equation}
Apart from  usual $D$-dimensional Poincar\'e symmetries $ISO(1, D-1)$,
i.e.\   translation and rotation which are represented by
the constant parameters $a^\mu$ and
$\omega_{\mu\nu}=-\omega_{\nu\mu}$, respectively,
we obtain the dilatation and the conformal boost whose infinitesimal
transformation parameters are given by $\lambda$ and
$c^\mu$.
Integrating the infinitesimal transformations,
the finite transformations could be obtained.
In addition to usual Poincar\'e transformations,
we could arrive at those for the dilatation and the conformal boost,
\begin{equation}
x'^\mu=\eee^{-\lambda}x^\mu,
\qquad
x'^\mu=
\frac{x^\mu-c^\mu x^2}
{1-2(cx)+c^2x^2}.
\label{conformal_boost}
\end{equation}
All of these transformations form the conformal group which is
isomorphic to the $SO(2, D)$.

In the gravity side, we solve the Killing equation
\begin{equation}
\delta g_{MN}=\nabla_M\varepsilon_N+\nabla_N\varepsilon_M=0,
\end{equation}
to find symmetries of the metric.
We are interested in the AdS part of the metric (\ref{ads_11}),
\begin{equation}
\dd s^2=L^2
\Big(u^2\big(-\dd t^2+\sum_{i=1}^3\dd x^{i2}
\big)
+\frac{1}{u^2}\dd u^2
\Big)
=L^2
\Big(
u^2\eta_{\mu\nu}\dd x^\mu\dd x^\nu+\frac{1}{u^2}\dd u^2
\Big).
\label{ads_12}
\end{equation}
Then, in this coordinate system,
we find the following generators ($\delta x^M=\varepsilon^M$ with
$x^M=(x^\mu, u)$)
which are isomorphic to $SO(2, 4)$;
the translation $\delta x^\mu=a^\mu, \ \delta u=0$,
the rotation $\delta x^\mu=\omega^\mu{}_\nu x^\nu, \ \delta u=0$,
the dilatation
\begin{subequations}
\begin{eqnarray}
\delta x^\mu
&=&
-\lambda x^\mu,
\label{dilatation_x}
\\
\delta u
&=&
\lambda u,
\label{dilatation_u}
\end{eqnarray}
\end{subequations}
and the conformal boost
\begin{subequations}
\begin{eqnarray}
\delta x^\mu
&=&
2(c_\nu x^\nu)x^\mu-c^\mu(x_\nu x^\nu)-\frac{c^\mu}{u^2},
\label{comformal_x}
\\
\delta u
&=&
-2(c_\nu x^\nu)u.
\label{conformal_u}
\end{eqnarray}
\end{subequations}

It is interesting to see that the transformation (\ref{comformal_x})
becomes the conformal boost in 4D (conformally) flat spacetime
if the limit $u\to\infty$ is taken.
Thus, the conformal symmetry in the 4D gauge theory
appears in the region $u\to\infty$ which is the boundary of
the AdS spacetime in the dual gravity theory.
The isometry group of AdS$_5$ acts as the conformal group on its 4D
boundary.
This is a realization of the holography.

The dilatation (\ref{dilatation_x}) and (\ref{dilatation_u})
or their finite forms of the transformations
$(x^\mu, u)\to (\eee^{-\lambda}x^\mu, \eee^{\lambda}u)$  with a
constant $\lambda$
indicate that the radial coordinate $u$ can
be understood as the energy scale of the boundary ($u\to\infty$)
4D theory \cite{uvir}.
Thus, the holographic dual may correspond to the
conformal field theory in the UV.

In the gauge theory side, there exists $SU(4)$ $R$-symmetry.
In the gravity side, there also exists $SO(6)\simeq SU(4)$ symmetry
which is isometry group of $S^5$ part of the geometry.
Thus, we conclude that the global symmetries of ${\cal N}=4$ super Yang-Mills
theory match with the isometries of the AdS${}_5\times S^5$
background.
This matching of the symmetry is one of the most powerful
checks of the correspondence.
Indeed, the symmetry matching plays an important role later.

\subsubsection{Coupling constants}

The claim (\ref{adscft}) without any restrictions on the
parameter space is known as the strongest conjecture.
Unfortunately, it is not easy to analyze the full type IIB string theory
in the near horizon background geometry through conventional
approaches.
We need to consider suitable approximations to make the problems
under control in the current knowledge.
Nevertheless, we could deduce many nontrivial properties from the
correspondence.

We here consider the relations between parameters of the two theories.
In Yang-Mills theory, there are two parameters i.e.\
the gauge coupling $g_{\rm YM}$ and $N_{\rm c}$ for the gauge group
$SU(N_{\rm c})$.
In the string theory side,
we have the string tension $\alpha'$ and the string coupling $g_{\rm s}$.
These are related through (\ref{d3_ym}) and (\ref{d3_l}),
\begin{equation}
g_{\rm YM}^2=2\pi g_{\rm s}
\quad
\mbox{and}
\quad
N_{\rm c}=\frac{L^4}{4\pi g_{\rm s}\alpha'^2}.
\end{equation}
The parameter introduced in (\ref{weak}) is
now interpreted as 't Hooft coupling $\lambda= N_{\rm c}g_{\rm s}
\propto N_{\rm c}g_{\rm YM}^2$
which is a relevant loop expansion parameter rather than
$g_{\rm YM}$ in the large $N$ gauge theory~\cite{th}.
In the gravity picture, (\ref{rr-brane_picture}) is applicable
for the large 't Hooft coupling (\ref{strong_lambda}) and
also large $N_{\rm c}$ for the string perturbation theory.
By using the AdS/CFT argument,
the corresponding gauge theory is a strongly coupled large $N_{\rm c}$
Yang-Mills theory.
This opens a new window to analyze the strongly correlating systems
by using the dual gravity theory via AdS/CFT correspondence.
We could describe the strongly coupled gauge theories in terms of the
weakly coupled classical gravity theories.

In general, it is hard to ``prove'' such a strong/weak duality mainly due to
the lack of nonperturbative techniques.
So far a huge number of ``tests'' for the correspondence have been
worked out and no contradiction has been found.
However, for tightly protected parts by the symmetry,
we could verify  the duality.
BPS states are good for this purpose, because, for instance,
expectation values of these do not depend on the coupling constant
$\lambda$.
We could directly compare two different theories.

\subsubsection{Partition function}

Since two pictures, i.e.\ gauge/gravity, describe
the same physical system,
the responses to perturbations should be same.

Quantitative correspondence can be given by the path integral formulation of
the two theories \cite{Gubser:1998bc,Witten:1998qj}.
Partition functions in two different pictures should be equated in the AdS/CFT
correspondence: 
\begin{eqnarray}
Z_{\rm CFT}(\mbox{\footnotesize 4D super Yang-Mills})
&\equiv&
 Z_{\rm string}(\footnotesize{\mbox{IIB on AdS}_5\times{\tiny{S^5}}})
\nonumber
\\
&\sim& Z({\mbox{\footnotesize{IIB supergravity on AdS}} _5\times S^5})
\nonumber
\\
&\sim&
\eee^{-S_{\mbox{\scriptsize supergravity}}
\big|_{\mbox{\scriptsize on-shell}}} \qquad (N_{\rm c}, \lambda \gg 1),
\label{pf}
\end{eqnarray}
where in the string side, we may evaluate the partition function as
the saddle point approximation of the supergravity action with
the large $N_{\rm c}$ and the large 't Hooft coupling.
For definiteness,
we have equated the relation in Euclidean signature.

Let us consider perturbations or excitations with a precise example.
We perturb the fields in the gravity picture which originate from
the type IIB superstring.
The response to the gauge field theory, for instance,  could be
observed in the perturbed action (\ref{dbi_exp}).
The perturbations produce the following interaction terms
which are added to the kinetic terms of the gauge theory,
\begin{equation}
S_{\rm int}
=-T_3(2\pi\alpha')^2\kappa
\!\int\!\dd^4x\big(-\frac{1}{4}\widehat{\Phi}F_{mn}^2
+\frac{1}{2}\widehat{h}_{mn}T^{mn}\big).
\label{source-operator}
\end{equation}
After taking the decoupling limit, the fields in the gravity sectors
are regarded as the external fields which are not dynamical
in the gauge field side.
These external fields act as the sources which couple
to operators in the gauge field theory.
In the interactions (\ref{source-operator}),
as the sources, the dilaton $\widehat{\Phi}(x^m)$
and the metric perturbations $\widehat{h}_{mn}(x^m)$
couple to the glueball $F_{mn}^2(x^m)$ and
the energy momentum tensor $T^{mn}(x^m)$
which would be operators in the gauge field theory.
Therefore under the perturbations,
the partition function of the gauge field theory
turns to be the generating function of the correlators
of some operators.

The sources $\widehat{\Phi}(x^m)$
and $\widehat{h}_{mn}(x^m)$ originate from the perturbation of the
fields in the bulk.
Thus, from the bulk gravity theory point of view,
the source defined on the D3-brane could be a boundary value
of the field in the bulk spacetime,
\begin{equation}
\widehat{\Phi}(x^m)=
\widehat{\Phi}(x^M)\big|_{\rm boundary}
\qquad \mbox{and} \qquad
\widehat{h}_{mn}(x^m)=
\widehat{h}_{mn}(x^M)\big|_{\rm boundary}.
\label{boundary_condition}
\end{equation}
In the bulk gravity theory we need to solve classical equations of
motion for $\widehat{\Phi}(x^M)$ and $\widehat{h}_{mn}(x^M)$
to evaluate the on-shell action.
The condition (\ref{boundary_condition}) can be understood as
the boundary conditions for these differential equations.

We could apply the idea to a more general case.
There may exist maps  between the fields $\varphi(x^M)$
in the bulk gravity side and the local gauge invariant
operators ${\cal O}(x^m)$ in the 4D gauge theory side.
With the argument above, the practical form of the conjecture (\ref{pf})
becomes
\begin{equation}
\eee^{W[\varphi_0]}\equiv
\big\langle\  \eee^{\scriptsize \!\int\!\dd^4x \varphi_0(x^m){\cal O}(x^m)}\
\big\rangle_{\rm CFT}
=\eee^{-S_{\mbox{\scriptsize
supergravity}}[\varphi_0(x^m)]\big|_{\mbox{\scriptsize on-shell}}}
\quad {\rm with} \quad
\varphi_0(x^m)=
\varphi(x^M)\big|_{\rm boundary},
\label{pf_1}
\end{equation}
which is known as Gubser-Klebanov-Polyakov-Witten (GKP-W) relation.
$W_{\mbox{\scriptsize CFT}}[\varphi_0]$ is the generating
functional for connected diagrams.
Suppose we have obtained a solution of the equation of
motion in the bulk spacetime with required boundary conditions.
We could compute the on-shell action by using the solution 
to express the on-shell action in terms
of the boundary value which is the source in the field theory point of view.
By differentiating the relation (\ref{pf_1}) with respect to the source
$\varphi_0(x^\mu)$,  we could obtain the correlation
function for corresponding operator in the gauge theory side,
\begin{equation}
\big\langle{\cal O}(x_1)\cdots{\cal O}(x_n)\big\rangle
=(-)^{n+1}\frac{\delta}{\delta\varphi_0(x_1)}\cdots
\frac{\delta}{\delta\varphi_0(x_n)}
S_{\mbox{\scriptsize supergravity}}[\varphi_0]
\big|_{\mbox{\scriptsize on-shell}}\Big|_{\varphi_0=0}.
\label{pf_3}
\end{equation}

In principle, by using the complete set of the Kaluza-Klein mass
spectrum of 10D type IIB supergravity on $S^5$~\cite{gunaydin_marcus, kks5},
we could analyze the dynamics of the perturbations around
AdS$_5\times S^5$ background.
However, one of the obstructions comes from the nature of the AdS$_5\times S^5$
background.
Since the radius of $S^5$, which is the same with that of AdS$_5$,
cannot be arbitrary small, we have to tackle with complicated Kaluza-Klein modes.
The next procedure we should pursue is to find out consistent truncations of the
uncontrollable modes.
However, except for a few sectors~\cite{smrs_98},
this direction is also complicated.

As an alternative, one could start from 5D theory.
There is a complete 5D supergravity theory
known as the $SO(6)$ gauged ${\cal N}=8$ supergravity
which contains the same supergravity multiplet as
the type IIB supergravity~\cite{gauged_sugra}.
It is widely accepted that
this supergravity theory may be a consistent truncation of the
type IIB theory on AdS$_5\times S^5$.
We here assume these effective 5D theories.

Before calculating correlation functions a la AdS/CFT correspondence
(\ref{pf_3}) through the gravitational theory,
let us briefly summarize the conformal field theory side in 4D.
In usual quantum field theories,
we classify local fields and operators ${\cal O}(x^\mu)$ by
their representations of the Poincar\'e group, while
in conformal field theories, we use the conformal group $SO(2, 4)$.
The conformal group has the little group $SO(1,1)\times SO(1, 3)$,
more conveniently compact group $SO(2)\times SO(4)$,
so that in addition to the spin representation, we have the scaling
dimension $\Delta$ which is the charge of $SO(2)$ subgroup.
We define the scaling dimension of a operator ${\cal O}_{\Delta}(x^\mu)$ as
\begin{equation}
{\cal O}_{\Delta}(x^\mu)\to
{\cal O}'_{\Delta}(x'^\mu)=\eee^{\lambda\Delta}{\cal O}_{\Delta}(x),
\label{scaling_op}
\end{equation}
under the scaling $x^\mu\to x'^\mu=\eee^{-\lambda}x^\mu$.
Conformal invariance determines the forms of the correlators of the
primary fields in terms of their scaling dimensions.
For example, for scalar primary fields, correlators are given by
\begin{eqnarray}
\big\langle{\cal O}_{\Delta_1}(x_1){\cal O}_{\Delta_2}(x_2)\big\rangle
&=&
\delta_{1, 2}\prod_{i<j}^2|x_{ij}|^{-\Delta},
\nonumber
\\
\big\langle{\cal O}_{\Delta_1}(x_1){\cal O}_{\Delta_2}(x_2)
{\cal O}_{\Delta_3}(x_3)\big\rangle
&=&
c_{123}\prod_{i<j}^3|x_{ij}|^{\Delta-2\Delta_i-2\Delta_j},
\label{cft_correlators}
\\
\big\langle{\cal O}_{\Delta_1}(x_1){\cal O}_{\Delta_2}(x_2)
{\cal O}_{\Delta_3}(x_3){\cal O}_{\Delta_4}(x_4)\big\rangle
&=&
c_{1234}(u,
v)\prod_{i<j}^4|x_{ij}|^{\frac{1}{3}\Delta-\Delta_i-\Delta_j},
\nonumber
\end{eqnarray}
where $c_{123}$ and $c_{1234}$ are undetermined coefficients and
 $x_{ij}=x_i-x_j$ and $\Delta=\Sigma_i\Delta_i$.
For scalar operators, the rotational invariance fixes the coordinate
dependence in terms of the distance $|x_{ij}|$.
Since the conformal boost in (\ref{conformal_boost}) gives the following
transformation
$$
 |x'_{ij}|^2=\frac{|x_{ij}|^2}
{\big(1-2(cx_i)+c^2x_i^2\big)\big(1-2(cx_j)+c^2x_j^2\big)},
$$
there exist $n(n-3)/2$ conformally invariant ratios
$\frac{|x_{ij}||x_{kl}|}{|x_{ik}|x_{jl}|}$ for $n \, (\ge4)$ point
functions.
The coefficient $c_{1234}$ in (\ref{cft_correlators}) depends
on these ratios for instance
$
u=\frac{|x_{12}||x_{34}|}{|x_{13}||x_{24}|},
v=\frac{|x_{14}||x_{23}|}{|x_{13}||x_{24}|}.
$
Now, let us consider the generating functional of the CFT side in
(\ref{pf_1}).
The conformal invariance of the action determines 
the scaling of the source field $\varphi_0(x^\mu)$,
\begin{equation}
\int\!\dd^4x'\varphi_0'(x'^\mu){\cal O}'_{\Delta}(x'^\mu)
=\eee^{(-4+\Delta)\lambda}\!\int\!\dd^4x \varphi_0'(x'^\mu)
{\cal O}_{\Delta}(x^\mu),
\quad \mbox{i.e.} \quad
\varphi'_0(x'^\mu)=\eee^{(4-\Delta)\lambda}\varphi_0(x^\mu).
\label{cft_source_dimension}
\end{equation}
This will be used to relate the scaling dimension $\Delta$ on
boundary CFT with the parameter in the bulk gravity theories.

\subsubsection{Correlation functions}

It is instructive to discuss the simplest case to understand the heart
of the AdS/CFT correspondence.
Let us consider a massive scalar field with mass $m$ in 5D AdS
spacetime.
We here work with the Euclidean Poincar\'e coordinate of AdS$_5$ spacetime
\begin{equation}
\dd s^2=\frac{L^2}{z^2}
\Big(\delta_{\mu\nu}\dd  x^\mu\dd x^\nu+\dd z^2\Big),
\label{ads5_poincare}
\end{equation}
which is given through $u\to L/z$ and $x^\mu\to x^\mu/L$ in the AdS
part in (\ref{ads_11}) with Euclidean signature.
The boundary and the Killing horizon are now located
at $z=0$ and $\infty$,  respectively.

In the AdS$_5$ background (\ref{ads5_poincare}),
the action of the massive scalar $\phi(x)$ with mass $m$ is
\begin{equation}
S=\frac{1}{2}\!\int\!\dd^5x\sqrt{g}
\big(g^{mn}\del_m\phi\del_n\phi
+m^2\phi^2\big)
=\frac{L^3}{2}\!\int_\epsilon^\infty\!\!\!\!\dd z
\!\int\!\dd^4x\Big(\frac{1}{z^3}\del_z\phi\del_z\phi
+\frac{1}{z^3}\delta^{\mu\nu}\del_\mu\phi\del_\nu\phi
+\frac{L^2m^2}{z^5}\phi^2\Big),
\label{action_scalar}
\end{equation}
where we have made the regularization $z=\epsilon\ (\to 0)$.
The equation of motion is given by
\begin{equation}
0=\frac{1}{\sqrt{g}}\del_m(\sqrt{g}g^{mn}\del_n\phi)-m^2\phi,
\quad {\rm i.e.}
\quad
0=\del_z\Big(\frac{1}{z^3}\del_z\phi\Big)
+\frac{1}{z^3}\delta^{\mu\nu}
\del_\mu\del_\nu\phi-\frac{L^2m^2}{z^5}\phi,
\label{eom_scalar_0}
\end{equation}
where we have imposed the Dirichlet boundary condition at the boundary
i.e.\ $\delta\phi(x)|_{\rm boundary}=0$, which is natural choice for
AdS/CFT correspondence.
Since the 4D Euclidean part is flat, we use the plane wave expansion,
\begin{equation}
\phi(x^\mu, z)=\!\int\!\frac{\dd^4x}{(2\pi)^4}\ \eee^{ik_\mu x^\mu}
f_k(z).
\label{fourier}
\end{equation}
Then, the equation of motion (\ref{eom_scalar_0}) becomes
the ordinary second order differential equation for the radial
coordinate $z$,
\begin{equation}
0=f''_k(z)-\frac{3}{z}f'_k(z)-k^2f_k(z)-\frac{L^2m^2}{z^2}f_k(z),
\label{eom_scalar_k}
\end{equation}
where the prime implies the derivative with respect to $z$.
One can solve the equation of motion exactly in this case,
\begin{equation}
f_k(z)=z^{2}\Big(A(k)K_\nu(kz)+B(k)I_\nu(kz)\Big),
\quad {\rm with} \quad
\nu=\sqrt{4+L^2m^2}.
\end{equation}
Here we consider the case that $\nu$ is not an integer.
The integration constants $A(k)$ and $B(k)$ should be fixed through
suitable boundary conditions.
One of these comes from the interior i.e.\ $z\to\infty$ (IR).
Since the Bessel functions $K_\nu(zk)$ and $I_\nu(kz)$ behave as
\begin{equation}
K_\nu(kz)\to
\eee^{-kz},
\qquad
I_\nu(kz)\to
\eee^{+kz},
\label{expansion_horizon}
\end{equation}
we impose the regularity condition at the IR, i.e.\ $B(k)=0$.

Let us now consider the UV behavior $z\to0$.
In this region, we expand the Bessel function and observe
\begin{eqnarray}
f_k(z)=A(k)z^2K_\nu(kz)
&\to&
A(k)z^2\times
\frac{1}{2}\Gamma\big(\nu\big)\Gamma\big(1-\nu\big)
\Bigg[
\bigg(\frac{kz}{2}\bigg)^{-\nu}
\sum_{n=0}\frac{(kz/2)^{2n}}{n!\  \Gamma\big(n+1-\nu\big)}
\nonumber
\\
&&
\hspace*{50mm}
-\bigg(\frac{kz}{2}\bigg)^{\nu}
\sum_{n=0}\frac{(kz/2)^{2n}}{n!\  \Gamma\big(n+1+\nu\big)}
\Bigg]
\nonumber
\\
&\equiv&
A(k)z^2 \Big[z^{-\nu}\Big\{a_0(k) + a_2(k)z^2+{\cal O}(z^4)\Big\}
+z^{\nu}\Big\{b_0(k)+b_2(k) z^2 +{\cal O}(z^4)\Big\}\Big].
\nonumber
\\
\label{UV_expansion}
\end{eqnarray}
We could relate the overall coefficient $A(k)$ as the boundary value.
We simply insert the solution to the expression  (\ref{fourier}) and
fix the normalization through
$\phi(x^\mu, z=\epsilon)\equiv \phi_{{\rm b}\epsilon}(x^\mu)$:
\begin{equation}
\phi(x^\mu, z)=\!\int\!\frac{\dd^4k}{(2\pi)^4}
\ \eee^{ik_\mu x^\mu}
A(k)z^2K_\nu(kz)
\quad
{\rm with}
\quad
A(k)=\frac{1}{\epsilon^2K_\nu(k\epsilon)}\!\int\!\dd^4x\ \eee^{-ik_\mu
x^\mu}\phi_{{\rm b}\epsilon}(x^\mu).
\label{solution_scalar}
\end{equation}
The on-shell action can be evaluated by putting the
solution (\ref{solution_scalar}) into the action (\ref{action_scalar}),
\begin{eqnarray}
S
&=&
\frac{L^3}{2}
\!\int\frac{\dd^4k\ \dd^4k'}{(2\pi)^4}
\delta^4(k+k')
\!\int_\epsilon^\infty\!\!\!\dd z
\frac{\dd}{\dd z}\Big(\frac{1}{z^3}f_k(z)f'_{k'}(z)\Big)
\nonumber
\\
&=&
-\frac{L^3}{2}\!\int\frac{\dd^4k \ \dd^4k'}{(2\pi)^4}
\delta^4(k+k')
\int\!\dd^4x\ \dd^4x'\ \eee^{-ik_\mu x^\mu}\eee^{-ik'_\nu x'^\nu}
\phi_{{\rm b}\epsilon}(x^\mu)\phi_{{\rm b}\epsilon}(x'^\mu)
\bigg(
\frac{1}{z^3}\frac{\big(z^2K_\nu(kz)\big)'}{z^2K_\nu(kz)}
\bigg)\bigg|_{z=\epsilon},
\nonumber
\\
\label{on-shell_10}
\end{eqnarray}
where we have used the equation of motion.
The last piece in the on-shell action (\ref{on-shell_10}) can be
estimated as\footnote{
We use the relations for the Bessel function $K_\nu(z)$:
$$
K_{\nu-1}(z)-K_{\nu+1}(z)=-\frac{2\nu}{z}K_\nu(z), \qquad
K_{\nu-1}(z)+K_{\nu+1}(z)=-2K'_\nu(z).
$$
}
\begin{eqnarray}
\bigg(
\frac{1}{z^3}\frac{\big(z^2K_\nu(kz)\big)'}{z^2K_\nu(kz)}
\bigg)\bigg|_{z=\epsilon}
&=&
\frac{2+\nu}{\epsilon^4}
-\frac{k}{\epsilon^3}
\frac{K_{\nu+1}(\epsilon k)}{K_{\nu}(\epsilon k)}
\nonumber
\\
&=&
\bigg\{\frac{2-\nu}{\epsilon^4} +{\cal O}(1/\epsilon^2)
\bigg\}
+\bigg\{
2\nu\underbrace{\bigg(-\frac{\Gamma\big(1-\nu\big)}
{\Gamma\big(1+\nu\big)}
\bigg(\frac{k}{2}\bigg)^{2\nu}
\bigg)}_{\displaystyle=b_0(k)/a_0(k)}\epsilon^{2\nu-4}
+{\cal O}(\epsilon^{2\nu-2})\bigg\},
\qquad
\label{on-shell_11}
\end{eqnarray}
where $a_0(k)$ and $b_0(k)$ are given in the asymptotic expansion around
$z\to\epsilon$ (\ref{UV_expansion}).

The first line of the expression (\ref{on-shell_11})
includes  divergent terms.
Since these coefficients are analytical functions of the 4D momentum
$k$,
we could remove them by adding
appropriate local counter terms in the
boundary without spoiling the equation of motion in the bulk~\cite{counter_term}.
The detail of this holographic renormalization program can be found in
the review~\cite{hrg}.
In the second line there exist non-analytic parts which could not
be removed through counter terms.
These terms contain physical information of 4D boundary theory.

We next perform the wave function renormalization
\begin{equation}
\phi_{{\rm b}\epsilon}(x^\mu)\epsilon^{\nu-2}\equiv\widetilde{\phi}_{\rm
 b}(x^\mu).
\label{source_reno}
\end{equation}
In this renormalization, the higher order terms in (\ref{on-shell_11})
vanish, such that we could access renormalized finite quantities.
Then, we can obtain the two point function of the corresponding operator
${\cal O}(x)$  which couples the scalar source $\phi(x)$ via
GKP-W relation (\ref{pf_3}),
\begin{eqnarray}
\big\langle{\cal O}(x){\cal O}(x')\big\rangle
&\equiv&
-\frac{\delta^2S}
{\delta\widetilde{\phi}_{\rm b}(x)\widetilde{\phi}_{\rm b}(x')}
=
-2L^3\nu\frac{\Gamma\big(1-\nu\big)}{\Gamma\big(1+\nu\big)}
\frac{1}{2^{2\nu}}
\!\int\!\frac{\dd^4 k}{(2\pi)^4}
\ \eee^{-ik(x-x')}k^{2\nu}
\nonumber
\\
&=&
\frac{2L^3}{\pi^2}\nu\frac{\Gamma\big(\nu+2\big)}
{\Gamma\big(\nu\big)}\frac{1}{|x-x'|^{2(2+\nu)}},
\label{two_point_function}
\end{eqnarray}
where we have used the formula (\ref{fourier_trans_formula})
to write the correlation function in  real space.
In the presence of the source term,
the one-point function can be also estimated as
\begin{equation}
\big\langle{\cal O}(x)\big\rangle_{\widetilde{\phi}_{\rm b}}
\equiv\frac{\delta S}{\delta\widetilde{\phi}_{\rm b}(x)}
=-L^3\!\int\!
\frac{\dd^4k}{(2\pi)^4}\ 2\nu A(k)b_0(k)\ \eee^{-ikx}.
\label{vev_source}
\end{equation}

Let us relate the result in terms of CFT.
Near the boundary (\ref{UV_expansion}),
we have
\begin{equation}
\phi(x^\mu, z)
=\!\int\!\frac{\dd^4x}{(2\pi)^4}
\ \eee^{ik_\mu x^\mu}
A(k)\Big(z^{2-\nu}\big(a_0(k)+{\cal O}(z^2)\big)
+z^{2+\nu}\big(b_0(k)+{\cal O}(z^2)\big)\Big).
\label{asymptotic_solution_s}
\end{equation}
We have identified the first part as the source term
(\ref{source_reno}), while the second term as the
VEV of the corresponding operator (\ref{vev_source}).
Since we  imposed the scale invariance of the CFT side
in the relation (\ref{pf_1}), i.e.\ (\ref{cft_source_dimension}),
we shall consider the scaling behavior in the gravity side.
Requiring the scale invariance of the asymptotic
solution (\ref{asymptotic_solution_s})
under the scaling $(x^\mu, z)\to (x'^\mu=\eee^{-\lambda}x^\mu,
z'=\eee^{-\lambda}z)$
in 5D and
(\ref{scaling_op}) and (\ref{cft_source_dimension}),
we find the condition
\begin{equation}
\Delta=2+\nu=2+\sqrt{4+L^2m^2},
\label{conformal_dimension_scalar}
\end{equation}
which relates the scaling dimension of 4D CFT to the mass of the 5D
gravity theory.
Comparing with the general argument from the CFT in (\ref{cft_correlators}),
the two point function  (\ref{two_point_function}) satisfies
the correct scaling behavior of the operator ${\cal O}(x^\mu)$ with
the dimension $\Delta=2+\nu$.
For one-point function, if we send the source as $\widetilde{\phi}_0=0$,
the one-point function (\ref{vev_source}) vanishes, since,
in this case,  the VEV is proportional to the source.
This is consistent from the CFT point of view in which the VEV
of the operator with non-zero scaling dimension should vanish.
Nevertheless, the relation (\ref{vev_source}) plays the central role for
the non-conformal application of the AdS/CFT
correspondence~\cite{bklt, klebanov_witten}.

Let us reconsider what we did in this simple example and further read
off some general structures.
Of course what we need is to solve the equation of motion like
(\ref{eom_scalar_k}),  which is the second order ordinary differential equation.
In order to tackle to such equations, it is standard to use the
Frobenius series expansions.
We can expand the solution locally around some point.
For instance, around the boundary $z=0$,
we set the solution as $f_k(z)=Cz^\alpha(1+{\cal O}(z))$ with a constant
$C$ and a exponent $\alpha$.
Inserting the solution into the equation of motion,
we get the relation which the exponent $\alpha$ should satisfy.
For the equation of motion (\ref{eom_scalar_k}), we get
\begin{equation}
0=\alpha(\alpha-4)-L^2m^2, \quad {\rm i.e.} \quad
\alpha=2\pm\sqrt{4+L^2m^2}=2\pm\nu
\equiv\Delta_\pm.
\label{exponents}
\end{equation}
It is obviously understood
that the asymptotic solution (\ref{asymptotic_solution_s})
is constructed in this way.
The conformal dimension $\Delta$ (\ref{conformal_dimension_scalar})
corresponds to the larger root of $\alpha$
i.e.\ $\Delta_+$ in (\ref{exponents}).
These two modes with different exponents $\Delta_\pm$ could be
the independent basis of the solution around the boundary:
\begin{equation}
\varphi(x^\mu, z)={\cal A}(x^\mu)F^{(I)}(z)+{\cal B}(x^\mu)F^{(II)}(z),
\end{equation}
with
\begin{eqnarray*}
F^{(I)}(z)
=
z^{\Delta_-}\big(1+{\cal O}(z)\big) + c(z)F^{(II)}(z)\log z, \qquad
F^{(II)}(z)
=
z^{\Delta_+}\big(1+{\cal O}(z)\big),
\end{eqnarray*}
where ${\cal A}(x^\mu)$ and ${\cal B}(x^\mu)$ are integration constants.
Here the term proportional to $\log z$ may be needed to make the basis
$F^{(I)}(z)$ and $F^{(II)}(z)$ linear independent,
only in the case that the difference of two exponents is an integer.
Coefficients in the expansion in the basis $F^{(I)}(z)$ and $F^{(II)}(z)$
including $c(z)$ could be obtained recursively\footnote{
In general, the term proportional to $z^{\Delta_+-\Delta_-}$ in
$F^{(I)}(z)$ could not be fixed.
}.
In the similar way, for the massive scalar case,
around the Killing horizon $z\to\infty$,
we can find two basis (\ref{expansion_horizon})
through the asymptotic form of the equation of motion
$0=f''_k(z)-k^2f_k(z)$.

What we should do next is the ``analytic continuation'' to obtain the full
solution via finding out the transformation matrix which could relate
the two local basis in the different regions.
For instance, the hypergeometric function may be constructed in this
way with the help of the integral representations.
In the most general cases,
there exist no analytic expressions so that one needs to do these
procedures by some numerical methods.

As we have done before,
the integration constant ${\cal A}(x^\mu)$ could be identified as the
source (c.f.\ (\ref{source_reno})),
while the other constant ${\cal B}(x^\mu)$ could be related to
the VEV of the corresponding operators (c.f.\ (\ref{vev_source})).
Imposing the boundary condition in the interior of the AdS spacetime,
${\cal B}(x^\mu)$ and ${\cal A}(x^\mu)$ could be related, in general, nonlocally.

In order to complete the discussion,
let us revisit the scaling dimension in more general ground.
The scaling dimension is given by the root of the equation (\ref{exponents}).
In the CFT point of view, for scalar operators,
there exists the bound $\Delta\ge 1$ by requiring
the unitary representation.
In the case $L^2m^2\ge0$,
$\Delta_+$ only satisfies this bound, and the dimension is given by
$\Delta=\Delta_+\ge 4$.
Interesting thing happens in AdS spacetime.
We can consider the negative mass satisfies the bound
$L^2m^2\ge -4$ which is known as
Breitenlohner-Freedman bound~\cite{bf_bound}.
Within this bound, the energy of the scalar field propagating the AdS spacetime
can be positive due to the curvature effect of the geometry.
When the mass decrees from 0 towards the BF bound,
the root $\Delta_-(<1 \ \mbox{for positive mass}^2)$
approaches to the unitary bound and
may eventually satisfy the unitary bound in the case
$-4\le L^2m^2\le -3$.
Therefore, in this mass parameter region, we have two independent solutions
which may correspond to two different boundary CFTs,
i.e.\  $1\le\Delta_-\le 2$ and $2\le\Delta_+\le 3$~\cite{klebanov_witten}.

Before closing this subsection,
we display similar relations to (\ref{exponents})
for lower spins cases
\begin{subequations}
\begin{eqnarray}
\mbox{vector} &\qquad& L^2m^2=(\Delta-1)(\Delta-3)
\label{massless_vector}
\\
\mbox{symmetric tensor}
&\qquad&
L^2m^2=\Delta(\Delta-4)
\end{eqnarray}
\end{subequations}
where $m$ is the mass of the bulk field and $\Delta$ corresponds
to the scaling dimension of the 4D operator.
The massless 5D gauge field couples the dimension three operator
which may correspond to a $U(1)$ conserved current in 4D CFT.
The graviton couples
to an operator with dimension four
which would be the energy-momentum tensor.

\section{AdS/CFT correspondence at finite temperature}

After the AdS/CFT correspondence, many modifications of the original
conjecture have been proposed and studied.
We treat the AdS/CFT correspondence as a special case of the general
bulk/boundary correspondence.
One of the directions is to modify the gravity side in a suitable
manner in order to study some features in the gauge theory side.
One of the attempts is that for the thermal
system~\cite{Witten:1998zw}.

\subsection{Thermal field theory}

Before embarking the application for AdS/CFT correspondence,
let us briefly summarize the thermal field theory and the black hole
thermodynamics.

We start by formulating the thermal field theories in equilibrium.
The thermodynamic quantities could be determined by the canonical
partition function at temperature $T$,
\begin{equation}
Z={\rm Tr} \ \eee^{-\frac{1}{T}{H_0}},
\label{thermal_partition_function}
\end{equation}
where $H_0$ is the Hamiltonian.

For free bosonic/fermionic gas,
we could compute the partition function (\ref{thermal_partition_function})
through the phase space integration,
\begin{equation}
\log Z=\mp V_3\!\int\!\frac{\dd^3p}{(2\pi)^3}
\log\big(1\mp\eee^{-\frac{1}{T}\omega(\vec{p})}\big),
\end{equation}
where $\mp$ correspond to boson and fermion, respectively, and
$\omega(\vec{p})$ is the dispersion relation, i.e.\
$\omega(\vec{p})=\sqrt{\vec{p}^2+m^2}$ for relativistic particle.
For later use, we display the results for the massless free
particles for canonical ensemble:
\begin{equation}
\log Z
=\left\{
\begin{array}{ll}
\displaystyle
V_3\frac{\pi^2}{90}T^3
& \mbox{boson}
\\
\displaystyle
V_3\frac{7\pi^2}{720}T^3
& \mbox{fermion}
\end{array}
\right.
\label{free_gass}
\end{equation}

We may evaluate the partition function
(\ref{thermal_partition_function}) by Euclidean path-integral
in the imaginary time formalism.
We analytically continue to Euclidean spacetime
and compactify the time direction with the thermal period $1/T$:
\begin{equation}
Z=\!\int\!{\cal D}\varphi \ \eee^{-S_{\rm E}[\varphi]},
\quad
{\rm with}
\quad
S_{\rm E}[\varphi]
=\!\int_0^{1/T}\!\!\!\!\dd\tau\!\int\!
\dd{\vec{x}}{\cal L}(\varphi(\tau, \vec{x})),
\end{equation}
where we need to impose periodic (anti-periodic) boundary conditions
for bosonic (fermionic) fields $\varphi(\tau, \vec{x})$,
which eventually break supersymmetry due to different types of
the mode expansions around the period.
The precise thermodynamic quantities can be found in the later
sections.

\subsection{Black hole thermodynamics}

One could proceed with the gravitational system
in the same spirit.
In $D$-dimensional Euclidean spacetime,
the evaluation of the partition function could be regarded as
that of the canonical ensemble,
\begin{equation}
Z=\!\int\!{\cal D}g\ \eee^{-S_{\rm E}[g]}
=\sum_n\ \eee^{-\frac{1}{T}E_n},
\end{equation}
where the action $S_{\rm E}[g]$
is typically Einstein-Hilbert term and
$E_n$ denotes  the eigenvalue of the Hamiltonian.

In the semiclassical approximation,
the most dominant contribution to the path-integral is from
the solution of the classical equation of motion i.e.\ Einstein
equation.
Since the black hole is one of these solutions,
the Einstein-Hilbert action evaluated at that configuration
could be considered as the leading contributions
to the free energy $F(= E -TS)$,
\begin{equation}
\eee^{-\frac{1}{T}F}\equiv Z\sim \eee^{-S_{\rm E}[g=g_{\rm black \
 hole}]},
\qquad
\mbox{i.e.} \qquad
F=TS_{\rm E}[g=g_{\mbox{\scriptsize black hole}}].
\label{free_energy_0}
\end{equation}
As we will discuss, the temperature $T$ could be defined as the
Hawking temperature, which is a natural consequence from the black hole
geometry.
In order to evaluate the on-shell action $S_{\rm E}$,
we need to regularize the action in general.
The typical procedure is to add suitable counter terms to
the pure gravitational action.
By using the regularized on-shell action,
we could obtain the thermal quantities,
for instance the energy $E$ and the entropy $S$ as
\begin{equation}
E=-T^2\frac{\del S_{\rm E}}{\del T},
\qquad
S=\frac{E}{T}-S_{\rm E}.
\label{energy_entropy}
\end{equation}

If we consider the charged black hole so-called Reissner-Nordstr\"om black hole,
we could access to the grand canonical partition function.
The Reissner-Nordstr\"om black hole is the solution of the
Einstein-Maxwell system,
\begin{equation}
S_{\rm E}[g, A]=S_{\rm EH}[g]+S_{\rm Maxwell}[g, A],
\label{rn_0}
\end{equation}
with
\begin{equation}
S_{\rm Maxwell}[g, A]
=-\frac{1}{16\pi G_{\rm N}^{(D)}}
\!\int_{\cal M}\!\!\dd^D x\sqrt{g}F_{mn}F^{mn}.
\end{equation}
The solution for the gauge potential is given by
\begin{equation}
A_{\rm RN}=A_t\dd t=\Big(-\mu+\frac{Q}{r^{D-3}}\Big)\dd t.
\end{equation}
The integration constants $\mu$ and $Q$ may
correspond to the chemical potential and the number density,
respectively.
Through the semiclassical evaluation of the
path-integral like (\ref{free_energy_0}),
we equate the grand canonical potential $\Omega(\mu, T)$
to the on-shell action,
\begin{equation}
\Omega=
T\Big(
S_{\rm EH}[g=g_{\rm RN}]
+S_{\rm maxwell}[g=g_{\rm RN}, A=A_{\rm RN}]
\Big)
=E -TS-\mu Q.
\label{grand_potential}
\end{equation}
%

\subsection{AdS-Schwarzschild black hole}

We here combine two path-integral formulations discussed in the
last two subsections in the context of the AdS/CFT correspondence.
We extend the identification of the
gauge/gravity theories in (\ref{pf}) for the thermal version.
The bulk geometry in the gravity side can be replaced by certain black
hole geometry.
We equate the canonical partition function (\ref{free_energy_0})
for the gravity to that of the thermal gauge theory.

Let us be back to the initial place for the correspondence in the string
theory framework.
For our purpose,  first we compactify 10D spacetime as $S^1\times R^9$.
In order to discuss a 4D world volume thermal gauge theory,
we consider D3-branes wrapping on the Euclidean time circle $S^1$.
In the context of AdS/CFT correspondence,
we need to know the corresponding supergravity solutions.
Here we use more general RR 3-brane solution (\ref{ads_s0})\footnote{
One can consider Euclidean version of the extremal
solution (\ref{ads_00}) with $S^1$ compactification.
However, by comparing the free energies,
one could see that the solution (\ref{ads_s0}) is energetically favored.
},
which corresponds to non-BPS D3-brane with renaming the parameter
$L\to\widetilde{L}$:
\begin{equation}
\dd s^2
=\frac{1}{\sqrt{\displaystyle 1+\frac{\widetilde{L}^{4}}{r^{4}}}}
\bigg\{
\Big(1-\frac{r_{\rm H}^{4}}{r^{4}}\Big)\dd \tau^2
+\sum_{i=1}^3\dd x^{i2}
\bigg\}
+\sqrt{1+\frac{\widetilde{L}^{4}}{r^{4}}}
\bigg\{\frac{\dd r^2}{\Big(\displaystyle1
-\frac{r_{\rm H}^{4}}{r^{4}}\Big)}
+r^2\dd\Omega^2_{5}
\bigg\},
\label{non_extremal_d3}
\end{equation}
where we work with the Euclidean time $\tau$ with the period $1/T$.
In order to fix the integration constant $\widetilde{L}$,
we use the identification (\ref{identification}) for the non-extremal
solution.
This identification gives the relation
\begin{equation}
\widetilde{L}^4(\widetilde{L}^4+r_{\rm H}^4)=L^8,
\label{relation_ll}
\end{equation}
where $L$ is the same as (\ref{d3_l}).
Taking the limit $r_{\rm H}\to 0$, the solution is obviously reduced to
the extremal D3-brane solution.

The metric (\ref{non_extremal_d3}) contains a coordinate
singularity at $r=r_{\rm H}$ which represents the event horizon.
Let us look at the horizon closely,
\begin{equation}
\dd s^2|_{r\to r_{\rm H}}
=\frac{4r_{\rm H}}{\sqrt{r_{\rm H}^4+\widetilde{L}^4}}
\big(r-r_{\rm H}\big)\dd\tau^2
+\frac{\sqrt{r_{\rm H}^4+\widetilde{L}^4}}
{4r_{\rm H}\big(r-r_{\rm H}\big)}\dd r^2+\cdots
=\frac{\sqrt{r_{\rm H}^4+\widetilde{L}^4}}{r_{\rm H}^2}
\Big(\dd\xi^2
+\frac{4r_{\rm H}^2}{r_{\rm H}^4+\widetilde{L}^4}
\xi^2\dd\tau^2\Big)
+\cdots,
\label{htd_0}
\end{equation}
where we have introduced the coordinate $\xi$ as the deviation from the
horizon $r=r_{\rm H}+\xi^2/r_{\rm H}$.
\begin{wrapfigure}{r}{4cm}
\begin{center}
\includegraphics[clip, width=4cm]{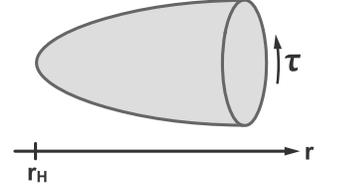}
\caption{$\tau$-$r$ surface around the horizon}
\label{tau_r}
\end{center}
\end{wrapfigure}
We can regard the Euclidean time coordinate $\tau$ as the angular
variable
$\theta\equiv\big(2r_{\rm H}/\sqrt{r_{\rm H}^4+\widetilde{L}^4}\big)\tau$.
Imposing the periodicity at $\xi=0$, i.e.\ removing the conical
singularity, we obtain
\begin{equation}
2\pi=\frac{2r_{\rm H}}{\sqrt{r_{\rm H}^4+\widetilde{L}^4}}\frac{1}{T},
\qquad {\mbox{i.e.}}
\qquad T=\frac{r_{\rm H}}{\pi\sqrt{r_{\rm H}^4+\widetilde{L}^4}}.
\label{hawking_temp_d3}
\end{equation}
This is the Hawking temperature which could be identified with that of
the gauge theory.
The previous D3-brane solution which is realized by $r_{\rm H}=0$ corresponds
to zero temperature, i.e.\ the ground state.

Let us now consider the decoupling limit.
This field theory limit may be taken to be $\alpha'\to 0$ with
keeping the temperature fixed in addition to the gauge
coupling constant.
By using the relations (\ref{relation_ll}), (\ref{hawking_temp_d3}),
and $L\propto\sqrt{\alpha'}$, we can estimate
the scaling behaviors
as $\widetilde{L}\propto\sqrt{\alpha'}$ and $r_{\rm H}\propto\alpha'$.
Eliminating $r_{\rm H}$ from the relations (\ref{relation_ll})
and (\ref{hawking_temp_d3}),  we could obtain the deviation from
extremality,
\begin{equation}
\widetilde{L}^4=L^4\Big(1-\frac{1}{2}(\pi^4L^4T^4)
+{\cal O}(L^8T^8)\Big).
\label{extremality}
\end{equation}
As we did in the D3-brane case, keeping the energy scale
$u=r/L^2$ i.e.\ $r\propto\alpha'$, we obtain the metric (\ref{non_extremal_d3}) as
\begin{equation}
\frac{\dd s^2}{L^2}
=u^2\bigg(\Big(1-\frac{u_{\rm H}^4}{u^4}\Big)\dd\tau^2
+\sum_{i=1}^3\dd x^{i 2}\bigg)
+\frac{1}{u^2}
\frac{\dd u^2}{\Big(\displaystyle 1-\frac{u_{\rm H}^4}{u^4}\Big)}
+\dd\Omega^2_5,
\label{non_extremal_d3_nh}
\end{equation}
where we have denoted $u_{\rm H}=r_{\rm H}/L^2$ which is order one in
the decoupling limit.
This near horizon geometry is (5D AdS-Schwarzschild black hole)$\times
S^5$ with a common radius $L$,
which provides the dual description of the boundary gauge
theory at finite temperature.

We can calculate thermodynamic quantities and verify the thermodynamic
relations.
In the near horizon limit, the mass (energy) can be calculated from
(\ref{tension_bp}) by using the relation (\ref{extremality}),
\begin{equation}
\frac{M}{V_3}=T_3N_{\rm c}+\frac{3}{8}\pi^2N_{\rm c}^2T^4,
\label{energy_d3t}
\end{equation}
where in the near horizon limit,
the temperature (\ref{hawking_temp_d3}) is given by
\begin{equation}
T=\frac{r_{\rm H}}{\pi L^2}.
\end{equation}
The first term is from the $N_{\rm c}$ D3-branes tension
and the second term corresponds to the energy density of the
field theory.
The entropy can be also calculated as the Bekenstein-Hawking entropy
in the context of the black hole thermodynamics,
\begin{equation}
S=\frac{\mbox{(Area)}}{4G_{\rm N}^{(D)}}.
\end{equation}
At the horizon, the metric (\ref{non_extremal_d3_nh}) has the
form
$
\dd s^2|_{u=u_{\rm H}}
=L^2u_{\rm H}^2\sum_{i=1}^3\dd x^{i 2}+L^2\dd\Omega^2_5.
$
Therefore, the area can be computed as
$
\mbox{Area}=\sqrt{(L^2u_{\rm H}^2)^3 L^{10}}\ V_3 V_{S^5},
$
and we obtain the entropy density
\begin{equation}
s=\frac{S}{V_3}=
\frac{\pi^2}{2}N_{\rm c}^2T^3.
\label{entropy_d3t}
\end{equation}
These energy and entropy could be also reproduced
through the evaluation of the on-shell
action (\ref{energy_entropy}) with identifying
the free energy of the black hole with that of the gauge theory.

Let us compare the result with that of the free gas approximation.
The physical  on-shell degrees of freedom for the field theory i.e.\
$SU(N_{\rm c})$ ${\cal N}=4$ super Yang-Mills which we have discussed
are $N_{\rm c}^2\times\big(2\ (\mbox{gauge field})
+6\ (\mbox{scalar})\big)$ for
the bosons and $N_{\rm c}^2\times 8\ (=N_{\rm c}^2\times 2\times 4)$ for
4 Weyl fermions.
Therefore, summing up the degrees of freedom for (\ref{free_gass}),
we find
\begin{equation}
\log Z=\frac{\pi^2}{6}V_3N_{\rm c}^2T^3.
\end{equation}
Since the entropy can be calculated via $S=\del(T\log Z)/\del T$,
we then conclude~\cite{Gubser:1996de}
\begin{equation}
s_{\mbox{\scriptsize black hole}}
=\frac{3}{4}s_{\mbox{\scriptsize free gas}}.
\end{equation}
The entropy for the strong coupling regime may be
reduced from the free system by the factor $3/4$.

In the results (\ref{energy_d3t}) and (\ref{entropy_d3t}),
the scaling $T^3$ could be understood by the conformal invariance.
The factor $N_{\rm c}^2$ indicates the theory is in the deconfined
phase where the color degrees of freedom could be visible.

\subsection{Finite density}

The AdS/CFT correspondence can be also extended to the case with
a finite charge density.
A $U(1)$ gauge symmetry in the AdS black hole background
may be dual to a global $U(1)$ symmetry in the gauge theory side.
Introducing the gauge field to the gravity side,
we could consider the grand potential (\ref{grand_potential})
and discuss the thermodynamics and related phase structure.

If we consider the gauge field in the bulk gravity theory,
the equation (\ref{pf}) may be the form,
\begin{equation}
\big\langle\ \eee^{\int\dd^4x A_\mu(x^m)J^\mu(x^\mu)}
\ \big\rangle_{\rm CFT}
=\eee^{-S_{\mbox{\scriptsize supergravity}}
[A_\mu(x^\mu)]\big|_{\mbox{\scriptsize on-shell}}}
\quad
\mbox{with}
\quad
A_\mu(x^\mu)=A_\mu(x^M)\big|_{\mbox{\scriptsize boundary}}.
\label{pf_10}
\end{equation}
In the left hand side in (\ref{pf_10}), the boundary value of the
bulk gauge field $A_\mu(x^\mu)$
couples to the current operator $J^\mu(x^\mu)$
which should be a conserved current due to the gauge invariance
of the coupling.
The boundary value of the time component of the gauge
field $A_t(x^\mu)$ could be identified with
the chemical potential $\mu$ associated with the density
in the gauge theory~\footnote{
In the presence of the black hole,
one could define the chemical potential in the gauge invariant way,
$\mu=\int_{r_{\rm H}}^\infty\dd rF_{rt}
=A_t(\infty)-A_t(r_{\rm H})=A_t(\infty)$,
where we have used the natural condition for the gauge potential
at the horizon $A_t(r_{\rm H})=0$.
 },
\begin{equation}
\mu=A_t(x^M)|_{\rm boundary}.
\label{chmical_potential}
\end{equation}
This could be understood in the following way.
Under the interpretation (\ref{chmical_potential}),
the left hand side of the equation (\ref{pf_10})
may become the grand canonical partition function
of the gauge theory
\begin{equation}
\big\langle\ \eee^{\int\dd^4x A_\mu(x^m)J^\mu(x^\mu)}
\ \big\rangle_{\rm CFT}
=\big\langle\ \eee^{\int_{0}^{1/T}\!\!\dd\tau
\mu Q}
\ \big\rangle_{\rm CFT}
=\int{\cal D}\phi \ \eee^{-\int_0^{1/T}\!\!\dd\tau(H_0-\mu Q)},
\end{equation}
where the charge $Q$ is defined by usual way $Q=\int\!\dd^3x J^t(x)$.

In the context of the AdS/CFT correspondence, the $U(1)$ gauge
symmetry can be naturally introduced~\cite{rnads_0}.
In the static non-extremal black 3-brane solution
(\ref{non_extremal_d3}),
we turn on three different angular momenta in the 6D
transverse space to the world volume.
Specifying the three rotation axes breaks the $SO(6)$ rotational
symmetry to Cartan parts $U(1)^3$.
As  in the usual AdS/CFT correspondence,
we next take the near horizon limit with keeping the rotational
parameters constant.
Then, the rotating 3-brane metric becomes a product spacetime
consisting of asymptotic AdS$_5$ and modified $S^5$.
Identifying the rotational parameters with charges,
the asymptotic AdS$_5$ part could be understood as the
black hole with three charges, known as
the STU AdS black hole~\cite{stu_bh}.
This is the solution of the 5D ${\cal N}=2$ gauged supergravity
which comes from the consistent truncation of the $SO(6)$ gauged
${\cal N}=8$ supergravity.
As we mentioned,  the $SO(6)$ gauged ${\cal N}=8$ supergravity
is considered as a consistent $S^5$ reduction of type IIB supergravity.
If the three charges are the same, the solution is simply reduced
to 5D Reissner-Nordstr\"om-AdS black hole which is the solution of
equations of motion given by the action
(\ref{rn_0}) with a negative cosmological constant.
Applying the AdS/CFT correspondence to this simple model,
one could discuss thermal physics especially in the hydrodynamic
limit~\cite{rnads_1}.

\section{D3/D7 model}

We have discussed the gauge/gravity duality and extended the
idea to describe the gauge theory with finite temperature and density.
In the gauge theory, so far, we have only considered the adjoint fields.
Two endpoints of open strings which correspond to point
charges in the fundamental/anti-fundamental representations have been
attached to the same branes,
so that they transform in the adjoint representation of the $SU(N_{\rm
c})$ gauge group.
In order to consider the fundamental matter with flavors,
we need to introduce open strings whose two endpoints, both of them, are not
attached to the color branes~\cite{karch_katz}.
This can be realized by introducing different types of D-branes.
Since the fundamental matter lives in 4D, this 4D spacetime
should be filled by these D$p$-branes $(p>3)$.
In this section we consider D3/D7 model which describes $SU(N_{\rm c})$
${\cal N}=4$
super Yang-Mills with
${\cal N}=2$ $N_{\rm f}$ quark hypermultiplets.

\begin{figure}[htbb]
\begin{minipage}{1.0\textwidth}
\begin{center}
\includegraphics[clip, width=6cm]{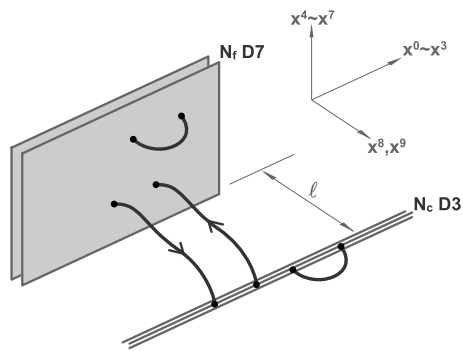}
\caption{D3/D7 configuration.}
\label{d3d7}
\end{center}
\end{minipage}
\end{figure}

We consider a stack of $N_{\rm c}$ D3-branes and
another stack of $N_{\rm f}$ D7-branes.
The D-brane configuration is given by Table \ref{braneprofile_d3d7} and
several open strings attached on the D-branes are visualized in Fig.\ref{d3d7}.
\begin{table}[htbb]
  	\begin{center}
  	\begin{tabular}{|>{\centering}p{20pt}*{9}{|>{\centering}p{13pt}}|p{13pt}|}
		\hline
		&$t$&$x^1$&$x^2$&$x^3$&$x^4$&$x^5$&$x^6$&$x^7$&$x^8$&$x^9$ \\
		\hline
		D3&$\bullet$&$\bullet$&$\bullet$&$\bullet$&&&&&& \\
		\hline
		D7&$\bullet$&$\bullet$&$\bullet$&$\bullet$
&$\bullet$&$\bullet$&$\bullet$&$\bullet$&& \\
		\hline
  	\end{tabular}
  	\caption{The brane configurations: the background D3- and the probe D7-branes}
	\label{braneprofile_d3d7}
  	\end{center}
\end{table}
As we have discussed,
we could see ${\cal N}=4$ super Yang-Mills sector
coming from open strings attached in the D3-branes.
Quarks are described by the open strings stretched
between D3- and D7-branes.
This is the case with  $n=4$ (number of (ND) directions)
in (\ref{open_r}) and
(\ref{open_ns}).
The R sector provides the fermion with mass $M_{\rm
q}=l/(2\pi\alpha')$,
while the NS sector provide the scalar with the same mass.
This indicates that the model has supersymmetry.
There also exist open string modes in D7-branes which provide the
$U(N_{\rm f})$ gauge theory in their world volume.
These adjoint representations are naturally considered
as the mesonic degrees of freedom.
These states are described by the fluctuation of the D7-branes.

Now we shall discuss the decoupling limit $\alpha'\to0$.
From (\ref{coupling_0}), the gauge coupling constants have different
$\alpha'$ scalings for D3- and D7-branes,
\begin{equation}
g_{\mbox{\scriptsize YM D7}}^2=(2\pi)^4\alpha'^2g_{\rm YM}^2,
\end{equation}
where $g_{\rm YM}^2$ is the D3-brane Yang-Mills coupling constant
given by (\ref{d3_ym}).
Taking the limit $\alpha'\to0$ while keeping the 4D Yang-Mills coupling
constant, $g_{\mbox{\scriptsize YM D7}}^2$ vanishes.
Therefore, the $U(N_{\rm f})$ gauge symmetry on $N_{\rm f}$ D7-branes
becomes global one.
Moreover, since the interaction terms between strings (D3-D7) and
those for (D7-D7) are proportional to $g_{\mbox{\scriptsize YM D7}}^2$,
the (D7-D7) sector could be decoupled to the other sectors (D3-D3)
and (D3-D7).
As a result, bifundamental fields which come from (D3-D7)/(D7-D3) strings
transform as the fundamental/antifundamental representations
of the $SU(N_{\rm c})$ gauge group and $U(N_{\rm f})$ global flavor group.

In order to discuss the gauge theory in the context of the AdS/CFT
correspondence, we need to have a dual gravity picture.
In general, it is hard to obtain supergravity solutions for these
coupled D-brane system.
Then one can use so-called the probe
approximation~\cite{karch_katz, babington, Kruczenski:2003uq}.
We here would like to treat D3-branes as the ``background'' and D7-branes as the
probe to this background.
Just like the quenched approximation in lattice QCD
in studying the effect of a test quark
in the given gluon background,
we assume that the probe flavor D7-branes do not affect
the geometry.
This could be achieved by $N_{\rm c}\gg N_{\rm f}$ such that
the gravitational backreaction could be
suppressed by $N_{\rm f}/N_{\rm c}$.
This could be easily observed in the Newton potential as in
(\ref{backreaction}),
$$
8\pi G_{\rm N}^{(10)}T_{tt}\Big|_{\rm D7}\propto\lambda_{\rm D7}
=N_{\rm f}g_{\rm s}=\lambda\frac{N_{\rm f}}{N_{\rm c}}.
$$
Therefore, we could separately solve the equations of motion for
the gravity sector.
As a result, D3-branes provide AdS$_5\times S^5$ background geometry,
and in this background we consider the consistent embedding of the D7-branes,
which is determined by the DBI action (\ref{dbi}).

\subsection{Thermodynamics}

In this section,
we briefly summarize D3/D7 system at finite temperature.
Following \cite{d3d7, d3d7_m},
we consider the ``black hole embedding'' of the probe D7-branes
in the D3-brane background in which we could introduce
the finite quark mass together with the chemical potential.

We start by introducing non BPS $N_c$ D3-branes.
The gravity dual descriptions is given by the metric
(\ref{non_extremal_d3_nh});
\begin{equation}
\dd s^2=
\frac{u^2}{L^2}
\Big(f_0(u)\dd t^2
+\sum_{i=1}^3\dd x^{i 2}
\Big)
+\frac{L^2}{u^2}
\Big(
\frac{\dd u^2}{f_0(u)}
+u^2\dd\Omega_5^2
\Big),
\qquad
\mbox{with}
\qquad
f_0(u)=1-\frac{u_{\rm H}^4}{u^4},
\label{D3geom_0}
\end{equation}
where we have rescaled the radial coordinate as $u\to u/L^2$ in the
expression (\ref{non_extremal_d3_nh}).
In the D3-brane background, the background dilaton is constant
$\eee^{\widetilde{\Phi}}=1$ in our notation.
Following the paper~\cite{babington, d3d7_m},
we introduce new coordinate $\varrho$ through
$\varrho^2=u^2+\sqrt{u^4-u_{\rm H}^4}$,
and rewrite the metric (\ref{D3geom_0}) as
\begin{equation}
\dd s^2
=
\frac{1}{2}\frac{\varrho^2}{L^2}
\left(
\frac{f^2(\varrho)}{\tilde f(\varrho)}\dd t^2
+\tilde{f}(\varrho)\sum_{i=1}^3\dd x^{i 2}
\right)
+
\frac{L^2}{\varrho^2}
\Big(\dd\varrho^2+\varrho^2 \dd\Omega_5^2\Big),
\ \
\mbox{with}
\ \
f(\varrho) = 1-\frac{u_{\rm H}^4}{\varrho^4}, \ \
\tilde{f}(\varrho)=1+\frac{u_{\rm H}^4}{\varrho^4}.
\label{D3geom}
\end{equation}
We work with the dimensionless coordinate
$\rho\equiv\varrho/u_{\rm H}$.
We refer to the horizon as $\rho=1$ and the AdS boundary
as $\rho\ (=\sqrt{2}u/u_{\rm H})\to\infty$.

Next, we introduce $N_{\rm f}$ D7-branes as the flavor branes
with the intersection given in Fig.\ref{d3d7}
where D7-branes are wrapping on $S^3$ of $S^5$.
Taking the probe approximation $N_{\rm f}\ll N_{\rm c}$,
we could consider the dynamics of $N_{\rm f}$ D7-branes
which can be described by the DBI action (\ref{dbi}) in the
10D background (\ref{D3geom}),
\begin{equation}
S_{\rm D7}=N_fT_7\!\int\!\dd^8\sigma
\sqrt{\ \det(g_{mn}+2\pi\alpha'F_{mn})},
\label{dbi_d7}
\end{equation}
where we do not consider
the fluctuations of the dilaton $\widehat{\Phi}(\sigma)$ and
the Kalb-Ramond field $B_{mn}(\sigma)$.
It is convenient to divide the transverse 6D part to
the D3-branes in (\ref{D3geom})
into two parts i.e.\ 4D and 2D whose coordinates are given by
spherical $(r, \Omega_3)$ and polar $(R, \varphi)$ coordinates,
respectively,
\begin{eqnarray}
(\dd\rho)^2 + \rho^2\dd\Omega_5^2
&=&
(\dd r)^2 + r^2\dd\Omega_3^2 + (\dd\!R)^2+R^2(\dd\varphi)^2
\nonumber
\\
&=&
(\dd\rho)^2 + \rho^2
\Big(
(\dd\theta)^2 + \sin^2\theta  \, \dd\Omega_3^2
+ \cos^2\theta \, (\dd\varphi)^2
\Big),
\label{met2}
\end{eqnarray}
where $r=\rho\sin\theta$,
$R=\rho\cos\theta$ with $0\le\theta\le\pi/2$
and $\rho^2=r^2+R^2$.
By construction,
we consider the case where the world volume coordinates of D7-branes
are given in the static
gauge i.e.\ $\sigma^m\equiv(t, x^i, \rho, \Omega_3)$.
Due to the symmetries for the translation in $(t, x^i)$ and the
rotation in $(\rho, \Omega_3)$, the embedding of the D7-branes
could  depend only on the radial coordinate $\rho$.
Since the rotational symmetry in $(R, \varphi)$ allows
to set $\varphi=0$, the embedding might be characterized by
$\chi(\rho)\equiv\cos\theta$ through $\theta(\rho)$ which is the angle
between two spaces $(r, \Omega_3)$ and $(R, \varphi)$.
The asymptotic value of the distance between D3 and D7-branes
which is given by $R(\rho)$ for large $\rho$ provides the quark mass
$M_{\rm q}$.

The induced metric on the D7-branes can be obtained as
\begin{equation}
\dd s^2_{\rm D7}
=
L^2\Bigg\{
\frac{\pi^2T^2}{2}\rho^2
\left(
\frac{f^2}{\tilde f}\dd t^2
+ \tilde{f} \sum_{i=1}^3\dd x^{i 2}
\right)
+
\frac{1}{\rho^2}
\left(
\frac{1-\chi^2 + \rho^2 \dot{\chi}^2 }{1-\chi^2}
\right)
\dd\rho^2
+
\big(1-\chi^2\big)\ \dd\Omega_3^2
\Bigg\},
\label{D3geom-induced}
\end{equation}
where the dot stands for the derivative with respective to $\rho$.
We also introduce the non-dynamical temporal component of the
gauge field $A_t(\rho)$ which incorporates the chemical potential and
the density at the AdS boundary.

By using the induced metric (\ref{D3geom-induced}) and the gauge
potential $A_t(\rho)$,
the DBI action (\ref{dbi_d7}) now becomes
\begin{equation}
S_{\rm D7}=\frac{\lambda N_cN_fT^3}{32}V_3
\!\int\!\dd\rho\
\rho^3\tilde{f}(1-\chi^2)
\sqrt{f^2(1-\chi^2+\rho^2\dot\chi^2)
-2\tilde{f}(1-\chi^2)\dot{\widetilde{A}_t^2}},
\label{dimensionless}
\end{equation}
where we have defined
$\widetilde{A}_t(\rho)\equiv 2\pi\alpha'A_t(\rho)/u_{\rm H}$
and the 't Hooft coupling $\lambda=g_{\rm YM}^2N_{\rm c}$.
Since there exist no $\widetilde{A}_t(\rho)$ terms in the action,
the equation of motion for $\widetilde{A}_t(\rho)$ can be
reduced to the following form with an integration
constant $\widetilde{d}$,
\begin{equation}
\widetilde{d}\equiv
\frac{\rho^3\tilde{f}^2(1-\chi^2)^2\dot{\widetilde{A}_t}}
{2\sqrt{f^2(1-\chi^2+\rho^2\dot{\chi}^2)-2\tilde{f}(1-\chi^2)
\dot{\widetilde{A}^2_t}}}.
\label{eom_ta}
\end{equation}
The equation of motion for $\chi(\rho)$ is given as
\begin{eqnarray}
0
&=&
\del_\rho
\Bigg\{
\frac{\rho^5f\tilde{f}(1-\chi^2)\dot{\chi}}
{\sqrt{1-\chi^2+\rho^2\dot{\chi}^2}}
\Bigg(1+\frac{8\widetilde{d}^2}{\rho^6\tilde{f}^3(1-\chi^2)^3}\Bigg)^{1/2}
\Bigg\}
\nonumber
\\
&&+\frac{\rho^3f\tilde{f}\chi}{\sqrt{1-\chi^2+\rho^2\dot{\chi}^2}}
\Bigg\{
\Big(3(1-\chi^2)+2\rho^2\dot{\chi}^2\Big)
\Bigg(1+\frac{8\widetilde{d}^2}{\rho^6\tilde{f}^3(1-\chi^2)^3}\Bigg)^{1/2}
\nonumber
\\
&&
\hspace*{34mm}
-\frac{24\widetilde{d}^2(1-\chi^2+\rho^2\dot{\chi}^2)}
{\rho^6\tilde{f}^3(1-\chi^2)^3}
\Bigg(1+\frac{8\widetilde{d}^2}{\rho^6\tilde{f}^3(1-\chi^2)^3}\Bigg)^{-1/2}
\Bigg\},
\quad\quad
\label{eom_chi}
\end{eqnarray}
where we have eliminated the gauge field $\widetilde{A}_t(\rho)$ by
using the relation (\ref{eom_ta}).
Near the boundary, asymptotic solutions of the equations of motion
(\ref{eom_ta}) and (\ref{eom_chi}) behave as
\begin{equation}
\widetilde{A}_t(\rho)
=
\widetilde{\mu}-\frac{\widetilde{d}}{\rho^2}+\cdots,
\qquad
\chi(\rho)
=
\frac{m}{\rho}+\frac{c}{\rho^3}+\cdots.
\label{aspt_a}
\end{equation}
The chemical potential $\mu$ can be defined as the boundary
value of $A_t(\rho)$,
while the quark mass $M_{\rm q}$ can be estimated through the asymptotic
value of the separation of D3 and
D7-branes i.e.\ $M_{\rm q}=\rho\chi(\rho)/(2\pi\alpha')$ at the AdS boundary.
Taking the rescaling of the gauge field $\widetilde{A}_t$ and
the coordinate $\rho$ into account, the integration constants
$\widetilde{\mu}$ and $m$ can be related to these values,
\begin{equation}
\widetilde{\mu}
=
\frac{2\pi\alpha'}{u_0}\mu
=\sqrt{\frac{2}{\lambda}}\frac{\mu}{T},
\qquad
m
=
2\pi\alpha'\frac{\sqrt{2}}{u_0}M_{\rm q}
=\frac{2}{\sqrt{\lambda}}\frac{M_{\rm q}}{T}.
\label{mMq}
\end{equation}
As we will estimate below\footnote{
For the dimension three operator,
we neglect contributions from squarks in the hypermultiplet.},
the remaining constants $\widetilde{d}$ and $c$ would be proportional
to the quark number density $n_{\rm q}$ and the quark
condensate $\langle\bar{\psi}\psi\rangle$, respectively.

It is easy to observe that the free part of a linearized equation of motion
for the scalar field $\chi(\rho)$ (\ref{eom_chi}), i.e.\
$
0=\del_\rho(\rho^5f\tilde{f}\dot\chi)+3\rho^3f\tilde{f}\chi
$
is equivalent to the Klein-Gordon equation in the
AdS-Schwarzschild background (\ref{D3geom})
with mass $m^2L^2=-3$.
By using the scaling argument for the scalar field
(\ref{exponents})\footnote{
Strictly speaking, we should move on to the extremal case
where we can define the boundary CFT.
},
it is confirmed that the scaling dimension of the corresponding operator
in the boundary theory may be three.
The similar argument could be confirmed for the massless gauge field.
The asymptotic solution of $\widetilde{A}_t(\rho)$ in (\ref{aspt_a})
supports the scaling
(\ref{massless_vector}) for a conserved current in 4D boundary.

In order to solve the nonlinear equations of motion (\ref{eom_ta})
and (\ref{eom_chi}), we need to use numerical methods.
Here we restrict to the black hole embedding in which
the D7-branes touch the horizon
since this might be thermodynamically favored configuration
in the system with finite density.
We impose boundary conditions at the horizon
as $\dot{\chi}(1)=0$, $\widetilde{A}_t(1)=0$ to remove singularities
and $\chi(1)=\chi_0$.
We fix $m$ and $\widetilde{\mu}$ which depend on $\chi_0$ and
$\widetilde{d}$ by matching the numerical solutions
with the asymptotic forms at the boundary.

We now consider the on-shell action which is related to
the partition function $Z$ of the field theory in the context of
AdS/CFT correspondence.
However the on-shell action contains UV divergences.
It is well-known that one can prepare local boundary counter terms
for probe D-branes in AdS spacetime by applying the holographic
renormalization~\cite{karch_06}.
In the D3/D7 system, taking the asymptotic solution (\ref{aspt_a})
into account, the relevant boundary counter terms take the
form~\cite{mmt_06, mmt_07},
\begin{equation}
S_{\rm ct}
=\frac{\lambda N_cN_fT^3}{32}V_3
\bigg\{
-\frac{1}{4}
\Big((\rho_{\rm max}^2-m^2)^2-4mc\Big)
\bigg\},
\end{equation}
where $\rho_{\rm max}$ is the cut-off for UV divergences
which may go to infinity after precise calculations.
It should be noticed that there exist finite contributions
in the counter terms.
Together with these counter terms,
we could obtain the regularized action
\begin{eqnarray}
S_{\rm D7 \ reg}
&=&
S_{\rm D7} + S_{\rm ct}
\nonumber
\\
&=&
\frac{\lambda N_cN_fT^3}{32}V_3
\Bigg\{
\int_1^{\infty}\!\!
\dd\rho
\bigg(\rho^3\tilde{f}(1-\chi^2)
\sqrt{f^2(1-\chi^2+\rho^2\dot{\chi}^2)
-2\tilde{f}(1-\chi^2)\dot{\widetilde{A}_t^2}}
\nonumber
\\
&&
\hspace*{41mm}
-\rho^3+m^2\rho\bigg)
-\frac{1}{4}\Big((m^2-1)^2-4mc\Big)
\Bigg\}.
\label{reg_action}
\end{eqnarray}
Since we are interested in the black hole embedding,
the integration supports from the horizon
to the AdS boundary.
Evaluating the on-shell action for (\ref{reg_action})
which is reduced to boundary values through the equation of motion,
we could observe that the quark condensate is proportional to
the integration constant $c$,
\begin{equation}
\langle\bar{\psi}\psi\rangle
=-\frac{T}{V_3}\frac{\del}{\del M_{\rm q}}\log Z
=-\frac{1}{8}\sqrt{\lambda}N_{\rm c}N_{\rm f}T^3c,
\label{qc}
\end{equation}
where we have used the asymptotic solutions (\ref{aspt_a})
and the relation (\ref{mMq}).
Since
we identify the grand potential as
$\Omega=-T\log Z$,
the quark number density can be also calculated through
the on-shell evaluation,
\begin{equation}
n_{\rm q}=-\frac{1}{V_3}\frac{\del\Omega}{\del\mu}
=\frac{1}{4}\sqrt{\frac{\lambda}{2}}
N_{\rm c}N_{\rm f}T^3\widetilde{d}.\label{D3D7nq}
\end{equation}
%

\subsection{Parton energy loss and quarkonium dissociation}

Now, we discuss a few probes of the QGP in this D3/D7 model;
Some general discussion on the QGP can
be found
in~\cite{Rischke:2003mt, McLerran:2003yx,Shuryak:2009zza,Satz:2011wf}.
We will consider the energy loss of partons and dissociation
of the (heavy) quarkonium.

\begin{figure}[htbb]
\centering
\mbox{%
\subfigure{%
\includegraphics[width=4.5cm]{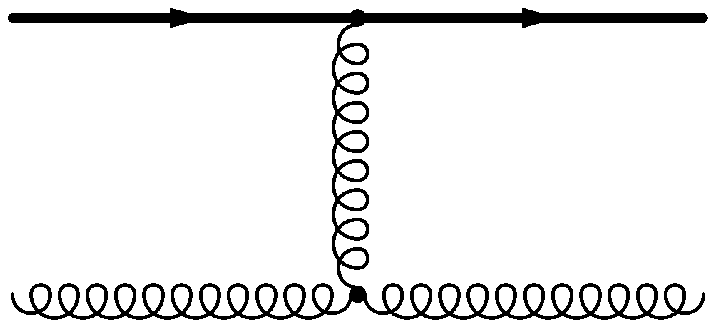}
}\quad
\subfigure{%
\includegraphics[width=4.5cm]{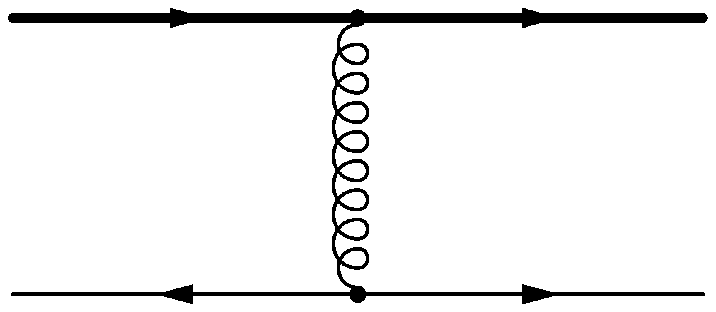}
}}
\caption{Typical diagrams for the  collisional energy loss.
Thick solid line is for energetic quarks traveling through medium, interacting with
 thermal quarks (thin solid lines) and gluons (helical lines). }\label{fig:cLoss}
\end{figure}

\begin{figure}[!hbtp]
\centering
\includegraphics[width=4.5cm]{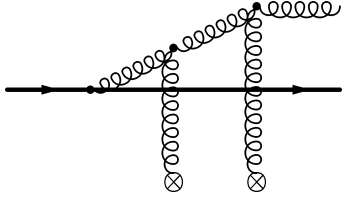}
\caption{A typical process for the  radiative energy loss.
Thick solid line denotes fast moving quarks through the QGP.
$\bigotimes$ is for colored static scattering sources
in the medium. }\label{fig:rLoss}
\end{figure}

The energy loss of partons is a useful probe of the QGP.
Though the energy loss is not a direct experimental observable,
it is  manifested in jet quenching, a signature of the QGP.
Jet quenching in AdS/CFT was initially discussed
in~\cite{Sin:2004yx,Liu:2006ug}.
Traveling through the dense medium, energetic partons will lose their energy.
There are two primary sources of the energy loss: collisional and radiative.
The collisional energy loss is due to the scattering
of the energetic partons with thermal quarks and
gluons in the QGP, see Fig.\ref{fig:cLoss} for a sample process,
while the radiative one is attributed to the
Bremsstrahlung during the interactions with the medium,
see Fig.\ref{fig:rLoss} for a typical diagram.
We refer to \cite{Peigne:2008wu,d'Enterria:2009am} for a review
on the physics of jet quenching.

Now we demonstrate how one can calculate the parton energy
loss in the D3/D7 model at finite temperature discussed above
following \cite{Herzog:2006gh,Gubser:2006bz}.
The energy loss per unit length of a quark moving through
the medium with velocity $v$ obtained in \cite{Herzog:2006gh} is given
by
\begin{equation}
	\frac{\dd E}{\dd x}=\frac{1}{v}\frac{\dd E}{\dd t}
=-\frac{\pi}{2}\sqrt{\lambda}T^2\frac{v}{\sqrt{1-v^2}}.
\end{equation}
Below, we will show how this result comes out.

\begin{figure}[!hbtp]
\centering
\includegraphics[width=5.5cm]{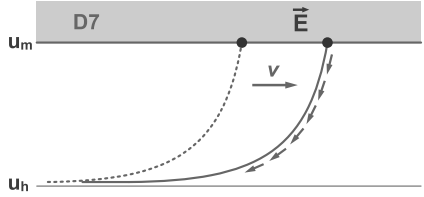}
\caption{A classical string configuration.
The downwards arrow indicates the energy (momentum)
flow along the string. }\label{fig:hELoss}
\end{figure}

$\mathcal{N}=4$ $SU(N_{\rm c})$ super Yang-Mills theory
with finite temperature has its gravity dual description
(\ref{D3geom_0}).
The probe D7-brane wraps on $S^3$ of $S^5$ and
fills all of the asymptotically AdS$_5$ down to a minimum radial value $u_m$.
To calculate the energy loss, we consider a classical open
string configuration whose one endpoint attached on the D7-brane
and another one stretches to the horizon of black hole, see Fig.\ref{fig:hELoss}.
This string corresponds to a quark whose mass  is related
to the minimum radius $u_m$.
The dynamics of this string can be described by the
Nambu-Goto action (\ref{nambu_goto})
\begin{equation}
S_{\textrm {NG}}
=
-\frac{1}{2\pi\alpha^\prime}
\!\int\!\dd\tau\dd\sigma\;
\sqrt{(\dot{X}\cdot X^\prime)^2-(\dot{X})^2(X^\prime)^2},
\end{equation}
where the dot and the prime stand for a derivative
with respect to $\tau$ and $\sigma$, respectively.
Without loss of generality, we can assume that the string
lives on three-dimensional slice of an asymptotically AdS$_5$.
The three-dimensional slice is described by the coordinates $(t,u,x)$,
where $x$ is one of the transverse coordinates $x^i$.
With the choice of a static gauge $t=\tau$ and $u=\sigma$,
the shape of the string is described by a single variable, $x=x(t,u)$.
Then, the determinant of the induced metric $g_{mn}$ becomes
\begin{equation}
g\equiv{\textrm {det}}g_{mn}
=-\Big(1-\frac{\dot{x}^2}{f(u)}+\frac{u^4 f(u)}{L^4}{x'}^2\Big).
 \end{equation}
The equation of motion for the string configuration reads
\begin{equation}
0=\del_u\bigg(\frac{u^4 f(u) x'}{\sqrt{-g}}\bigg)
-\frac{L^4}{f(u)}\del_t\bigg(\frac{\dot{x}}{\sqrt{-g}}\bigg).
\label{EOML}
\end{equation}
For simplicity, we will consider the string moving with constant
velocity $v$, which corresponds to quarks moving under a constant
electric field in gauge theory side.

With the ansatz $x(t,u)=x(u)+vt$ and the condition that
the action should be real, one can obtain the solution of (\ref{EOML})
as
\begin{equation}
x(t,u)=x_0+\frac{L^2 v}{2 u_{\rm H}}
\left(\frac{\pi}{2}-\tan^{-1}\left(\frac{u}{u_{\rm H}}\right)
-\coth^{-1}\left(\frac{u}{u_{\rm H}}\right)\right)+vt.
 \end{equation}
With this solution one can easily calculate the energy (or momentum)
flow along the string which is given by the canonical momentum densities
\begin{equation}
\pi^\sigma_M\equiv
\frac{\partial\mathcal{L}}{\partial {X'}^M}
=-\frac{G_{MN}}{2\pi\alpha^\prime}
\frac{(\dot{X}\cdot X^\prime)\dot{X}^N-\dot{X}^2{X'}^N}{\sqrt{-g}},
 \end{equation}
where $\mathcal{L}$ is the Lagrangian of the Nambu-Goto action.
If we consider $M=t$ case, we obtain the energy flow
along the string to the black hole horizon,
or minus of the energy loss of the quark,
\begin{equation}	
 \frac{\dd E}{\dd t}
=\frac{\pi}{2}\sqrt{\lambda}\,T^2\frac{v^2}{\sqrt{1-v^2}}.
\end{equation}

Based on the idea that off-shell gluons in a thermal medium
can be viewed as a string with its both endpoints passing
through the horizon of an AdS black hole,
the energy loss of an energetic gluon in a thermal plasma
was studied in \cite{Gubser:2008as}.
An interesting idea of heavy quark energy loss through the Cherenkov
radiation of mesons by quarks was explored in the gauge/gravity
duality \cite{CasalderreySolana:2009ch}.
For more on energy loss including recent developments,
we refer to \cite{Chernicoff:2012gu,Arnold:2012uc} and references therein.

Now, we move onto another probe of the QGP.
The thermal dissociation of the (heavy) quark bound states,
which might be due to Debye screening in the QGP,
will convey signals of onset of new state of matter (QGP).
There are a few ways to study heavy quarkonium properties at high temperature.
 An obvious and simple way is to consider linearized equations
of motion for fluctuations of the D-brane embeddings.
 In this case, however, one has to be cautious about the fact
that the bound states in D$p$/D$q$ configurations are deeply bound,
while heavy quarkonia are shallow bound states \cite{mmt_07}.
Holographic meson melting through the quasinormal modes
of D7-brane fluctuations is studied in \cite{Hoyos:2006gb}.
One can also solve Schr\"odinger  equation with temperature-dependent
interquark potentials derived from a holographic model \cite{Hou:2009qj}
or calculate the holographic finite-temperature spectral function
of (heavy) quarkonia \cite{Myers:2007we,Erdmenger:2007ja}.

\section{D4/D8/$\overline{\rm\bf D8}$ model}

One of the major goals in the gauge/gravity duality is
to find holographic models which capture features of QCD,
in other words, to find a dual geometry for QCD.
Witten proposed a construction of the holographic dual of 4D pure
$SU(N_{\rm c})$ Yang-Mills theory~\cite{Witten:1998zw}.
The confinement/deconfinement could be discussed in geometrical manner.
In order to introduce the flavors, D8/$\overline{\mbox D8}$-branes
would be considered in this section.
In addition to the confinement/deconfinement,
this D4/D8/$\overline{\mbox{D8}}$ model so-called
Sakai-Sugimoto model~\cite{Sakai:2004cn} nicely
describes the nonabelian chiral symmetries.
In this section, we briefly sketch the D4-brane background and
D8/$\overline{\mbox{D8}}$ flavor branes.
We yield precise calculations of some physical quantities to
the D4/D6 model in the next section,
since technical details are given in the parallel way.

\subsection{D4-brane background}

We start by considering $N_{\rm c}$ D4-branes in Type IIA string theory.
The $N_{\rm c}$ D4-branes provides 5D $SU(N_{\rm c})$ super Yang-Mills
theory in the world volume with coupling constant
\begin{equation}
g_{\rm YM \,5}^2=(2\pi)^2\alpha'^{\frac{1}{2}}g_{\rm s}.
\label{gauge_c_5}
\end{equation}
Since the 5D coupling constant $g_{\rm YM \, 5}$ is dimensionful,
at high energies, we need to ask UV completion to a description
in 11D M-theory~\cite{imsy}.
However, this is not relevant in our discussion, since we will be interested
in low energies below $1/g_{\rm YM \, 5}^2$.

In order to discuss the gravity dual,
we here consider the black 4-brane solutions given in (\ref{sp}) with $p=4$
and renaming $(r, r_{\rm H}, L)\to (U, U_{\rm T}, \widetilde{R})$,
\begin{subequations}
\begin{eqnarray}
\dd s^2
&=&
\frac{1}{\sqrt{\displaystyle 1+\frac{\widetilde{R}^{3}}{U^{3}}}}
\bigg\{
-\Big(1-\frac{U_{\rm T}^{3}}{U^{3}}\Big)\dd t^2
+\sum_{i=1}^3\dd x^{i\, 2}
+\dd\tau^2
\bigg\}
+\sqrt{1+\frac{\widetilde{R}^{3}}{U^{3}}}
\bigg\{\frac{\dd U^2}{\Big(\displaystyle1
-\frac{U_{\rm T}^{3}}{U^{3}}\Big)}
+U^2\dd\Omega^2_{4}
\bigg\},
\label{non_extremal_d4_metric}
\\
\eee^{\widetilde{\Phi}}
&=&
\Big(1+\frac{\widetilde{R}^3}{U^3}\Big)^{-1/4},
\qquad
F_4
=\dd C_3=
\frac{8\pi^3N_{\rm c}\alpha'^{\frac{3}{2}}}{V_{S^4}}\epsilon_4,
\label{non_extremal_d4_dilaton}
\end{eqnarray}
\end{subequations}
where $\epsilon_4
=\sqrt{g_{S^4}}\dd\theta^1\dd\theta^2\dd\theta^3\dd\theta^4$ is
the invariant volume form of $S^4$ with the coordinates $\theta^i$.
The extra direction of the 5D world volume is labeled by $\tau$.
The identification of the charges (\ref{identification}) leads a
relation
\begin{equation}
\widetilde{R}^3(\widetilde{R}^3+U_{\rm T}^3)=R^6,
\qquad \mbox{with} \qquad
R^3=\pi N_{\rm c}g_{\rm s}\alpha'^{\frac{3}{2}}.
\end{equation}
The Hawking temperature is given by
\begin{equation}
T=\frac{3}{4\pi}\sqrt{\frac{U_{\rm T}}{\widetilde{R}^3+U_{\rm T}^3}}.
\label{ht_5}
\end{equation}
As we did in the D3-brane background, we take the decoupling limit
$\alpha'\to0$ with keeping the field theory parameters,
the gauge coupling constant
(\ref{gauge_c_5}), the temperature (\ref{ht_5}), and the energy
$U/\alpha'$ finite.
With the scaling behavior of the parameters;
$g_{\rm s}\propto\alpha'^{-\frac{1}{2}}$,
$R\propto\alpha'^{\frac{1}{3}}$,
$\widetilde{R}\propto\alpha'^{\frac{1}{3}}$,
$U_{\rm T}\propto\alpha'$, and $U\propto \alpha'$,
we obtain the near horizon geometry as the decoupling limit,
\begin{subequations}
\begin{eqnarray}
\dd s^2
&=&
\left(\frac{U}{R}\right)^{3/2}
\!\!\Big\{-f(U)\dd t^2+\sum_{i=1}^3\dd x^{i\, 2}+\dd\tau^2\Big\}
+\left(\frac{R}{U}\right)^{3/2}
\!\!\Big\{\frac{\dd U^2}{f(U)}+U^2\dd\Omega_4^2\Big\},
\label{d4ft}
\\
\eee^{\widetilde{\Phi}}
&=&
\Big(\frac{U}{R}\Big)^{3/4},
\qquad
F_4=\dd C_3
=\frac{8\pi^3N_{\rm c}\alpha'^{\frac{3}{2}}}{V_{S^4}}\epsilon_4,
\label{d4_dilaton}
\end{eqnarray}
\end{subequations}
with
Hawking temperature
\begin{equation}
T=\frac{3}{4\pi}\sqrt{\frac{U_{\rm T}}{R^3}},
\end{equation}
where $f(U)=1-(U_{\rm T}/U)^3$.
It should be noted that the effective string coupling becomes finite
\begin{equation}
g_{\rm s}^{\rm  eff}(U)=g_{\rm s}\eee^{\tilde{\Phi}(U)}
=\Big(\frac{g_{\rm YM\, 5}}{2\pi}\Big)^{3/2}
\frac{(U/\alpha')^{3/4}}{(\pi N_{\rm c})^{1/4}}.
\label{string_coupling_d4}
\end{equation}
We could obtain a  relation in which the supergravity approximation
is valid $g_{\rm s}^{\rm eff}(U)\ll1$.
In the case of $U_{\rm T}=0$ in the background (\ref{d4ft}),
the naked singularity appears at $U=0$
where the scalar curvature diverges.
Therefore, in this gravity dual we could not discuss zero temperature
gauge theory.

However,
we can find another nontrivial solution with the same
asymptotics with (\ref{d4ft}) which is dual to the boundary field theory
at zero/low temperature.
This is obtained by applying the double Wick rotations i.e.\
$t\to i\tau$, $\tau\to i t$  to the metric
(\ref{non_extremal_d4_metric}).
Taking the decoupling limit, its near horizon geometry
is given by~\cite{Witten:1998zw},
\begin{equation}
\dd s^2
=
\left(\frac{U}{R}\right)^{3/2}
\!\!\Big\{-\dd t^2+\sum_{i=1}^3\dd x^{i\, 2}+f(U)\dd\tau^2\Big\}
+\left(\frac{R}{U}\right)^{3/2}
\!\!\Big\{\frac{\dd U^2}{f(U)}+U^2\dd\Omega_4^2\Big\},
\quad
\mbox{with}
\quad
f(U)=1-\left(\frac{U_{\rm KK}}{U}\right)^{3}.
\label{d4}
\end{equation}
Except for renaming $U_{\rm T}$ to $U_{\rm KK}$ which is a free
parameter at this moment,
the other parameters are the same as before.
The dilaton $\widetilde{\Phi}(x)$ and the RR field $F_4(x)$ are identical to
(\ref{d4_dilaton}).

We now compactify one spatial dimension $\tau$
to $S^1$ with radius $1/M_{\rm KK}$.
Then, 4D Yang-Mills coupling constant $g_{\rm YM}$ may
be related to the 5D coupling,
\begin{equation}
g_{\rm YM}^2=g_{\rm YM\, 5}^2\frac{M_{\rm KK}}{2\pi}.
\label{gauge_c_4}
\end{equation}
We impose the anti-periodic boundary conditions
for the world volume fermions on this circle,
so that the supersymmetry is explicitly broken.
Not only fermions are massive in tree-level, but also scalar fields
on the branes acquire masses
through fermion one-loop.
Therefore, in energies below $M_{\rm KK}$,
we could expect to have the pure 4D $SU(N_{\rm c})$ Yang-Mills
theory with the coupling constant (\ref{gauge_c_4}).

There exists a singular point at $U=U_{\rm KK}$ in the metric
(\ref{d4}).
We could apply what we did in the derivation of the Hawking temperature
(c.f.\ (\ref{htd_0}) and Fig.\ref{tau_r}).
Since the period of $\tau$ is $2\pi/M_{\rm KK}$,
a conical singularity at $U=U_{\rm KK}$ may be removed
by imposing the relation
\begin{equation}
M_{\rm KK}^2=\frac{9U_{\rm KK}}{4R^3}.
\end{equation}
This makes the $(U, \tau)$ submanifold a cigar-like form
with a tip at $U=U_{\rm KK}$
where the $\tau$-circle shrinks to zero.
The radial direction $U$ smoothly terminates at $U=U_{\rm KK}$
and there exist no singularities.
The bulk string theory parameters $U_{\rm KK}$, $g_{\rm s}$, and
$R$ are related to those of gauge theory $M_{\rm KK}$, $g_{\rm YM}$,
and $N_{\rm c}$ through
\begin{equation}
U_{\rm KK}=\frac{2\lambda\alpha'M_{\rm KK}}{9},
\qquad
g_{\rm s}=\frac{\lambda}{2\pi N_{\rm c}\sqrt{\alpha'}M_{\rm KK}},
\quad
R^3=\frac{\lambda\alpha'}{2M_{\rm KK}},
\quad
\mbox{with}
\quad
\lambda=g_{\rm YM}^2N_{\rm c}.
\label{paraMatchin}
\end{equation}

Let us consider the validity regime of supergravity
approximation.
We impose that the spacetime curvature should be much smaller
than the string length scale.
Since the maximum value of the curvature of the background (\ref{d4})
can be directly calculated as the order $(R^3 U_{\rm KK})^{-1/2}$,
the condition should be
\begin{equation}
(R^3U_{\rm KK})^\frac{1}{2}\gg \alpha',
\quad \mbox{i.e.} \quad
\lambda\gg 1.
\label{strong}
\end{equation}
As in the case of D3-brane, the gravity dual may be used  for
4D gauge theory with large 't Hooft coupling.
By using the effective string coupling (\ref{string_coupling_d4}),
we could estimate the critical value of $U$ where the dilaton is order
one,
\begin{equation}
U_{\rm cutoff}=\frac{(\pi N_{\rm c})^{\frac{1}{3}}M_{\rm KK}\alpha'}
{(2\pi)^2g_{\rm YM}^2}.
\end{equation}
We simply adapt the large $N_{\rm c}$ limit in the context of this paper.

It should be noted that
the Kaluza-Klein modes on the D4-brane do not
decouple within the supergravity approximation~\cite{Witten:1998zw}.
The masses of glueballs are of the same order as $M_{\rm KK}$,
since $M_{\rm KK}$ is the only parameter in IR.
Nevertheless, the qualitative features of the glueball spectrum agree with
lattice calculations~\cite{glueball}.

\subsection{Confinement/deconfinement phase transition}

Having the dimensionful parameter $U_{\rm KK}$ or
$M_{\rm KK}$,
some interesting phenomena could be discussed.
Indeed, we could describe the confinement/deconfinement phase transition
in terms of the geometries which is referred as Hawking-Page
transition~\cite{hawking_page}.
At finite temperature, there are two gravity
backgrounds i.e.\ Euclidean versions of (\ref{d4ft}) and (\ref{d4})
in which each of them has two circles.
Comparing the free energies defined by the thermal partition function,
it has been shown that
there is a first order phase transition between two of them at
$T=M_{\rm KK}/(2\pi)$,
where these two become essentially the same due to $U_{\rm KK}=U_{\rm T}$.
At lower temperatures $T< M_{\rm KK}/(2\pi)$,
the Euclidean version of (\ref{d4}) is energetically favored, while at
higher temperatures $T>M_{\rm KK}/(2\pi)$ the background
(\ref{d4ft}) is favored~\cite{Witten:1998zw,asy_07}.
This can be understood intuitively by comparing the two periods of
circles i.e.\ $1/T$ for $t$ and $2\pi/M_{\rm KK}$ for $\tau$.
The metric having the smaller circle which can shrink to zero is
always chosen.
The computation for the free energies of the two phases shows
that their $N_{\rm c}$-dependence is $N_{\rm c}^0$ and $N_{\rm c}^2$ for
the low temperature and the high temperature phases, respectively.
This means that in the low (high) temperature phase, the gauge degrees
of freedom are confined (deconfined)~\cite{Witten:1998zw,asy_07}.
In the previous D3-brane background, due to the conformal nature,
we are always in the deconfined phase observed in the results
(\ref{energy_d3t}) and (\ref{entropy_d3t}).
An interesting recent development regarding phase transition is reported
in \cite{Morita}.

\subsection{D8/$\overline{\mbox{D8}}$ flavor branes}

We here briefly discuss the Sakai-Sugimoto model.
We take D8 and $\overline{\mbox{D}8}$ branes as the flavor branes
and put $N_{\rm f}$ pairs of them with
the intersection in  Table \ref{braneprofile01}.
 \begin{table}[bthb]
  	\begin{center}
  	\begin{tabular}{|>{\centering}p{35pt}*{9}{|>{\centering}p{13pt}}|p{13pt}|}
		\hline
		&$t$&$x^1$&$x^2$&$x^3$&$(\tau)$&$U$
&$\theta^1$&$\theta^2$&$\theta^3$&~$\theta^4$
\\
		\hline
		D4&$\bullet$&$\bullet$&$\bullet$&$\bullet$&$\bullet$&&&&&
\\
		\hline
		D8/$\overline{\mbox
	 D8}$&$\bullet$&$\bullet$&$\bullet$&$\bullet$&&$\bullet$&$\bullet$&$\bullet$
&$\bullet$&$\bullet$
\\
		\hline
  	\end{tabular}
  	\caption{The brane configurations: the background D4- and the
probe D8/$\overline{\mbox{D8}}$-branes}
	\label{braneprofile01}
  	\end{center}
 \end{table}
$N_{\rm f}$ D8-branes and $N_{\rm f}$ $\overline{{\rm D}8}$-branes
are placed at the positions $\tau=l/2$ and $\tau=-l/2$
with a separation distance $l$, respectively,
These positions are generally functions of the world volume coordinates
and determined by the equation of motion of the effective action of
these branes.

The low energy effective theory is described by the lightest modes
of open strings stretched between (D4, D4), (D4, D8)
and (D4, $\overline{{\rm D}8}$).
As we discussed in the D3/D7 model, the gauge theories in the
D8 and $\overline{{\rm D}8}$-brane world volumes which come from
open strings in (D8, D8) and
($\overline{{\rm D}8}$,  $\overline{{\rm D}8}$) are decoupled,
so that the symmetries in the gauge theory side are reduced to
the global symmetries $U(N_{\rm f})_{\rm L}$ and $U(N_{\rm f})_{\rm R}$.
Open strings in (D4, D4) give pure $SU(N_{\rm c})$ Yang-Mills as before.
Those in (D4, D8) and (D4, $\overline{{\rm D}8}$)
 which are in the case $n=6$ in the (\ref{open_r}) and
(\ref{open_ns}) provide massless fermion in R sector.
The NS sector only gives massive modes.
Since D8 and $\overline{{\rm D}8}$
branes, whose relative angle is $\pi$,  have the opposite ways for GSO projection,
those strings describe left-handed
and right-handed chiral fermions which transform in the fundamental
representation of both of $SU(N_{\rm c})$ color group and
$U(N_{\rm f})_{\rm L}$ and $U(N_{\rm f})_{\rm R}$ flavor group,
respectively.
As in the configuration in Table \ref{braneprofile01},
there are no common transverse
directions where strings with finite length can live and give the mass, and so
these fermions are massless.
It should be mentioned that open strings in (D8, $\overline{{\rm D}8}$)
could produce the tachyon field.
The mass of these string modes might be estimated from (\ref{open_ns}) as
\begin{equation}
m^2=-\frac{1}{2\alpha'}+\Big(\frac{\rm distance}{2\pi\alpha'}\Big)^2\, .
\label{tachyon_mass}
\end{equation}
If the distance between D8 and $\overline{{\rm D}8}$ is large enough,
the system could be stable.

In the Sakai-Sugimoto model, we can view the spontaneous
breaking of nonabelian chiral symmetry with a new insight.
At a large radial position which corresponds to a high energy
regime, D8 and $\overline{{\rm D}8}$ branes are well separated and
one can see the full $U(N_{\rm f})_{\rm L}\times U(N_{\rm f})_{\rm R}$
chiral symmetry.
However, in the cigar-like background (\ref{d4}), as the energies are
reduced i.e.\ going into the bulk,
D8 and $\overline{{\rm D}8}$ branes have no place to end and consequently
join together into the $N_{\rm f}$ continuous D8 branes at some point in
IR while keeping an asymptotic separation $l$ between D8 and
$\overline{{\rm D}8}$ branes fixed at large $U$.
Therefore D8-branes are embedded in the bulk by forming  ``U-shape''.
Here only the diagonal $U(N_{\rm f})$ symmetry remains unbroken.
This is the geometrical realization of the spontaneous chiral
symmetry breaking.

Let us consider the finite temperature case where we can discuss
the confinement/deconfinement transition.
At low temperature, the relevant geometry is given by the Euclidean
version of (\ref{d4}) with the thermal circle for $t$.
As we discussed above, the chiral symmetry is broken
since in this regime the D8-branes make the U-shape.
At high temperature, the corresponding background is (\ref{d4ft}).
This geometry is not the cigar type in the $(\tau, U)$-plane
which enforces D8-branes to form U-shape.
In this case there are two solutions:
the U-shape as before and a ``parallel''.
In the case of the parallel, the D8 and  $\overline{{\rm D}8}$ branes end on
the black hole horizon separately and consequently there is a full
$U(N_{\rm f})_{\rm L}\times U(N_{\rm f})_{\rm R}$ chiral symmetry.
Comparing the free energies of these configurations,
it can be shown that in the low temperature regime the U-shaped embedding
is favored, which implies that the chiral symmetry is broken.
At high temperature, the parallel embedding is preferred, and therefore
the broken symmetry is restored~\cite{asy_07}.
\begin{figure}[htbb]
\begin{minipage}{1.0\textwidth}
\begin{center}
\includegraphics[clip, width=9cm]{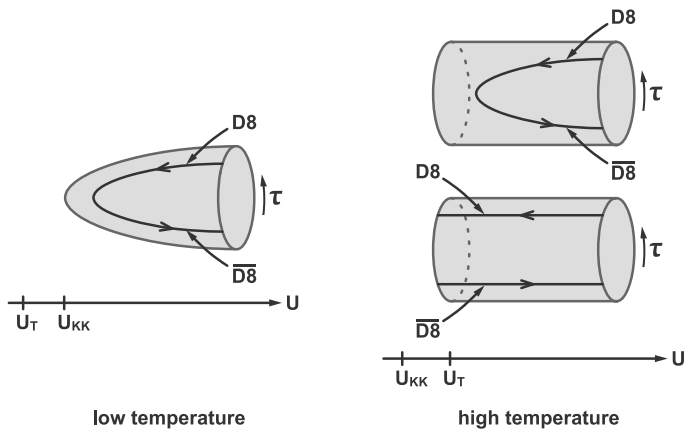}
\caption{D8/$\overline{\mbox{D8}}$ configuration on confined background (left) and
deconfined background (right)}
\label{d4d8d8bar}
\end{center}
\end{minipage}
\end{figure}

The baryon chemical potential and density can be also naturally
introduced in the model,
which endows the phase diagram with rich structures  in the deconfined
phase~\cite{density}.

Motivated by the mass relation (\ref{tachyon_mass}),
it has been shown that (D8, $\overline{\mbox{D8}}$)
bifundamental scalar mode could participate into the
model in the name of ``tachyon'' and give the quark mass
and the chiral condensate~\cite{d4d8_tachyon}.
Moreover, it has been suggested that the tachyon condensate provides
the mechanism of chiral symmetry breaking.
Alternative approaches to introduce the quark mass into the model have
been considered in~\cite{d4d8_mass}.

\subsection{Holographic baryons and nuclear force \label{sec:nnP}}

As the applications of the Sakai-Sugimoto model in low energy QCD phenomenology,
we summarize a few works on holographic baryons and two-nucleon potentials.

\vskip 0.3cm
\noindent  \textbf{Holographic baryons}\\
Roughly speaking, there are three methods
to describe the property of baryons in the Sakai-Sugimoto model.
We list some of results from each approach.

An immediate way is to start from a 4D effective action of mesons
derived from the Sakai-Sugimoto model and do the conventional
Skyrmion analysis to obtain the hedgehog soliton solution.
In \cite{Nawa:2006gv}, with the 4D effective action of pions
and $\rho$-mesons obtained from the Sakai-Sugimoto model,
some interesting properties of baryons were calculated.
In Table \ref{T1}, we show the table II in \cite{Nawa:2006gv}.
Here $M_{\textrm {HH}}$ is the ground state Skyrmion mass, and
 $\sqrt{\langle r^2 \rangle}$ is the root-mean-square radius of the Skyrmion.
 \begin{table}[h]
 \begin{center}
 \caption{Some results  based on the 4D effective action
derived from the Sakai-Sugimoto model  and from a conventional
Skyrmion approach as in Table II of \cite{Nawa:2006gv}.}
 \vskip 0.3cm
 \begin{tabular}{>{\centering}p{0.20\textwidth}ccc}
		\hline
	&~~~Results ~from~ \cite{Nawa:2006gv}
~~~&~~~~Skyrmion~approach~~~~&~~~~~Experiment~~~~~\\
	\hline
	$f_\pi$ & 92.4 MeV (input) & 64.5 MeV & 92.4 MeV \\
	$m_\rho$ & 776.0 MeV (input) & - & 776.0 MeV \\
	$e$ & 7.32 & 5.44 & - \\
	$E_{\textrm {ANW}}\equiv\frac{f_\pi}{2e}$ & 6.32 MeV & 5.93 MeV & - \\
	$r_{\textrm {ANW}}\equiv\frac{1}{e f_\pi}$ & 0.29 fm & 0.56 fm & - \\
	$M_{\textrm {HH}}$ & 834.0 MeV & 864.3 MeV & - \\
	$\sqrt{\langle r^2 \rangle}$ & 0.37 fm & 0.80 fm & 0.60 $\sim$ 0.80 fm \\
	$M_N$ & - & 938.9 MeV (input) & 938.9 MeV \\
	$M_\Delta-M_N$ & - & 293.1 MeV (input) & 293.1 MeV \\
	\hline
	 \end{tabular}\label{T1}
 \end{center}
 \end{table}

Secondly, based on the fact that the size of
the instanton soliton, $\sim 1/(M_{\rm KK}\sqrt{\lambda})$,
is much smaller than the typical length scale of
the effective theory derived from the Sakai-Sugimoto model,
$\sim 1/M_{\rm KK}$,
one can build up a 5D effective action
of baryons \cite{Hong:2007kx}.
We will revisit this model soon.

Finally, one can obtain the instanton soliton
in 5D which is dual to the 4D Skyrmion \cite{Hata:2007mb}.

In Table \ref{T2}, we quote
a table in \cite{Hashimoto:2008zw}
in which static properties of neutrons and protons
are extensively studied based on the 5D instanton soliton approach.

 \begin{table}[h]
 \begin{center}
 \caption{Some results  based on the 5D instanton soliton
approach in the Sakai-Sugimoto model
and from a conventional Skyrmion approach as in \cite{Hashimoto:2008zw}.}
 \vskip 0.3cm
 \begin{tabular}{>{\centering}p{0.15\textwidth}ccc}
	\hline
	&~instanton~soliton~approach~
&~~~~Skyrmion~~~~&~~~~~experiment~~~~~\\
	\hline
	${\langle r^2 \rangle}_{I=0}^{1/2}$ & 0.742 fm & 0.59 fm & 0.806 fm \\
	${\langle r^2 \rangle}_{M,I=0}^{1/2}$ & 0.742 fm & 0.92 fm & 0.814 fm \\
	${\langle r^2 \rangle}_{E,p}$ & (0.742 fm)$^2$ & $\infty$ & (0.875 fm)$^2$ \\
	${\langle r^2 \rangle}_{E,n}$ & 0 & $-\infty$ & $-0.116$ fm$^2$ \\
	${\langle r^2 \rangle}_{M,p}$ & (0.742 fm)$^2$ & $\infty$ & (0.855 fm)$^2$ \\
	${\langle r^2 \rangle}_{M,n}$ & (0.742 fm)$^2$ & $\infty$ & (0.873 fm)$^2$ \\
	${\langle r^2 \rangle}_A^{1/2}$ & 0.537 fm & - & 0.674 fm \\
	$\mu_p$ & 2.18 & 1.87 & 2.79 \\
	$\mu_n$ & $-1.34$ & $-1.31$ & $-1.91$ \\
	$\left|{\mu_p}/{\mu_n}\right|$ & 1.63 & 1.43 & 1.46 \\
	$g_A$ & 0.734 & 0.61 & 1.27 \\
	$g_{\pi NN}$ & 7.46 & 8.9 & 13.2 \\
	$g_{\rho NN}$ & 5.80 & - & 4.2 $\sim$ 6.5 \\
	\hline
 \end{tabular}\label{T2}
 \end{center}
 \end{table}

For some recent discussion on baryon properties
in the Sakai-Sugimoto model,
we refer to \cite{Rho:2009ym}.

\vskip 0.3cm
\noindent\textbf{Nucleon-nucleon potentials from holography}\\
A basic but pretty much essential problem
in nuclear physics is to understand the nuclear force.
So far innumerable attempts have been made
to construct the nucleon-nucleon potential based
on one-boson exchange picture, chiral effective theory, etc.
See \cite{Epelbaum:2005pn}
for some reviews on the nuclear forces.

In this section,
we briefly review the work of \cite{Kim:2009sr}
in which the one-boson exchange model is used to calculate the nuclear force.
See \cite{Kim:2008iy,Hashimoto:2009ys}
for nucleon-nucleon potentials based on the five-dimensional
instanton soliton approach,
where  a trial soliton configuration that is supposed to describe a
two-nucleon system is used to obtain the potential.
We start with the four dimensional nucleon-meson
action \cite{Hong:2007kx,Kim:2009sr}
obtained from the Sakai-Sugimoto model
\begin{equation}
\label{Lag:pointB}
\int\!\dd^4x{\cal L}_4
=
\int\!\dd^4x
\big(-i\bar {\cal N}
\gamma^\mu\partial_\mu {\cal N}
-im_{\cal N}\bar {\cal N}{\cal N}
+{\cal L}_{\rm vector} +{\cal L}_{\rm axial}\big).
 \end{equation}
${\cal L}_{\rm vector}$ and ${\cal L}_{\rm axial}$ denote
the cubic interaction terms among vector and axial-vector mesons and baryons:
\begin{subequations}
\begin{eqnarray}
{\cal L}_{\rm vector}
&=&
-i\bar{\cal N} \gamma^\mu \beta_\mu
 {\cal N}-\sum_{k \ge 1}g_{V}^{(k)} \bar {\cal N} \gamma^\mu  v_\mu^{(2k-1)}
 {\cal N}+\sum_{k \ge 1}g_{dV}^{(k)} \bar {\cal N} \gamma^{\mu\nu}
\partial_\mu v_\nu^{(2k-1)}
 {\cal N},
\label{tesor:VNN}
\\
{\cal L}_{\rm axial}
&=&
-\frac{i g_A}{2}\bar {\cal N}  \gamma^\mu\gamma^5
\alpha_\mu {\cal N} -\sum_{k\ge 1} g_A^{(k)}
\bar {\cal N} \gamma^\mu\gamma^5
v_\mu^{(2k)} {\cal N}
+\sum_{k\ge 1}g_{dA}^{(k)} \bar {\cal N} \gamma^{\mu\nu}
\gamma^5\partial_\mu v_\nu^{(2k)}
 {\cal N},
 \end{eqnarray}
\end{subequations}
where $\alpha_\mu$ ($\beta_\mu$) contain odd (even) number of pions.
Before we discuss the nucleon-nucleon potential in this model,
we show a few interesting predictions of the model.
The tensor (or magnetic) interaction between the isospin singlet
vector mesons
and  nucleon extracted from (\ref{tesor:VNN}) is zero,
i.e.\ $g_{dV}^{(k)}=0$ for isospin
singlet mesons such as $\omega$ \cite{Kim:2009sr},
which is consistent with empirical values.
For instance $g_{V}^{(1)}/g_{dV}^{(1)}=0.1\pm 0.2$
for $\omega$-mesons \cite{EW88}.
Another robust result is the following relation between isospin
singlet coupling $g_V^{(k)}$ with $V=\omega$
and triplet one $g_V^{(k)}$ with $V=\rho$ \cite{Hong:2007ay},
\begin{eqnarray}
	\frac{g_\omega^{(k)}}{g_\rho^{(k)}}\simeq N_{\rm c} +\delta(k).
 \end{eqnarray}
If one sets $N_{\rm c}=3$ and considers the lowest mode,
one finds that $g_\omega^{(1)}/g_\rho^{(1)}\simeq 3.6$;
empirical values of the ratio are around 4 to 5, see \cite{EW88} for example.
Recently, from the effective action in (\ref{Lag:pointB})
four-nucleon contact interactions are derived,
and the low energy constant for each contact term is calculated \cite{Kim:2011xu}.

From now on we focus on the result of \cite{Kim:2009sr}
in the large $N_{\rm c}$ and large $\lambda$ limit.
The leading large $N_{\rm c}$ and large $\lambda$ scaling
of the cubic coupling is classified in \cite{Kim:2009sr}.
For instance, for pseudo-scalars ($\varphi=\pi,\eta'$) we have
\begin{eqnarray}
\frac{g_{\pi\CN\CN}}{2m_\CN} M_{\rm KK}
&=&
\frac{g_A^{\rm triplet}}{2f_\pi}M_{\rm KK}
\simeq \frac{2\cdot 3\cdot\pi}{\sqrt{5}}
\times \sqrt{\frac{N_{\rm c}}{\lambda}},
\nonumber \\
\frac{g_{\eta' \CN\CN}}{2m_\CN} M_{\rm KK}
&=&
\frac{N_{\rm c} g_A^{\rm singlet}}{2f_\pi}
M_{\rm KK}\simeq \sqrt{\frac{3^9}{2}} \pi^2
\times \frac{1}{\lambda N_{\rm c}} \sqrt{\frac{N_{\rm c}}{\lambda}}.
 \end{eqnarray}
 The leading contributions arise from the following cubic couplings
\begin{equation}
\frac{g_{\pi{\cal NN}}M_{\rm KK}}{2m_{\cal N}}
\sim g_{\omega^{(k)}\cal NN}
\sim \frac{\tilde g_{\rho^{(k)}\cal NN}M_{\rm KK}}{2m_{\cal N}}
\sim g_{a^{(k)}\cal NN} \sim \sqrt{\frac{N_{\rm c}}{\lambda}}.
 \end{equation}
Now, we adopt the conventional one-boson exchange potential approach
to obtain the holographic nucleon-nucleon potential.
We take the form of the one-boson exchange potentials obtained
in nuclear physics for various mesons,
which is well summarized in \cite{EW88},
and use the meson masses and coupling constants calculated
from the Sakai-Sugimoto model.
For instance, the one pion exchange potential is given by
\begin{equation}
\label{pion0}
V_{\pi}^{\rm holographic}
=\frac{1}{4\pi}
\left(\frac{g_{\pi{\cal NN}}M_{\rm KK}}{2m_{\cal N}}\right)^2
\frac{1}{M_{\rm KK}^2r^3}\;S_{12}\;\vec\tau_1\cdot\vec\tau_2,
\end{equation}
where $S_{12}=3(\vec\sigma_1\cdot\hat r)(\vec\sigma_2\cdot\hat r)
-\vec\sigma_1\cdot\vec\sigma_2$.

The potential obtained in this way
shows a similar tendency with the empirical ones.
The minimum of the potential for large $\lambda$ and $N_{\rm c}$
in \cite{Kim:2009sr} is at
about $r\simeq 5.5/M_{\rm KK}\sim 1.1 ~{\rm fm}$
with $M_{\rm KK}\sim 1~{\rm GeV}$, which is close to the minimum
position of empirical two nucleon potentials.
This minimum position does not change much
if one considers the potential with finite values
of $\lambda$ and $N_{\rm c}$ \cite{Kim:2009sr}.
However, the depth of the potential at the minimum position depends
strongly on the value of $\lambda$ and $N_{\rm c}$,
and it is supposed to be much shallow compared
to the empirical one due to lack of scalar (or two-pion) exchange contributions.
This lack of scalar attraction at the intermediate range of
the nuclear force is one of the essential problem
to be resolved towards a realistic nucleon-nucleon potential in holographic QCD.

\section{D4/D6 model}

Though the Sakai-Sugimoto model goes well with  many hadronic
phenomena including a spontaneous (nonabelian) chiral symmetry
breaking, it is not easy to introduce quark masses
into the model since no space is available between
D4- and D8$/\overline{\textrm{D}8}$-branes.
Furthermore,  this model does not have attractive forces
mediated by a scalar field which plays some role in low energy QCD phenomenology.
Introducing  different type of flavor branes gives a chance
to improve these aspects: D4/D6 model \cite{Kruczenski:2003uq}.
Since the D4/D6 model uses the same D4 background, its dual gauge
theory description is  similar to that of the Sakai-Sugimoto model.
On the other hand, D6 flavor branes can also have non-trivial
embeddings on the D4-brane background as D3/D7 model, and
so it tells us some interesting aspects of various QCD (-like) phenomena.
However, the penalty we have to pay is that the D4/D6 model
cannot realize nonabelian chiral symmetry like $SU(2)_{\rm L}\times
SU(2)_{\rm R}$, but only
$U(1)$ axial symmetry.

In D4/D6 model we use the same background (\ref{d4})
as in the Sakai-Sugimoto model,
\begin{equation}
\label{D4Conf}
	\dd s^2
=\left(\frac{U}{R}\right)^{3/2}\!\!\!
\Big\{-\dd t^2+\sum_{i=1}^3\dd x^{i\, 2}+f(U)\dd\tau^2\Big\}
+\left(\frac{R}{U}\right)^{3/2}\!\!
\left(\frac{U}{\xi}\right)^2\!\Big\{\dd\xi^2+\xi^2\dd\Omega_4^2\Big\},
 \end{equation}
where, for convenience, the dimensionless coordinate $\xi$
has been introduced; $\dd\xi^2/\xi^2=\dd U^2/f(U)U^2$ and
$(U/U_{\rm KK})^{3/2}=(\xi^{3/2}+\xi^{-3/2})/2$.

 \subsection{D6 flavor brane}

To introduce the fundamental quarks, we
put $N_{\textrm f}$ D6-branes on the D4-brane
background \cite{Kruczenski:2003uq}, see
Table \ref{braneprofile1} for the brane configuration.
 \begin{table}[h]
  	\begin{center}
  	\begin{tabular}{|>{\centering}p{20pt}*{9}{|>{\centering}p{13pt}}|p{13pt}|}
		\hline
		&$t$&$x^1$&$x^2$&$x^3$&($\tau$)
&$\rho$&$\psi^1$&$\psi^2$&$y$&~$\phi$ \\
		\hline
		D4&$\bullet$&$\bullet$&$\bullet$&$\bullet$&$\bullet$&&&&& \\
		\hline
		D6&$\bullet$&$\bullet$&$\bullet$&$\bullet$&&$\bullet$
&$\bullet$&$\bullet$&& \\
		\hline
  	\end{tabular}
  	\caption{The brane configurations: the background D4- and the probe D6-branes}
	\label{braneprofile1}
  	\end{center}
 \end{table}
For the sake of convenience, the radial directions
of the transverse part are decomposed as $\xi^2=\rho^2+y^2$.
Then, the background metric is rewritten as
 \begin{equation}
\label{reD4Conf}
	 \dd s^2
=\left(\frac{U}{R}\right)^{3/2}\!\!\!
\Big\{-\dd t^2+\sum_{i=1}^3\dd x^{i\, 2}+f(U)\dd\tau^2\Big\}
+\left(\frac{R}{U}\right)^{3/2}\!\!
\left(\frac{U}{\xi}\right)^2
\!\Big\{\dd\rho^2+\rho^2\dd\Omega_2^2
+\dd y^2+y^2\dd\phi^2\Big\}.
 \end{equation}
Open strings with two endpoints on (D4, D4) are considered
as pure $SU(N_{\textrm c})$ Yang-Mills theory as the Sakai-Sugimoto model.
On the other hand, the other kind of open strings ending on (D4, D6)
is in a fundamental representation with finite masses.
A chiral $U(1)_{\textrm A}$ symmetry is described
by a rotation in $(y, \phi)$-plane in this model.
This chiral symmetry is broken spontaneously by non-vanishing
chiral condensate which corresponds to the asymptotic
slope \cite{Kruczenski:2003uq}, see below for more on this.

We assume that the separation $y$ between D4- and D6-branes depends
only on $\rho$ with a fixed angle: $y=y(\rho)$ and $\phi=\phi_0$.
In addition, we set $\tau=\tau_0$, that is, D6-brane is localized
on the compactified circle $S^1$.
The induced metric on D6-brane is then
\begin{equation}
\label{inducedD6}
\dd s_{\textrm {D6}}^2
=\left(\frac{U}{R}\right)^{3/2}\!\!\!
\Big\{-\dd t^2+\sum_{i=1}^3\dd x^{i\,2}\Big\}
+\frac{R^{3/2}U^{1/2}}{\xi^2}
\Big\{\big(1+{y^\prime}^2\big)\dd\rho^2+\rho^2\dd\Omega_2^2\Big\},
 \end{equation}
where the prime stands for the derivative with respect to $\rho$.
The embedding of D6-brane is determined by $y(\rho)$.
From the DBI action (\ref{dbi}) for D6-brane without the bulk Kalb-Ramond field and
the world volume gauge field,
\begin{equation}
\label{D6DBI1}
S_{\textrm {D6}}
=-T_6\!\int\!\dd^7\sigma\,\eee^{-\widetilde{\Phi}}
\sqrt{-\textrm{det}g_{mn}(\sigma)}
=-\tau_6\!\int\!\dd t\,\dd\rho\,
\rho^2\big(1+1/\xi^3\big)^2\sqrt{1+{y^\prime}^2},
\quad
\mbox{with}
\quad
\tau_6=\frac{T_6V_3V_{S^2}U_{\rm KK}^3}{4},
\end{equation}
we can obtain the equation of motion for $y(\rho)$,
\begin{equation}
\label{D6eom1}
-6\rho^2\left(y/\xi^5\right)\left(1+1/\xi^3\right)\sqrt{1+{y^\prime}^2}
=\frac{\dd}{\dd\rho}\bigg(\frac{\rho^2\left(1+1/\xi^3\right)^2
		 y^\prime}{\sqrt{1+{y^\prime}^2}}\bigg).
 \end{equation}
This equation of motion can be solved numerically
with given boundary conditions.
For large $\rho\rightarrow\infty$,
the equation of motion becomes much simpler form
$ 0=(\rho^2 y^\prime)^\prime$,
therefore, asymptotic behavior of the embedding is
\begin{equation}
\label{asymptoticD6}
y(\rho)\simeq y_\infty+\frac{c}{\rho}+\cdots.
 \end{equation}
Here $y_\infty$ and $c$ are related to the bare quark
mass $M_{\rm q}$ and the chiral condensate $\langle\bar{\psi}\psi\rangle$,
respectively.

 \subsection{Meson spectrum: fluctuations of D6-branes} \label{mesonMass}

The open string modes whose endpoints,
both of them, live on $N_{\textrm f}$
probe-branes are in the adjoint representation of
$U(N_{\textrm f})$ flavor symmetry.
These modes can be described by the fluctuation around
the classical embedding solutions of D$p$-branes.
Hence, we can consider meson spectra as fluctuations of the probe
D6-branes
\begin{equation}
y(\sigma^m)
=\bar{y}(\rho)+\delta y(\sigma^m),
\qquad \textrm{and} \qquad
\phi(\sigma^m)=\bar{\phi}+\delta\phi(\sigma^m),
 \end{equation}
 where $\sigma^m$ are the world volume coordinates
$\sigma^m=(t,\vec{x},\rho,\psi_1,\psi_2)$.
 Here $\bar{y}(\rho)$ and $\bar{\phi}=\phi_0$ denote the classical
solutions of each equations of motion.
Each fluctuation corresponds to scalar and pseudo-scalar meson, respectively.
Then, the induced metric is  written as
\begin{eqnarray}
\label{inducedD6fluctuation}
\dd s_{\textrm {D6}}^2
&=&\left(\frac{U}{R}\right)^{3/2}\!\!\!
\Big\{-\dd t^2+\sum_{i=1}^{3}\dd x^{i\, 2}\Big\}
+\frac{R^{3/2}U^{1/2}}{\xi^2}
\Big\{\big(1+\bar{y}'^2\big)\dd\rho^2+\rho^2\dd\Omega_2^2\Big\}
\nonumber \\
&&
+\frac{R^{3/2}U^{1/2}}{\xi^2}
\Big\{2\bar{y}^\prime(\del_m\delta y) \ \dd\rho\, \dd\sigma^m
+\big(
(\del_m\delta y)(\partial_n\delta y)
+\left(\bar{y}+\delta y\right)^2
(\partial_m\delta\phi)(\partial_n\delta\phi)\big)\dd\sigma^m\dd\sigma^n
\Big\}.
\quad
 \end{eqnarray}
Taking it into account that the bulk metric components also
contain some fluctuations,
we obtain a Lagrangian density of the DBI action of D6-brane
up to quadratic order
\begin{eqnarray}
\mathcal{L}_{\rm D6}
&\simeq&
-\rho^2\sqrt{1+\bar{y}'^2}
\nonumber \\
&&\times\Bigg\{
1+3\left(\frac{8\bar{y}^2-\bar{\xi}^2}{\bar{\xi}^{10}}
+\frac{4\bar{y}^2-\rho^2}{\bar{\xi}^7}\right){\delta y}^2
-\frac{\bar{y}^\prime}{1+\bar{y}'^2}
\left(1+\frac{1}{\bar{\xi}^3}\right)
\frac{\bar{y}}{\bar{\xi}^5}\delta y\delta y^\prime
\nonumber
\\
&&
\hspace*{6mm}
+\frac{1}{2\left(1+\bar{y}'^2\right)^2}
\left(1+\frac{1}{\bar{\xi}^3}\right)^2
\left(\frac{R^3\left(1+\bar{y}'^2\right)}{\bar{U}\bar{\xi}^2}
\partial_\mu\delta y\partial^\mu\delta y
+\left(\delta y^\prime\right)^2
+\frac{1+\bar{y}'^2}{\rho^2}\partial_i\delta y\partial^i\delta y\right)
\nonumber \\
&&
\hspace*{6mm}
+\frac{\bar{y}^2}{2\left(1+\bar{y}'^2\right)}
\left(1+\frac{1}{\bar{\xi}^3}\right)^2
\left(\frac{R^3\left(1+\bar{y}'^2\right)}{\bar{U}\bar{\xi}^2}
\partial_\mu\delta\phi\partial^\mu\delta\phi
+\left(\delta\phi^\prime\right)^2
+\frac{1+\bar{y}'^2}{\rho^2}\partial_i\delta\phi\partial^i\delta\phi\right)
\Bigg\},
 \end{eqnarray}
where $\mu$ runs through our 4D coordinates, $i$ stands for the
coordinates $\psi^i$, and the prime implies the derivative with respect
to $\rho$.
The linearized equations of motion are, for $\delta y$
\begin{eqnarray}
0&=&\frac{9}{2^{4/3}M_{\textrm {KK}}^2}
\frac{\rho^2}{\bar{\xi}^3}\left(1+\frac{1}{\bar{\xi}^3}\right)^{4/3}
\!\!\!\partial_\mu\partial^\mu\delta y
+\left(1+\frac{1}{\bar{\xi}^3}\right)^2\partial_i\partial^i\delta y
\nonumber \\
&&
+\sqrt{1+\bar{y}'^2}\,\frac{\partial}{\partial\rho}
\left(\frac{\rho^2}{(1+\bar{y}'^2)^{3/2}}
\left(1+\frac{1}{\bar{\xi}^3}\right)^2\delta y^\prime\right)
-\sqrt{1+\bar{y}'^2}\,\frac{\partial}{\partial\rho}
\left(\frac{\bar{y}^\prime \rho^2}{\sqrt{1+\bar{y}'^2}}
\left(1+\frac{1}{\bar{\xi}^3}\right)
\frac{\bar{y}}{\bar{\xi}^5}\right)\delta y
\nonumber \\
&&-6\rho^2(1+\bar{y}'^2)
\left(\frac{8\bar{y}^2-\bar{\xi}^2}{\bar{\xi}^{10}}
+\frac{4\bar{y}^2-\rho^2}{\bar{\xi}^7}\right)\delta y,
 \end{eqnarray}
and, for $\delta\phi$
\begin{equation}
0
=
\frac{9}{2^{4/3}M_{\textrm {KK}}^2}
\frac{\bar{y}^2\rho^2}{\bar{\xi}^3}
\left(1+\frac{1}{\bar{\xi}^3}\right)^{4/3}
\!\!\!\!\!\partial_\mu\partial^\mu\delta\phi+\bar{y}^2
\left(1+\frac{1}{\bar{\xi}^3}\right)^2
\!\!\partial_i\partial^i\delta\phi
+\frac{1}{\sqrt{1+\bar{y}'^2}}
\frac{\partial}{\partial\rho}
\left(\frac{\bar{y}^2\rho^2}{\sqrt{1+\bar{y}'^2}}
\left(1+\frac{1}{\bar{\xi}^3}\right)^2\!\!\delta\phi^\prime\right).
\end{equation}
These linearized equations of motion can be solved numerically
using the shooting method \cite{Kruczenski:2003uq}.

 \subsection{Baryon vertex: compact D4-branes}

\begin{figure}[!hbtp]
\begin{center}
\includegraphics[width=5cm]{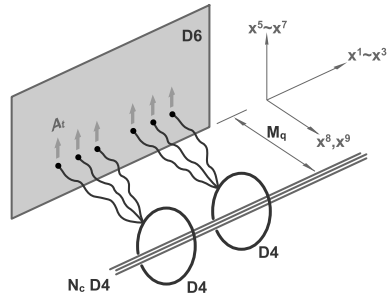}
\caption{D4/D6 and compact D4}
\label{fig:d4d6}
\end{center}
\end{figure}

Now we introduce a baryon in our holographic description
through a baryon vertex \cite{Witten:1998xy}.
The baryon vertex in 4D corresponds to the compact D4-branes
wrapping $S^4$ transverse to the background D4-branes
(See Table \ref{braneprofile2} and Fig.\ref{fig:d4d6}).
\begin{table}[h]
\label{braneprofile2}
  	\begin{center}
  	\begin{tabular}{|>{\centering}p{20pt}*{9}{|>{\centering}p{13pt}}|p{13pt}|}
		\hline
		&$t$&$x^1$&$x^2$&$x^3$&($\tau$)&$\xi$&$\theta$&$\theta^1$
&$\theta^2$&$\theta^3$ \\
		\hline
		D4&$\bullet$&$\bullet$&$\bullet$&$\bullet$&$\bullet$&&&&& \\
		\hline
		c\,D4&$\bullet$&&&&&&$\bullet$&$\bullet$&$\bullet$&~$\bullet$ \\
		\hline
  	\end{tabular}
  	\caption{The brane configurations: the background D4- and the compact D4-branes}
  	\end{center}
\end{table}

The induced metric on the compact D4-brane is
\begin{equation}
\label{inducedD4}
\dd s_{\textrm {D4}}^2
=-\left(\frac{U}{R}\right)^{3/2}\!\!\dd t^2
+R^{3/2}U^{1/2}
\bigg\{\bigg(1+\frac{\dot{\xi}^2}{\xi^2}\bigg)\dd\theta^2
+\sin^2\theta\,\dd\Omega_3^2\bigg\},
 \end{equation}
where $\,\dot{\xi}=\partial\xi/\partial\theta$.
In addition to the DBI action with the world volume gauge potential
$A=A_t(\theta)\dd t$,
we consider the WZ term (\ref{wz_p}) with $N_{\rm c}$ fundamental
strings.
By using the RR flux $F_4(\sigma)$ in (\ref{d4_dilaton}),
the action of the compact D4-brane is given
by~\cite{Callan:1999zf},
\begin{eqnarray}
\label{D4DBI}
S_{\textrm {D4}}
&=&
-T_4\!\int\!\dd^5\sigma\,\eee^{-\widetilde{\Phi}}
\sqrt{-\textrm{det}\left(g_{mn}+2\pi\alpha^\prime F_{mn}\right)}
+\mu_4\!\int\!(2\pi\alpha'A)\wedge F_4
\nonumber \\
&=&
-\tau_4\!\int\!\dd t\,\dd\theta\,\sin^3\theta
\bigg(\sqrt{\left(1+1/\xi^3\right)^{4/3}
\left(\xi^2+\dot{\xi}^2\right)-\widetilde{F}^2}-3\widetilde{A}_t\bigg)
\equiv
\tau_4\!\int\!\dd t\, \dd\theta\,\mathcal{L} ,
 \end{eqnarray}
with
$$
\tau_4=\frac{1}{2^{2/3}}T_4V_{S^3} R^3 U_{\textrm {KK}},
\quad
\widetilde{A}_t=\frac{2^{2/3}(2\pi\alpha')}{U_{\rm KK}}A_t,
\quad
\textrm{and}
\quad
\widetilde{F}
=\partial_\theta\widetilde{A}_t.
 $$
The dimensionless displacement is defined as
\begin{equation}	
\frac{\partial\mathcal{L}}{\partial\widetilde{F}}
=\frac{\sin^3\theta\:\widetilde{F}}
{\sqrt{\big(1+1/\xi^3\big)^{4/3}(\xi^2+{\dot{\xi}}^2)-\widetilde{F}^2}}
\equiv-D(\theta).
 \end{equation}
In terms of the dimensionless displacement,
the equation of motion for the gauge field is expressed as
\begin{equation}
\partial_\theta D(\theta)=-3\sin^3\theta,
\end{equation}
and the solution is given by
\begin{equation}
D(\theta)=D_0+3\Big(\cos\theta-\frac{1}{3}\cos^3\theta\Big).
 \end{equation}
The integration constant $D_0$ will be determined later.
Now, we rewrite the Lagrangian as
\begin{equation}
\label{LegendreD4}
\bar{\mathcal{L}}[\xi,\dot{\xi};\theta]
=\sqrt{(\xi^2+\dot{\xi}^2)\big(1+1/\xi^3\big)^{4/3}
\left(D(\theta)^2+\sin^6\theta\,\right)}.
 \end{equation}
The equation of motion from this  Lagrangian determines
the configuration of D4-brane which depends only on the one
parameter $\xi_0$, the position of the compact D4 at $\theta=0$.
This rewritten Lagrangian can be regarded as Hamiltonian
for the baryon vertex since it is obtained
by the procedure similar to the Legendre transformation.
Then, the free energy of the compact D4-brane is
\begin{equation}
\mathcal{F}_{\textrm {D4}}
=\tau_4\int \dd\theta\,\bar{\mathcal{L}}
=\tau_4\int \dd\theta\,\sqrt{(\xi^2+\dot{\xi}^2)
\big(1+1/\xi^3\big)^{4/3}\left(D(\theta)^2+\sin^6\theta\,\right)}.
\end{equation}
Considering its symmetry, the simplest situation we can imagine
is  that the $N_{\rm c}$ fundamental strings are attached only at the
poles of this D4-brane,  i.e.\  $\theta=0$ and $\theta=\pi$.
At this point, the shape of the cusp is determined
by the number of the strings attached.
Using (\ref{LegendreD4}), we can fix the number explicitly.
The tension at the cusp is obtained by the variations of
the energy functional with respect to the position of the cusp as
\begin{equation}
f_{\textrm {D4}}
=\frac{\delta}{\delta U_c}
\Big[\tau_4\!\int\!\dd\theta\,\bar{\mathcal{L}}\,\Big]
\Big|_\textrm{fixing other values}
=\frac{N_{\textrm c}T_{\rm F}\,|D_0\pm 2|}{4}\;
\frac{1+1/\xi_c^3}{1-1/\xi_c^3}\;
\frac{\dot{\xi}_c}{\sqrt{\xi_c^2+{\dot{\xi}_c}^2}},
\end{equation}
where $|D_0+2|$ and $|D_0-2|$ correspond
to $\theta=0$ and $\theta=\pi$, respectively, and $\xi_c=\xi(0)$ or $\xi(\pi)$.
Now we set $D_0=-2$, that is, all fundamental strings are
attached at the north pole.
Here $T_{\rm F}$ is the tension of the fundamental string.
Through the numerical calculations, it is known that the force
$f_{\textrm {D4}}$ at the cusp is always smaller than
$N_{\textrm c}\times T_{\rm F}$~\cite{Kim:2009ey}.
The baryon vertex alone cannot remain stable
because the attached fundamental strings pull up
the north pole to infinity~\cite{Callan:1999zf}.
One way to make it stable is to introduce  probe branes
which are connected to the baryon vertex through
the $N_{\textrm c}$ fundamental strings.

Since the endpoint of fundamental strings plays
a role of the source for the gauge field on the probe D6-brane,
we can turn on the $U(1)$ gauge field $A=A_t(\rho)\dd t$ on the D6-brane (see
Fig.\ref{fig:d4d6}).
The DBI action for this brane is
\begin{eqnarray}
\label{D6DBI2}
S_{\textrm {D6}}
&=&
-T_6\!\int\!\dd^7\sigma\,\eee^{-\widetilde{\Phi}}
\sqrt{-\textrm{det}\left(g_{mn}+2\pi\alpha^\prime F_{mn}\right)}
\nonumber \\
&=&
-\tau_6\!\int\!\dd t\,\dd\rho\,\rho^2
\left(1+1/\xi^3\right)^{4/3}\sqrt{\left(1+{y^\prime}^2\right)
\left(1+1/\xi^3\right)^{4/3}-\widetilde{F}^2}\, ,
\end{eqnarray}
where we have defined
$\widetilde{A}_t(\rho)=(2^{2/3}(2\pi\alpha')/U_{\rm KK})A_t(\rho)$
and
$\widetilde{F}(\rho)=\del_\rho\widetilde{A}_t(\rho)$.
From the equation of motion for gauge fields,
the conserved charge $\widetilde{Q}$ is defined as
\begin{equation}
\label{constraintQ}
\frac{\partial\mathcal{L}}{\partial\widetilde{F}}=
\frac{\rho^2\left(1+1/\xi^3\right)^{4/3}\widetilde{F}}
{\sqrt{\left(1+{y^\prime}^2\right)\left(1+1/\xi^3\right)^{4/3}
-\widetilde{F}^2}}\equiv\widetilde{Q}.
 \end{equation}
This conserved quantity $\widetilde{Q}$ is related to the number
of point sources (fundamental strings) $Q$ by
\begin{equation}
\widetilde{Q}
=\frac{U_{\textrm {KK}}}{2^{2/3}(2\pi\alpha^\prime)\tau_6}\,Q.
 \end{equation}
We can rewrite the Lagrangian with  respect to $\widetilde{Q}$ with
the constraint $\partial_\rho\widetilde{Q}=0$,
\begin{equation}
\label{LegendreD6}	
\bar{\mathcal{L}}[y,y^\prime;\rho;\widetilde{Q}]
=\sqrt{\left(1+{y^\prime}^2\right)
\left(1+1/\xi^3\right)^{4/3}
\left(\rho^4\left(1+1/\xi^3\right)^{8/3}+\widetilde{Q}^2\right)}.
 \end{equation}
Again, this Lagrangian plays a role of Hamiltonian.
The force of D6-brane at the cusp is given by
\begin{equation}
f_{\textrm {D6}}
=
\frac{\delta}{\delta U_c}
\Big[\tau_6\!\int\!\dd\rho\,\bar{\mathcal{L}}\,\Big]
\Big|_\textrm{fixing other values}
=
2^{2/3}\tau_6\;\frac{1}{U_{\textrm {KK}}}\;
\frac{1+1/\xi_c^3}{1-1/\xi_c^3}\;
\frac{y^\prime_c}{\sqrt{1+{y^\prime_c}^2}}\;\widetilde{Q}
=
QT_{\rm F}\;\frac{1+1/\xi_c^3}{1-1/\xi_c^3}\;
\frac{y^\prime_c}{\sqrt{1+{y^\prime_c}^2}}.
\end{equation}
Since the tension of the fundamental string is larger than
that of D$p$-branes, the fundamental strings shrink and two branes,
on which the string was attached, meet at a single point.
To be a stable configuration, the forces at the cusp should be balanced
\begin{equation}
\frac{Q}{N_{\textrm c}} f_{\textrm {D4}}=f_{\textrm {D6}}(Q).
 \end{equation}
Once its embedding is determined, the total free energy of this D4/D6
system is obtained as a sum of each brane
 \begin{equation}
\mathcal{F}_{\textrm {tot}}
=\frac{Q}{N_{\textrm c}}\mathcal{F}_{\textrm {D4}}
+\mathcal{F}_{\textrm {D6}}(Q).
 \end{equation}

Now we consider multi-flavor cases.
For simplicity, we take $N_{\textrm f}=2$.  This can be described
by adding one more probe brane: D4/D6/D6.
Since we have two probe branes, the force balancing condition is
modified to
\begin{equation}
\frac{Q}{N_c}f_{\textrm {D4}}=f_{\textrm {D6}}^{(1)}(Q_1)+f_{\textrm
 {D6}}^{(2)}(Q_2),
\end{equation}
where $Q=Q_1+Q_2$.
With this constraint, we find minimum energy configuration.
The total energy of the system is given by
\begin{equation}
\mathcal{F}_{\textrm {tot}}
=\frac{Q}{N_{\textrm c}}\mathcal{F}_{\textrm {D4}}
+\mathcal{F}_{\textrm {D6}}(Q_1)+\mathcal{F}_{\textrm {D6}}(Q_2).
 \end{equation}
%

\subsection{Nuclear symmetry and transition to strange matter}

Now we consider a few interesting physics in the D4/D6 model:
the nuclear symmetry energy \cite{Li:2008gp} 
and a toy model study on nuclear to strange matter transition.

\vskip 0.3cm
\noindent  \textbf{Symmetry energy}\\
The nuclear symmetry energy is defined as
the energy (per nucleon) difference between
the isospin symmetric nuclear matter and the pure neutron matter.
To describe the symmetry energy,
we define an asymmetric factor $\tilde{\alpha}\equiv(N-Z)/(N+Z)$, $Z(N)$
is the number of protons (neutrons).
The energy density per nucleon is written as
\begin{equation}
E(\rho,\tilde{\alpha})
\simeq E(\rho,0)+E_{\textrm {sym}}(\rho)
\tilde{\alpha}^2+\cdots,
\qquad \mbox{where} \qquad
E_{\textrm {sym}}
=\frac{1}{2}\frac{\partial^2 E}{\partial\tilde{\alpha}^2}
\Big|_{\tilde{\alpha}=0}.
\end{equation}
Since proton and neutron (or up and down quarks) have
almost the same masses, we consider a case where two D6-branes
have the same asymptotic values.
In terms of  $\tilde{\alpha}$, the total free energy is expressed as
\begin{equation}
\mathcal{F}_{\textrm {tot}}
=\frac{Q}{N_{\textrm c}}\mathcal{F}_{\textrm {D4}}
+\mathcal{F}_{\textrm {D6}}\left[
\left(\frac{1+\tilde{\alpha}}{2}\right)Q\right]
+\mathcal{F}_{\textrm {D6}}
\left[\left(\frac{1-\tilde{\alpha}}{2}\right)Q\right].
 \end{equation}
Then, the symmetry energy follows
\begin{equation}
\label{symE}
E_{\textrm {sym}}
=\frac{2\tau_6}{N_{\textrm B}}\int \dd\rho\;\frac{\rho^4
\sqrt{1+{y^\prime}^2}\left(1+1/\xi^3\right)^{10/3}
\widetilde{Q}^2}{\left(4\rho^4 \left(1+1/\xi^3\right)^{8/3}
+\widetilde{Q}^2\right)^{3/2}},
\end{equation}
where $N_{\textrm B}=Q/N_{\textrm c}$ is the baryon
number~\cite{Kim:2010dp}.
Two main messages here are:
(1) the symmetry energy keeps increasing with density,
which is independent of the choice of the value of the t' Hooft
coupling and $M_{\textrm {KK}}$,
(2) $E_{\textrm {sym}}\sim\rho^{1/2}$, which is very insensitive
to the value of the 't Hooft coupling and $M_{\textrm {KK}}$.
The symmetry energy was also calculated in some other
holographic studies~\cite{Park:2011zp}. 

\vskip 0.3cm
\noindent  \textbf{Nuclear to strange matter transition}\\
With the D4/D6 type models, one can study a transition
from nuclear to strange matter \cite{Kim:2009ey}.
For this, two D6-branes with different asymptotic values are considered.
At low density, the lower brane is attached to the baryon vertex.
This means that at low density, Fermi sea consists of only light quark (up or down).
As densities go up, there exist critical density
at which both of the D6-branes are connected to the baryon vertex,
meaning that intermediate (strange) quarks start to pile up
in the Fermi sea, in addition to the light quarks.

\section{Recent development: sample works\label{sampleW}}

To demonstrate the versatility of the holographic QCD approach
and its recent contributions to QCD (or QCD-like) phenomenology,
we compile here some of recent results from various studies based
on top-down or bottom-up models.

\begin{itemize}
\item[-] Various hadronic form-factors were calculated  in diverse
approaches~\cite{Gutsche:2012bp}.

\item[-] The QCD trace anomaly was investigated in a holographic 
QCD model with the bulk dilaton field~\cite{Goity:2012yj}.

\item[-] In \cite{Cai:2012xh}, asymptotically
AdS black hole solutions of an Einstein-Maxwell-Dilaton (EMD) 
system was constructed and QCD (or its cousin) phase diagram 
at finite temperature and chemical potential 
was extensively studied in the newly found AdS black hole background.

\item[-] A mean field analysis for the bulk fermion field 
was proposed in \cite{Harada:2011aa}. 
Four-dimensional mean field of
the fermion bilinear was used in an AdS/QCD model study~\cite{Kim:2007xi}.

\item[-] The authors of \cite{Koile:2011aa} evaluated 
the structure function of deep inelastic scattering from scalar 
mesons and polarized vector mesons in the D3/D7 model 
and the Sakai-Sugimoto model.

\item[-] A study on thermal QCD at large $N_c$ was 
performed in \cite{Mia:2011iv}.
In \cite{Veschgini:2010ws}, some critical issues in building 
up a gravity dual model of QCD at finite temperature were discussed.
Following the approaches described in \cite{Alanen:2009xs, Gursoy:2008za},
the author of ~\cite{Megias:2010ku} studied QCD thermodynamics 
comprehensively.

\item[-] An interesting attempt to build a bridge between 
holographic QCD and lattice Monte Carlo studies of the true QCD vacuum
has been made in \cite{Thacker:2010zk}.

\item[-] Based on the light front holography 
approach~\cite{Brodsky:2003px,Brodsky:2006uqa}, nucleon 
generalized parton distributions were investigated in \cite{Vega:2011fz}.

\item[-] In \cite{Colangelo:2011xk}, the  vertex function of anomalous 
two vector and one axial-vector currents was calculated 
in the soft wall model~\cite{Karch:2006pv} with the Chern-Simons term.

\item[-] Quark number susceptibility, a good probe for QCD phase 
transition, was calculated in an improved soft wall model~\cite{Cui:2011wb}.

\item[-] Detailed description on the photon-hadron high energy 
scattering in a gravity dual description is given in~\cite{Nishio:2011xz}.

\item[-] In \cite{Domokos:2011dn}, tensor fields were incorporated 
into the hard wall model~\cite{Erlich:2005qh,DaRold:2005zs,Hirn:2005vk},
and their properties were extensively studied.

\item[-] The chiral phase transition in an external magnetic 
field at finite temperature and chemical potential was 
studied in the Sakai-Sugimoto model. It is shown that for small
temperatures the magnetic field lowers the critical chemical potential
for chiral symmetry restoration~\cite{Preis:2010cq} contrary 
to the magnetic catalysis in free space.

\item[-] In \cite{Hashimoto:2010wv}, thermalization of mesons 
with a time-dependent baryon number chemical potential was analyzed
in a D3/D7 model.

\item[-] In \cite{Pahlavani:2010zzb}, the binding energy 
of a holographic deuteron and tritium was calculated 
in the Sakai-Sugimoto  model.
\end{itemize}

\section{Closing Remarks}

Starting from a concise summary on string theory,
we tried to present a bird's eye view on
holographic QCD for the non-experts,
focusing more on the basic materials needed to understand the approach.
We did not intend to demonstrate widely what holographic QCD
can do for nuclear and hadron physics.
Therefore, we showed only some out of innumerable results
from holographic QCD models,
see \cite{Erdmenger:2007cm,McGreevy:2009xe,Kim:2011ey,
CasalderreySolana:2011us,deTeramond:2012rt}
for a review.

Despite  tremendous efforts and some partial successes,
it is clear that the holographic QCD approach towards realistic
systems in nature is still  far from its final form.
There can be numerous facts that support this statement,
but we will touch on only two of them.
The first obvious one is that the AdS/CFT holds at large $N_{\rm c}$
and mostly stringy holographic QCD models assume
the probe approximation $N_{\rm c}\gg N_{\rm f}$.
Therefore, the holographic QCD with its present form
can describe, at best, an extreme of nonperturbative aspects of QCD.
There is no guarantee that extrapolation down to $N_{\rm c}=3$ is
justifiable unless physical quantities at hand belong to
some universality classes.
However, thorough understanding of the extreme end of QCD would give
some insight on nuclear-hadron physics and offer a qualitative
guide to realistic QCD.
For a recent attempt to overcome the large $N_{\rm c}$
issue (or go beyond the probe approximation), one may
see \cite{Benini:2007gx}. 
The second issue is about the missing of essential ingredients
in a holographic QCD model, which is sometimes related
to the large $N_{\rm c}$ problem, to describe a realistic system.
To address this issue, we give one example.
It is well known that scalars with a mass of a few hundreds
MeV (or two-pions) are essential to describe stable nuclear matter
and to explain the intermediate-range attraction in nuclear forces.
While, holographic QCD models on the market are mostly lack
of this scalar attraction. Still we have no clear answer to
this problem, but there have been  some attempts to improve
this aspect \cite{Dymarsky:2010ci}. 

Apart from the issues mentioned above, there can be some more
reasons to believe that holographic QCD in its present form
may not be able to describe QCD-related realistic phenomena.
However, now is certainly too early to abandon the ship.
Surely to tide over fundamental issues like the probe limit,
we may have to wait long for a (would-come) breakthrough in string theory.
Meanwhile, holographic QCD practitioners are to explore more
to equip with essential parts suitable for QCD-related realistic systems.

\section*{Acknowledgments}

We thank Deokhyun Yi for providing the figures.
YK  expresses his sincere gratitude  to  Yunseok Seo, Sang-Jin Sin,
and Piljin Yi for collaborations in top-down models.
We acknowledge the Max Planck Society (MPG),
the Korea Ministry of Education, Science, and Technology (MEST),
Gyeongsangbuk-Do and Pohang City for the support of the Independent Junior
Research Group at the Asia Pacific Center for Theoretical Physics (APCTP).

\appendix

\section{Formulae and conventions}
\renewcommand{\theequation}{A.\arabic{equation}}

We work with $D$-dimensional spacetime with signature $(-,+,\cdots,+)$.
The line element is given by
\begin{equation}
\dd s^2=G_{MN}(x)\dd x^M\dd x^N.
\end{equation}
The Christoffel symbol is defined by
\begin{equation}
\Gamma^M{}_{NL}
=\frac{1}{2}G^{MP}
\Big(
\del_NG_{PL}+\del_LG_{PN}-\del_PG_{NL}
\Big).
\end{equation}
The covariant derivative on the tensor
$T^{N_1 \cdots N_p}{}_{L_1\cdots L_q}(x)$ is defined by
\begin{eqnarray}
\nabla_MT^{N_1 \cdots N_p}{}_{L_1\cdots L_q}
=
\del_MT^{N_1 \cdots N_p}{}_{L_1\cdots L_q}
&+&
\Gamma^{N_1}{}_{MP}T^{PN_2\cdots N_p}{}_{L_1\cdots L_q}
+({\rm all \ upper \ indices})
\nonumber
\\
&-&
\Gamma^{P}{}_{ML_1}T^{N_1 N_2\cdots N_p}{}_{PL_2\cdots L_q}
-({\rm all \ lower \ indices}).
\end{eqnarray}
The Riemann tensor is given by
\begin{equation}
R^P{}_{LMN}
=\del_M\Gamma^P{}_{NL}
+\Gamma^P{}_{MQ}\Gamma^Q{}_{NL}-(M\longleftrightarrow N).
\end{equation}
Then we can define the Ricci tensor and the scalar curvature
$R_{MN}=R^P{}_{MPN}$ and $R=G^{MN}R_{MN}$, respectively.

If one performs the Weyl rescaling
$G_{MN}(x)=\eee^{\alpha\Phi(x)}\widetilde{G}_{MN}(x)$ with a constant
$\alpha$, various quantities transform as follows:
\begin{eqnarray}
\sqrt{-G}
&=&
\eee^{\alpha D\Phi/2}\sqrt{-\widetilde{G}},
\nonumber
\\
R(G)
&=&
\eee^{-\alpha\Phi}
\Big(\widetilde{R}(\widetilde{G})
+\frac{\alpha^2}{4}(D-2)(D-1)
\widetilde{G}^{MN}\del_M\Phi\del_N\Phi
+\alpha\frac{D-1}{\sqrt{-\widetilde{G}}}
\ \del_M\big(\sqrt{-\widetilde{G}}\widetilde{G}^{MN}\del_N\Phi\big)
\Big), \qquad
\label{weyl_rescaling_0}\\
\sqrt{-G}R(G)
&=&
\eee^{\frac{\alpha(D-2)\Phi}{2}}
\sqrt{-\widetilde{G}}
\Big(
\widetilde{R}(\widetilde{G})
-\frac{\alpha^2}{4}(D-2)(D-1)\widetilde{G}^{MN}\del_M\Phi\del_N\Phi
\Big)
\nonumber
\\
&&
+\del_M\Big(\alpha(D-1)\eee^{\frac{\alpha(D-2)\Phi}{2}}\sqrt{-\widetilde{G}}
\widetilde{G}^{MN}\del_N\Phi\Big),
\nonumber
\end{eqnarray}

We introduce the differential form.
The zero-form is just a scalar $\phi(x)$ and one-form $A(x)$ is a vector
which is expanded by the basis $\dd x^\mu$,
\begin{equation}
A_1(x)\equiv A_\mu(x)\dd x^\mu.
\end{equation}
Rank $p$ antisymmetric tensor, i.e.\ $p$-form can be defined
as
\begin{equation}
A_p(x)\equiv\frac{1}{p!}
A_{\mu_1\cdots\mu_p}\dd x^{\mu_1}\wedge\cdots\wedge\dd x^{\mu_p},
\end{equation}
where the wedge product $\wedge$ is defend by the totally
antisymmetric tensor product, for example,
$$
\dd x^\mu \wedge\dd x^\nu\equiv\dd x^{\mu}\otimes\dd x^{\nu}
-\dd x^\nu\otimes\dd x^\mu=-\dd x^\nu\wedge\dd x^\mu.
$$
The exterior derivative $\dd=\dd x^\lambda\del_\lambda$ acting on
$p$-form is given by
\begin{equation}
\dd A(x)\equiv\frac{1}{p!}
\big(\del_\lambda A_{\mu_1\cdots\mu_p}(x)\big)
\dd x^\lambda\wedge\dd x^{\mu_1}\wedge\cdots\wedge\dd x^{\mu_p}.
\end{equation}
It is easy to observe $\dd^2=0$.
The differential form is naturally defined over a manifold.
If the dimensions of manifold is $k$, then the only nonzero integration
is given by that over the $k$-form,
\begin{equation}
\int_{M_k}\! A_{k}(x)
=\!\int_{M_k}\!
A_{1\cdots k}(x)\dd x^1\wedge\cdots\wedge\dd x^{k}
\equiv\!\int_{M_k}\!A_{1\cdots k}(x)\dd^{k}x,
\end{equation}
where it is not required to introduce the metric of the manifold.
The number of  the independent basis of $p$-form in $D$-dimensional spacetime
is the same as
that of $(D-p)$-form.
One can relate these through the Hodge duality operation $*$,
which plays an important role for the electric/magnetic duality:
\begin{equation}
*\big(
\dd x^{\mu_1}\wedge\cdots\wedge\dd x^{\mu_p}\big)
\equiv
\frac{1}{(D-p)!}
\epsilon^{\mu_1\cdots\mu_p}{}_{\nu_1\cdots\nu_{D-p}}
\dd x^{\nu_1}\wedge\cdots\wedge\dd x^{\nu_{D-p}},
\label{hodge_dual}
\end{equation}
where $\epsilon_{\mu_1\cdots\mu_D}(x)$ is the Levi-Civita tensor
which is normalized by $\epsilon_{1\cdots D}(x)\equiv\sqrt{-G(x)}$ and
$\epsilon^{1\cdots D}(x)=-1/\sqrt{-G(x)}$.
The dual operation satisfies
\begin{equation}
** A_p=(-)^{p(D-p)+1}A_p.
\end{equation}
The $D$-dimensional invariant volume form is given by
$$
*1=\dd^Dx\sqrt{-G}.
$$
One can define the inner product of forms over $M$,
\begin{equation}
\int_M\! A_p\wedge*B_p
=\frac{1}{p!}\!\int_M\!\dd^Dx\sqrt{-G}A_{\mu_1\cdots\mu_p}
B^{\mu_1\cdots\mu_p}.
\end{equation}
We also use the convention
$$
|F_p|^2\equiv\frac{1}{p!}F_{M_1\cdots M_p}F^{M_1\cdots M_p}.
$$
%

\section{AdS$_{\bm{d+1}}$}
\renewcommand{\theequation}{B.\arabic{equation}}

Anti-de Sitter space (AdS) appears as a vacuum solution of the Einstein equation
with the negative cosmological constant.
It is useful to consider $d+1$-dimensional AdS$_{d+1}$ space
as an embedded Lorentzian submanifold in
$d+2$-dimensional flat spacetime with metric
\begin{equation}
\eta_{MN}={\rm diag}(-,+,+,\cdots, +,-),
\label{metric_{d+2}}
\end{equation}
and coordinates $X^M, \ (M=0, 1, \ldots, d+1)$.
The submanifold is then defined through the condition
\begin{equation}
-(X^0)^2+(X^1)^2+(X^3)^2+\cdots+(X^d)^2-(X^{d+1})^2=-l^2,
\label{hypersurface}
\end{equation}
where the parameter $l$ would be the radius of AdS space.
It is obvious that the defined space equips with an $SO(2, d)$ symmetry
which is the isometry (the symmetry of the metric) of the AdS$_{d+1}$
space. In addition, the space is homogeneous and isotropic.
It is useful to introduce coordinates $x^m, (m=0, 1, \ldots, d)$
in AdS$_{d+1}$ space and define
an induced metric $g_{mn}(x)$ as
\begin{equation}
g_{mn}(x)=G_{MN}(X)\del_mX^M\del_nX^N,
\end{equation}
where $G_{MN}(X)$ is the metric of the entire space given by
(\ref{metric_{d+2}}) in the present case.

\subsection{Global coordinates}

We first consider the global coordinates
$(\tau, \rho, x^i), \ (i=1, \ldots, d)$ covering all of the hypersurface
which satisfies the condition (\ref{hypersurface}),
\begin{equation}
\begin{array}{rcl}
X^0
&=&
l\cosh\rho\ \cos\tau,
\\
X^i
&=&
l\sinh\rho\ x^i,
\\
X^{d+1}
&=&
l\cosh\rho\ \sin\tau,
\end{array}
\end{equation}
with $0\le\tau\le2\pi, \ \rho\ge0$ and $(x^1)^2+\cdots+ (x^d)^2=1$.
The induced metric is given by
\begin{equation}
\dd s^2
=l^2
\Big(
-\cosh^2\!\!\rho\ \dd\tau^2
+\dd\rho^2
+\sinh^2\!\!\rho\ \dd\Omega^2_{d-1}
\Big),
\label{metric_global_coordinates}
\end{equation}
where we have introduced the spherical
coordinates $\theta^i, \ (i=1, \ldots, d-1)$ on the unit $S^d$
such as,
\begin{equation}
\dd\Omega^2_{d-1}
=
(\dd\theta^1)^2
+\sin^2\theta^1(\dd\theta^2)^2
+\sin^2\theta^1\sin^2\theta^2(\dd\theta^3)^2
+\cdots
+\sin^2\theta^1\cdots\sin^2\theta^{d-2}(\dd\theta^{d-1})^2,
\end{equation}
with $0\le\theta^1, \cdots, \theta^{d-2}\le \pi$ and
$0\le\theta^{d-1}\le 2\pi$.
It is useful to write down the volume of unit $S^{d-1}$,
\begin{equation}
V_{S^{d-1}}
\!=\!
\int_0^{2\pi}\!\!\!\!\dd\theta^{d-1}
\!\int_0^\pi\!\!\!\!\dd\theta^{d-2}
\cdots
\!\int_0^\pi\!\!\!\!\dd\theta^1
\sqrt{g_{S^{d-1}}}
=\frac{2\pi^{d/2}}{\Gamma(d/2)},
\label{vs(d-1)}
\end{equation}
where
$$
\sqrt{g_{S^{d-1}}}
=(\sin^{d-2}\theta^1)(\sin^{d-3}\theta^2)
\cdots(\sin^2\theta^{d-3})(\sin\theta^{d-2}).
$$
We display relevant volumes of $S^{d-1}$ in the paper,
$$
V_{S^3}=2\pi^2, \qquad
V_{S^4}=\frac{8\pi^2}{3}, \qquad 
V_{S^5}=\pi^3.
$$ 
Since the time like circle of $\tau$ breaks the causality,
we need to move on to the universal covering space to extend the
circle to $-\infty<\tau<\infty$.
We always refer to AdS space as this universal covering space.

We further introduce new coordinate $\theta$ as
$
\sinh\rho=\tan\theta, \ (0\le\theta\le\pi/2).
$
Then the metric (\ref{metric_global_coordinates}) becomes
\begin{equation}
\dd s^2
=
\frac{l^2}{\cos^2\theta}
\Big(
-\dd\tau^2+\dd\theta^2+\sin^2\theta \ \dd\Omega^2_{d-1}
\Big).
\label{metric_global_coordinates_01}
\end{equation}
Except for the conformal factor, the spacelike hypersurface
at $\tau=\mbox{const.}$ in (\ref{metric_global_coordinates_01}) gives
the half of $S^d$.
Its equator $(\theta=\pi/2)$ corresponds to the boundary of AdS.
Along the equator, (\ref{metric_global_coordinates_01}) describes
the conformally compactified $d$-dimensional Minkowski spacetime.
This is important for the AdS$_{d+1}$/CFT$_d$ correspondence.

\subsection{Poincar\'e coordinates}

We next introduce the Poincar\'e
coordinates $(x^\mu, u)$ with $\mu=0, \underbrace{1, \cdots,  d-1}_i$,
\begin{equation}
\begin{array}{rcl}
X^0
&=&
\displaystyle
\frac{1}{2u}
\Big(
1+u^2(L^2+\eta_{\mu\nu}x^\mu x^\nu)
\Big),
\\
X^i
&=&
Lux^i,
\\
X^d
&=&
\displaystyle
\frac{1}{2u}
\Big(
1-u^2(L^2-\eta_{\mu\nu}x^\mu x^\nu)
\Big),
\\
X^{d+1}
&=&
Lux^0.
\end{array}
\end{equation}
Then, the induced metric is given by
\begin{equation}
\dd s^2
=L^2
\Big(u^2\eta_{\mu\nu}\dd x^\mu\dd x^\nu
+\frac{1}{u^2}\dd u^2\Big).
\end{equation}
Further transforming the coordinates as $u\to L/z$ and $x^\mu\to
x^\mu/L$, we arrive at
\begin{equation}
\dd s^2=\frac{L^2}{z^2}
\Big(\eta_{\mu\nu}\dd x^\mu\dd x^\nu+\dd z^2\Big).
\end{equation}
In this coordinate system,
the boundary and the horizon correspond to $z=0$ and $z=\infty$,
respectively.

\end{document}